\documentclass[twocolumn,floats,eqsecnum,pra,showpacs]{revtex4-1}

\usepackage{amsmath}
\usepackage{amsfonts}
\usepackage{graphicx}

\usepackage{tikz,pgf}
\usetikzlibrary{decorations,arrows,shapes}
\usetikzlibrary{calc}

\newcommand {\be}{\begin{equation}}
\newcommand {\ee} {\end{equation}}
\newcommand {\bea}{\begin{eqnarray}}
\newcommand {\eea} {\end{eqnarray}}

\newcommand{\ket}[1]{\left| #1 \right>}
\newcommand{\bra}[1]{\left< #1 \right|}

\newcommand{\hilb}{\mathcal{H}}

\begin{document}


\title{Edge state inner products and real-space entanglement spectrum \\of trial quantum Hall states}
\author{J. Dubail$^1$, N. Read$^1$, and E.H. Rezayi$^2$}
\affiliation{$^1$ Department of Physics, Yale
University, P.O. Box 208120, New Haven, Connecticut 06520-8120, USA\\
$^2$ Department of Physics, California State University, Los Angeles, California 90032, USA}

\date{July 30, 2012; Revised: November 9, 2012}

\begin{abstract}
We consider the trial wavefunctions for the fractional quantum Hall effect that are given by conformal blocks, and construct their associated edge excited states in full generality. The inner products between these edge states are computed in the thermodynamic limit, assuming generalized screening ({\it i.e.} short-range correlations only) inside the quantum Hall droplet, and using the language of boundary conformal field theory (boundary CFT). These inner products take universal values in this limit: they are equal to the corresponding inner products in the bulk 2d chiral CFT which underlies the trial wavefunction. This is a bulk/edge correspondence; it shows the equality between equal-time correlators along the edge and the correlators of the bulk CFT up to a Wick rotation. 

This approach is then used to analyze the entanglement spectrum of the ground state obtained with a bipartition $A \cup B$ in real-space. Starting from our universal result for inner products in the thermodynamic limit, we tackle corrections to scaling using standard field-theoretic and renormalization group arguments. We prove that generalized screening implies that the entanglement Hamiltonian $H_E = -\log  \rho_A$ is isospectral to an operator that is local along the cut between $A$ and $B$. We also show that a similar analysis can be carried out for particle partition.
We discuss the close analogy between the formalism of trial wavefunctions given by conformal blocks and Tensor Product States, for which results analogous to ours have appeared recently. Finally, the edge theory and entanglement spectrum of $p_x \pm i p_y$ paired superfluids are treated in a similar fashion in the appendix.
\end{abstract}


\maketitle


\section{Introduction}

\subsection{Motivation}

The Fractional Quantum Hall Effect (FQHE) is an archetype of strongly interacting many-body electronic systems. As the filling fraction is varied, various fully gapped phases of matter can be observed experimentally. These topological phases of matter \cite{toporder} exhibit spectacular collective behavior, such as localized excitations with quantized fractional charge,  abelian (and possibly non-abelian) fractional statistics, or protected gapless edge modes. Our understanding of the FQHE mostly relies on the ``variational'' approach pioneered by Laughlin \cite{Laughlin}. His celebrated wavefunction is not the ground state of a physically realistic Hamiltonian (say, electrons in a magnetic field with Coulomb repulsion). Yet, it describes accurately the FQHE at some specific filling fractions and it is commonly accepted that it belongs to the same topological phase as the real physical system.

Since Laughlin's contribution, several types of trial wavefunctions have been proposed for the FQHE at various filling fractions. These include for example hierarchical states \cite{HaldaneHierarchy}, composite fermion wavefunctions \cite{CF}, or the important family of states given by conformal blocks introduced by Moore and Read (MR) \cite{MR} (states corresponding to Jack polynomials have also been proposed later \cite{HaldaneBernevig}, they can actually be included in the latter family \cite{EstienneSantachiara}). In this paper we focus on these MR trial wavefunctions and their {\it quantum entanglement} properties.

The idea that quantum entanglement \cite{nielsenchuang} can help characterize different topological phases of matter has emerged over the past years. This approach has provided valuable new insights, while traditional methods based on symmetry breaking and local order parameters are widely accepted to fail \cite{toporder}. For instance, the topological entanglement entropy \cite{Zanardi,KitaevPreskill,LevinWen} of the ground state of a fully gapped Hamiltonian is one robust measure of quantum entanglement in a topological phase in two dimensions. More precisely, a bipartition of the quantum system is defined when the Hilbert space factors into two parts $\hilb = \hilb_A \otimes \hilb_B$. For most physical systems, a natural bipartition is given by a cut in position space in continuous systems or between a row or a plane of sites in lattice models. The object of interest is then the reduced density matrix $\rho_A = {\rm tr}_{\hilb_B} \left| \psi \right> \left< \psi \right|$. The von Neumann entropy of $\rho_A$ scales linearly with the area of the cut between parts $A$ and $B$ \cite{arealaw}, and it contains a universal order $O(1)$ piece: the topological entanglement entropy. At the end of Ref. \cite{KitaevPreskill}, Kitaev and Preskill (KP) pointed out that the topological entanglement entropy arises naturally if one assumes that the reduced density matrix $\rho_A$ has the form of the thermal density matrix of a $1+1$-dimensional gapless chiral theory along the cut.

The spectrum of the ``entanglement Hamiltonian'' $H_E = - \log \rho_A$ is called the entanglement spectrum (ES). The eigenvalues of $H_E$ are called pseudoenergies. Li and Haldane (LH) studied the ES of quantum Hall systems numerically in Ref. \cite{HaldaneLi}, and observed that it contains a low-lying part in which the multiplicities are related in a universal way to the ones of the conformal field theory (CFT) describing the low-energy edge excitations. This low-lying universal part is usually well-separated from the rest of the ES by an entanglement gap (see also the extended discussion in \cite{Ronny}). LH suggested that this low-lying part could be used as a diagnostic tool when comparing two ground state wavefunctions. In addition, they observed that for some specific trial states, such as the Laughlin wavefunction, the entanglement gap goes to infinity, leaving only the low-lying universal part. Although the work of LH relies on a bipartition that is a cut in momentum space (the so-called ``orbital partition'' \cite{Schoutens1}), which is not local \cite{DRR}, their main observations remain valid for the more natural cut in real-space \cite{DRR, RSPBernevig, RSPSimon}. The real-space ES is expected to be the spectrum of a local field theory along the cut, in agreement with the insightful suggestion of KP: not only the multiplicities but also the eigenvalues of $H_E = -\log \rho_A$ are the ones of a (perturbed) CFT along the cut, when the length of the cut is large compared to the mean particle spacing (which plays the role of an UV cutoff). The conjectured locality of the ES, also dubbed ``scaling property'' in \cite{DRR}, is discussed in greater detail in section \ref{sec:intro_entanglement}.

The purpose of this paper is to provide an analytic framework that explains why these properties of the real-space ES hold for the large family of MR trial wavefunctions. It involves a general construction of the space of edge excitations, and a precise analysis of the inner product in that Hilbert space. As a byproduct, we will arrive at the important conclusion that, assuming generalized screening, {i.e.} short-range correlations only in the bulk (this will be discussed below), there is an isometric isomorphism (in the thermodynamic limit) between the Hilbert space of the gapless edge excitations and the Hilbert space of the CFT used to construct the ground state trial wavefunction. This result is a precise ``bulk/edge correspondence'', which, stated loosely, says that the edge CFT and the bulk CFT are the same up to a Wick rotation. In particular, this rules out the possibility of using non-unitary CFTs to construct FQHE trial wavefunctions, as those {\it cannot} be consistent with generalized screening. Some version of this correspondence has long been expected \cite{MR}, although no precise statement about a general relation between the inner products in the space of edge states and those in the CFT has ever appeared in the literature. In Ref. \cite{WenEdge}, Wen provided an important argument that implies such a relation in the particular case of the Laughlin wavefunction, relying on the plasma mapping \cite{Laughlin}; the formalism we develop in this paper is strongly inspired by his.
We will then extend our ideas to tackle the real-space entanglement of the ground state. Let us emphasize that we will work only with wavefunctions, and do not address any question related to (physical) Hamiltonians in this paper.

Previous steps towards an analytic understanding of the ES in quantum Hall systems include direct calculations for the integer quantum Hall effect \cite{IQHE_Klich,IQHE1,DRR} or other free fermion systems \cite{Fidkowski,Turner,DR} and rigorous results on the multiplicities for a large class of trial wavefunctions \cite{Maria,Benoit}. Some general arguments for a correspondence between the entanglement and edge spectra have been proposed previously. In \cite{QKL}, it is suggested to start from two pieces $A$ and $B$ of a topological phase which both support (counter-propagating) gapless chiral edge states, and then to glue them along the edge by switching on an interaction Hamiltonian that couples the systems $A$ and $B$. Restricting the analysis to a coupling between the edge states only (the possibility of exciting the system in the bulk is neglected), standard renormalization group (RG) arguments yield the thermal form of the reduced density matrix suggested by KP. Despite its simplicity, this ``cut-and-glue'' argument relies neither on a specific wavefunction nor on a precise choice of the bipartition of the Hilbert space, and is therefore intrinsically different from the approach we adopt in this paper. Another approach \cite{Swinglestuff} emphasizes geometric aspects and makes use of Lorentz invariance to obtain a general argument, which again is very different from our approach in this paper.

At the most fundamental level, the ultimate explanation for the entanglement-edge
correspondence should be
something like the following (part of which also appears as a small part of the argument
of Ref.\
\cite{Swinglestuff}): if the effective low-energy field theory of the topological phase is some
Chern-Simons gauge theory, then to obtain a reduced density matrix, the degrees of freedom
of a subregion that are traced out must be gauge-invariant. The reduced density matrix is
then
gauge invariant. It can be represented field-theoretically by a functional integral, which
clearly must involve the same Chern-Simons theory in the interior of the remaining region
$A$.
In order to be gauge invariant, boundary degrees of freedom are required \cite{witten},
and
the Hilbert space of these is the same as that of a physical edge (that is, the quantum
numbers and multiplicities agree). More generally, the appearance of the edge degrees of
freedom is necessitated by
gauge-invariance, or in other words to absorb the effects of ``anomalies'' (in the
field-theoretic sense)
in the bulk theory, just as in the case of a physical edge. The space of
low-(pseudo-)energy degrees of freedom
required to accomplish this is robust. The degeneracy of the pseudoenergies of these
states
is resolved only by subleading non-universal effects, that contain an ultraviolet cutoff.
These effects
produce an effective local entanglement Hamiltonian when the partition is carried out in
a local fashion in real space,
and this is precisely what we obtain in our analysis of trial quantum Hall wavefunctions.
In our work,
the use of trial wavefunctions that are conformal blocks takes the place of the gauge
theory, and the
role of ultraviolet cutoff is played by the particle spacing.

\subsection{Fractional quantum Hall effect and \\the lowest Landau level}
\label{sec:FQHE}

Let us first recall some standard notations for the many-body problem in the lowest Landau level (LLL). One considers $N$ indistiguishable (spinless) charged particles in a two dimensional
surface pierced by a normal and uniform magnetic field. In this paper, this surface
is either the plane or the sphere $S^2$ \cite{HaldaneHierarchy}. In both cases, we use complex coordinates to parametrize the surface: the plane is simply parametrized by $z = x+i y \in \mathbb{C}$, while for the sphere of radius $R_{S^2}$, we use the stereographic coordinate $z = 2R_{S^2} \, e^{-i \phi} \, \tan \theta/2$, where $(\theta,\phi)$ are the spherical polar coordinates.

It is well-known that wavefunctions in the lowest Landau level (LLL) correspond to analytic functions \cite{Laughlin}. We therefore consider wavefunctions
\begin{equation}
	\Psi(z_1,\dots,z_N)
\end{equation}
which are analytic in $z_1,\dots,z_N$, and satisfy the right statistics (either bosonic or fermionic) under the exchange of the $z_i$'s. On top of these requirements, $\Psi(z_1,\dots,z_N)$ must be normalizable:
\begin{equation}
	\label{eq:normalizable}
	\frac{1}{N!} \int_{\mathbb{C}^N}\, \prod_{i=1}^N e^{V(z_i,\bar{z}_i)} d^2 z_i \, \left| \Psi(z_1,\dots,z_N) \right|^2  \, <\, +\infty \, .
\end{equation}
The measure $e^{V(z_i,\bar{z}_i)} d^2 z_i$ depends on the surface which we are considering. It can be computed directly by solving the Landau problem for a single particle (see \cite{HaldaneHierarchy} for the spherical case). The notation $e^{V(z_i,\bar{z}_i)}$ comes from the fact that $V(z_i,\bar{z}_i)$ can be viewed, in the plasma mapping \cite{Laughlin}, as an electrostatic potential created by a background charge (we will come back to this point in part \ref{sec:screening}):
\begin{equation}
	\label{eq:potential}
	V(z,\bar{z}) = \left\{  \begin{array}{lcl}  -(N_\Phi+2) \log \left[1+|z|^2/(2 R_{S^2})^2 \right]  & & {\rm (sphere)} \\ \\
-|z|^2/2\ell_B^2  &&  {\rm (plane)} \end{array}  \right.
\end{equation} 
where $\ell_B$ is the magnetic length in the plane, and $N_\Phi$ is the number of
magnetic fluxes which pierce the sphere.

The Hilbert space corresponding to the LLL is finite dimensional in the sphere, but not in the plane. Indeed, for a single particle, a
basis of {\it normalizable} analytic functions $\Psi(z)$ is provided by the monomials $z^m$, where $m \geq 0$ can be any integer for the plane, while $m \in \left\{0,1,\dots,N_\Phi \right\}$ for the sphere. Keeping that remark in mind, in the rest of this paper, our notations allow us to treat the plane
and the sphere on equal footing.

The sphere and the plane both enjoy rotational invariance. The angular momentum in the plane
is written $L_{\hat{z}}$. Note that $z$ is the complex coordinate in the plane while $\hat{z}$ stands for a unit vector perpendicular to it. The basis of monomials $z^m$ are
eigenstates of $L_{\hat{z}}$
\begin{equation}
	L_{\hat{z}}\cdot z^m \;=\; m \, z^m \, .
\end{equation}
Meanwhile, with these conventions, the angular momentum $L_{\hat{z}}^{S^2}$ around the vertical
axis of the sphere (also written $\hat{z}$) acts on the monomials as
\begin{equation}
	L^{S^2}_{\hat{z}}\cdot z^m \;=\; \left( N_\Phi/2-m \right) \, z^m
\end{equation}
so the angular momentum on the sphere can always be related to the one in the plane. In particular, for $N$ particles $L_{\hat{z}}^{S^2} = N N_\Phi/2- L_{\hat{z}}$.

\subsection{Moore-Read construction}

\label{sec:blocks}
We now recall how an Ansatz for the wavefunction $\Psi(z_1,\dots,z_N)$ can be obtained by looking at conformal blocks in certain 2d chiral CFTs \cite{MR}. For a recent discussion of the MR construction, and its implications for non-abelian statistics, see \cite{Read2009}. For basic CFT material, see \cite{BYB,SaleurItzyksonZuber,Mussardo}. Let us consider the wavefunction
\begin{equation}
	\label{eq:wfblock}
	\Psi(z_1,\dots,z_N) \; = \; \frac{1}{\sqrt{Z_N}} \left< N \right| \prod_{i=1}^N a(z_i) \left|0\right>
\end{equation}
where $a(z)$ is a local operator ({\it i.e.} a primary field) in a given chiral CFT, acting on the vacuum $\left| 0 \right>$. The charged vacuum $\left< N \right|$ will be defined precisely below; it carries a charge that is opposite to the total charge of $\prod a(z_i)$, in order for the correlator (\ref{eq:wfblock}) to be non-zero. The chiral CFT and the field $a(z)$ must be chosen such that $\Psi(z_1,\dots,z_N)$ is single-valued: this implies that there must be a single fusion channel when one fuses the field $a(z)$ with itself. The field $a(z)$ can therefore be called ${\it abelian}$. Of course, for consistency, the function $\Psi(z_1,\dots,z_N)$ must also be analytic and have the correct statistics, which requires additional properties for the field $a(z)$. For later convenience, the factor $Z_N$ is defined such that the wavefunction $\Psi(z_1,\dots,z_N)$ given by the formula (\ref{eq:wfblock}) is normalized for the norm (\ref{eq:normalizable}).

In the MR construction, the chiral CFT is chosen as the (tensor) product of two sectors ${\rm CFT}_{U(1)} \otimes {\rm CFT}_{\psi}$, and the field $a(z)$ is a product of a vertex operator in the $U(1)$ {\it charge sector} and an abelian field $\psi(z)$ in the {\it statistics sector} ${\rm CFT}_{\psi}$:
\begin{equation}
	a(z) \, = \, e^{i \varphi(z)/\sqrt{\nu}} \times \psi(z) \; .
\end{equation}
In this expression, $\varphi(z)$ is a free chiral boson with propagator $\left< \varphi(z) \varphi(w) \right> \,=\, - \log(z-w)$, and normal ordering is implicitly assumed in the exponential. The $U(1)$ symmetry is generated by the transformations $\varphi(z) \mapsto \varphi(z) + {\rm const}$. The correlator of the vertex operators $\left<N\right| \prod e^{i \varphi(z_i)/\sqrt{\nu}} \left|0\right>$ has to be invariant under these shifts, so the out vacuum $\left<N \right|$ must carry a $U(1)$-charge proportional to $N$
\begin{equation}
	\label{eq:outvacuum}
	\left<N \right| \, = \,\lim_{z \rightarrow \infty} \frac{1}{z^{N^2/\nu}} \left< 0 \right| e^{-i N\varphi(z)/\sqrt{\nu}} \; .
\end{equation}
This definition is standard in radial quantization of a CFT in the plane (see \cite{BYB,SaleurItzyksonZuber,Mussardo}). 
With this out vacuum $\left< N\right|$, the correlator of the vertex operators is non-zero. It is equal to the Laughlin-Jastrow factor $\prod_{i=1}^N (z_i-z_j)^{1/\nu}$, leading to a trial state with filling fraction $\nu$. The correlator in the statistics sector $\left< \prod \psi(z_i) \right>$ depends on the choice of the CFT for the field $\psi(z)$. For example, the Laughlin wavefunction itself corresponds to the simplest case when $\psi(z)$ is the identity operator. Other possible choices of $\psi(z)$ include: a free fermion with propagator $1/(z_i-z_j)$, leading to the MR (Pfaffian) wavefunction \cite{MR} or minimal Fateev-Zamolodchikov parafermions \cite{FateevZamo} which give the Read-Rezayi (RR) series \cite{RR}. Other choices of the field $\psi(z)$ correspond to other wavefunctions which have appeared in the literature, for example the ones expressed in terms of Jack polynomials \cite{HaldaneBernevig}. 

Like the $U(1)$ charge sector, the statistics sector is associated with some underlying symmetry, for example a $\mathbb{Z}_2$ symmetry in the case of a Majorana field generated by $\psi(z) \mapsto - \psi(z)$ (and more generally, a $\mathbb{Z}_k$ symmetry for parafermions). In that case, our definition of the out vacuum (\ref{eq:outvacuum}) must be completed by the insertion of a $\mathbb{Z}_2$-charge ($\mathbb{Z}_k$-charge) when $N$ is odd ($N \neq 0 {\, \rm mod \,} k$), in order for the correlator $\left<N \right| \prod a(z_i) \left| 0\right>$ to be non-zero. The CFT for the statistics sector is always a rational one (there is a finite number of primary fields, which form a closed algebra under the operator product expansion).

\subsection{Entanglement of the trial states}
\label{sec:intro_entanglement}

Because of their particular structure, the trial states given by conformal blocks have some very specific entanglement properties, which are inherited from the underlying CFT. In order to sketch some of these features, we first need to define the bipartition of the Hilbert space $\hilb = \hilb_A \otimes \hilb_B$ for which one computes the reduced density matrix $\rho_A$, or alternatively the Schmidt decomposition
\begin{equation}
	\label{eq:Schmidt}
	\left| \left. \Psi \right>\right> \, = \, \sum_{i} e^{-\xi_i/2} \left|\left. \Psi_{i}^A \right>\right> \otimes \left| \left. \Psi^B_{i} \right> \right>,
\end{equation}
The ES depends on the bipartition and is, by definition, the set of pseudoenergies $\xi_i$'s \cite{HaldaneLi}. Throughout our paper we will use the notation with double rightangle for kets in the $2+1$ physical Hilbert space, while simple rightangles are reserved for states in the CFT. The ground state $\left|\left. \Psi \right>\right> $ corresponds to the wavefunction (\ref{eq:wfblock}).

\subsubsection{Real-Space Partition (RSP)}

A natural way to partition a system of itinerant particles on some manifold $M$ is to do a cut in position space. The single-particle Hilbert space $\hilb_1$ is the $L^2$ space of all normalizable functions on $M$ (in this section we do not require any analyticity condition). If we divide the manifold as $M = M_A \cup M_B$, then $\hilb_1$ admits a decomposition as a direct sum $\hilb_1 = \hilb_{1A} \oplus \hilb_{1B}$, where $\hilb_{1A}$ ($\hilb_{1B}$) is the subspace of functions with support in $M_A$ ($M_B$). This simply means that any function $f({\bf r})$ can be written as $f^A({\bf r})+f^B({\bf r})$, with $f^A({\bf r})=0$ ($f^B({\bf r})=0$) if ${\bf r} \notin M_A$ (${\bf r} \notin M_B$). This decomposition induces a corresponding bipartition of the $N$-particle space as
\begin{equation}
	\hilb_N \, = \, \bigoplus_{N_A=0}^N \hilb_{N_A,A} \otimes \hilb_{N_B,B} ,
\end{equation}
where $N_A$ ($N_B$) is the number of particles in part $A$ ($B$). This bipartition is called real-space partition (RSP).

In quantum Hall systems the RSP \cite{DRR,RSPBernevig,RSPSimon} is obtained by dividing the complex plane $\mathbb{C}$ where the coordinates $z_i$ are defined (see section \ref{sec:FQHE}) into two complementary parts $\mathbb{C} = A \cup B$. As is customary in the literature, the bipartition of the plane $\mathbb{C}=A\cup B$ is usually taken such that the subsystem $A$ is rotationally invariant (and simply-connected for simplicity): $A$ is then a disc of radius $R$ centered on the origin. The trial state $\left|\left.\Psi \right>\right>$ is usually an eigenstate of the angular momentum $(L^A_{\hat{z}} + L^B_{\hat{z}}) \left|\left. \Psi \right>\right> = L_{\hat{z}}\left|\left. \Psi \right>\right>$, so the angular momentum of part $A$, noted $L^A_{\hat{z}}$, is a good quantum number. The bipartition of the Hilbert space $\hilb_N$ with $N$ particles thus takes the form
\begin{equation}
	 \hilb_N \, = \, \bigoplus_{N_A=0}^N \bigoplus_{L_{\hat{z}}^A}  \hilb_{N_A, L_{\hat{z}}^A} \otimes  \hilb_{N_B, L_{\hat{z}}^B},
\end{equation}
with $N_A+N_B=N$, and the different eigenvectors/eigenvalues in the Schmidt decomposition (\ref{eq:Schmidt}) can be classified according to the number of particles $N_A$ and to the angular momentum $L_{\hat{z}}^A$.

In general, for RSP there is a non-degenerate lowest pseudoenergy $\xi(N_{A0},L_{\hat{z}0}^A)$ at some values $N_{A0}$ and $L_{\hat{z}0}^A$ which depend on the system size $N$. We define $\Delta \xi$, $\Delta N_A$ and $\Delta L_{\hat{z}}^A$ by subtracting off these values.

\subsubsection{Scaling property of the entanglement spectrum}

One can make a general conjecture, that is expected to hold for any ground state wavefunction in a $2+1d$ topological phase, provided that the (connected) correlation functions of local operators evaluated in that ground state are all short-range: {\it in each topological sector (i.e. for a fixed total anyonic charge in part $A$), the entanglement Hamiltonian $H_E = -\log \rho_A$ is isospectral to a ``pseudo-Hamiltonian'' that is local along the cut between $A$ and $B$}. Equivalently, in \cite{DRR}, this conjecture, dubbed ``scaling property'', was stated as follows: {\it for all $\Delta N_A$ and $\Delta L_{\hat{z}}^A$, as $N \rightarrow \infty$, the set of $\Delta \xi$'s approach the energy levels of a ``pseudo-Hamiltonian'' that is the integral of a sum of local operators in a $1+1 d$ theory defined along the cut between $A$ and $B$}. In particular, in phases of matter such as the FQHE states, there is a low-lying part (as observed by LH) that corresponds to a gapless sector in the $1+1d$ theory. In general, the theory also has gapped excitations, which come with pseudoenergies larger than or equal to the entanglement gap. These gapped excitations are associated with Schmidt eigenstates that differ from the ``cut ground state'' ({\it i.e.} the Schmidt eigenstate $\left|\left. \Psi^A_{N_{A0},L_{\hat{z}0}^A} \right>  \right>$ that has the smallest pseudoenergy), not only along the cut, but also far from the cut. Such states contribute with amplitudes that decay rapidly (exponentially) with the distance to the cut, hence the presence of an entanglement gap. The pseudoenergy levels that lie above the entanglement gap correspond to a mixture of gapped and gapless excitations. We note that for quantum Hall systems, this type of scenario has been discussed in the case of the orbital partition in \cite{MixtureES}.

 For some trial FQHE states, such as the Laughlin or MR (Pfaffian) states, the entanglement gap goes to infinity, leaving only the gapless low-energy part. One of the purpose of this paper is to explain why these wavefunctions, and more generally all the wavefunctions given by conformal blocks, exhibit this particular feature. Then we will also explain why, for these trial wavefunctions, the {\it locality} of the ES follows from the fact that all correlations are short-range, a property which is sometimes called {\it generalized screening} in reference to Laughlin's plasma mapping, as in \cite{Read2009}. We will find that, for the trial wavefunctions given by the MR construction, the ES is the spectrum of the Hamiltonian of a perturbed $1+1d$ CFT, and that this CFT is the one that underlies the ground state wavefunction. It is also the theory that describes the physical edge excitations, as we will show. In the simplest case where the CFT is perturbed only by irrelevant operators, the pseudoenergies are given by the eigenvalues of
\begin{equation}
	\label{eq:ES_explicit}
	\Delta \xi \, = \, \frac{v}{R} L_0 + \dots
\end{equation}
where $L_0 = \Delta L_{\hat{z}}^A + O(\Delta N_A)$ is the Virasoro generator of dilatations and rotations for the CFT in the plane. $R$ is the radius of part $A$ (for a circular region with perimeter $2\pi R$) and $v$ is some non-universal ``velocity''. This velocity has the dimension of a length, and is of the order of the mean particle spacing close to the cut, which is the natural UV cutoff in our problem. Thus the ratio $v/R$ in our scaling is always of order $1/\sqrt{N_A}$. The ellipses in (\ref{eq:ES_explicit}) are terms of higher order in $v/R \sim 1/\sqrt{N_A}$, which come from perturbing operators that are more irrelevant. The precise dependence of the eigenvalues of $L_0$ on $\Delta N_A$ depends on the details of the CFT. For the Laughlin wavefunction at filling fraction $\nu$, one has $L_0 = \Delta L_{\hat{z}}^A + \frac{(\Delta N_A)^2}{2\nu}$. More generally, the ES has to be discussed case-by-case, depending on what FQHE state one is dealing with. Depending on the CFT that underlies the state, some perturbing operators may or may not be present in the ``pseudo-Hamiltonian'' that gives the ES. It is also worth emphasizing that, just like for the physical energy spectrum in the presence of a real edge, the wavefunctions in the MR construction (with both the $U(1)$ charge sector and a non-trivial statistics sector) usually have an ES with {\it two} branches of excitations (rather than one), with different velocities $v_{U(1)}$ and $v_\psi$. More details about the perturbing operators are given in part \ref{sec:ES}, where we derive our main results on the ES.

Another important aspect that we want to point out in this paper is the striking similarity between these particular trial wavefunctions and other wavefunctions that are being used in condensed matter and in quantum information: the Matrix and Tensor Product States (MPS and TPS). In that context, ground state entanglement properties have long been studied, with questions that are analogous to the ones that we tackle here. We will come back to this point in part \ref{sec:MPS}.

\subsubsection{Particle Partition (PP)}

Another bipartition is natural for systems of identical itinerant particles: the {\it particle partition} (PP) \cite{PPSchoutens}. It is obtained by assigning a fictitious ``pseudospin'' degree of freedom for each particle. We label $\left|\left. A\right>\right>$ or $\left|\left. B\right>\right>$ the two orthogonal pseudospin states. A spinless ground state is mapped into the larger space that includes pseudospin by assigning to each particle $j$ the pseudospin state $\left( \left|\left.A \right>\right>_j + \left|\left. B\right>\right>_j \right)/\sqrt{2}$. The bipartition of this larger space is simply $\hilb_A \otimes \hilb_B$, where all the particles with pseudospin $\left|\left.A \right>\right>_j$ ($\left|\left.B\right>\right>_j$) constitute part $A$ ($B$). This definition of PP \cite{DRR}, which includes different particle numbers, is an extension of the original one \cite{PPSchoutens}.

In PP, rotational invariance is inherited from the invariance of the full system. In particular, on the sphere $S^2$ the total angular momentum $(L^A)^2$ is also a good quantum number. In that case, the Schmidt eigenvalues/eigenvectors can be organized in $SO(3)$ multiplets. It can be shown easily that the Schmidt rank is the same in each $L_{\hat{z}}^A$ subsector both for RSP and PP \cite{DRR,RSPBernevig}.
For PP, one can define $L^A_0$ as the maximum total angular momentum for the subsystem $A$, and $N_{A0}$ the corresponding number of particles. Then we can define the quantum numbers $\Delta N_A$ and $\Delta L^A$ (PP) by subtracting off these values.

Unlike the RSP, the PP is not a local bipartition, so there is no reason to expect that the ES should be the spectrum of a local operator. Overall, if one looks at the whole spectrum, the PP is not local. However, for large $L_{\hat{z}}^A$, we will see that the Schmidt eigenstates $\left|\left. \Psi^A_{i} \right>\right>$ actually correspond to the ground state configuration of $N_A$ particles localized in a circular cap centered on the north pole (and $N_B$ particles in a cap centered on the south pole), and of edge excitations above this ground state. In that sense, and for $N_A / N \rightarrow 1/2$ when $N,N_A \rightarrow \infty$, the part $A$ ($B$) corresponds to the northern (southern) hemisphere. Thus, for these states, parts $A$ and $B$ may roughly be seen as spatial region, just like in the case of RSP (this limit is also considered in \cite{Maria}). This observation can be made precise, and it has the consequence that the {\it particle ES}---the ES with PP---at large $L_{\hat{z}}^A$ can be analyzed with the same techniques as for RSP. It leads to a similar result, namely that this part of the particle ES actually corresponds to the spectrum of a local operator along the ``cut'' (the equator). Of course, as we highlighted in the last paragraph, the $SO(3)$ rotational invariance of PP implies that the pseudoenergies depend on $\Delta L^A$ rather than on $\Delta L^A_{\hat{z}}$, so this local operator needs to be different form (\ref{eq:ES_explicit}), which is valid only for RSP. The particle ES will be discussed in greater detail in section \ref{sec:ES_PP}.

\subsection{Structure of the paper}

Our paper is organized as follows. In section \ref{sec:MRmore}, we define some complementary notations for the trial wavefunctions given by conformal blocks, and in section \ref{sec:edge} we argue that there is a natural and straightforward way of constructing the edge states which correspond to those. More precisely, we exhibit a linear mapping from the space of states in the CFT to the space of wavefunctions for the edge states. In part \ref{sec:screening}, we give a detailed discussion of the screening property, and relate it to a boundary conformal field theory formalism. As a consequence, we derive one of the main result of this paper (section \ref{sec:edge_conjecture}), which states that, assuming screening, the quantum-mechanical inner products between the edge states are identical (in the thermodynamic limit) to the inner products in the conformal field theory. In other words, the linear mapping from section \ref{sec:edge} becomes an isometric isomorphism when the number of particles goes to infinity. We provide some numerical checks of this result in section \ref{sec:numerical}. For a finite number of particles, the corrections to scaling for the inner product between the edge states can be tackled with RG arguments, in the framework of perturbed boundary CFT. This is discussed in section \ref{sec:corrections_overlaps}. Then, in part \ref{sec:ES}, we relate our results for the edge states to the entanglement spectrum, and apply them to prove the ``scaling property'' conjectured in \cite{DRR}. We give a detailed scaling analysis of the different contributions that can appear in the ``pseudo-Hamiltonian'', therefore leading to precise predictions for the ES of the trial states given by conformal blocks. The Laughlin state is treated in full details as an example, for RSP and for PP. Finally, in part \ref{sec:MPS}, we emphasize the close relation between the wavefunctions given by conformal blocks and Tensor Product States, and discuss the relevance of our results in that context.

\section{Trial wavefunctions for the edge excitations}

\label{sec:part_edge}

\subsection{More structure for the CFT space}
\label{sec:MRmore}

\subsubsection{The chiral algebra}
In general, there is a field $a^\dagger(z)$ in the CFT such that the operator product expansion
of $a^\dagger$ with $a$ is
\begin{equation}
	a(z_1) a^\dagger(z_2) \, \underset{z_1 \rightarrow z_2}{\sim} \, \frac{1}{(z_1-z_2)^{2 h_a}} \, + \, \dots
\end{equation}
Here we have introduced the conformal dimension $h_a$ of the field $a(z)$. Since the field $a(z)$ is chiral, $h_a$ is also its spin. In the quantum Hall literature, the conformal dimension $h_a$ is sometimes refered to as the {\it spin per particle} \cite{Read2009}. It is also related to the ``shift'' on the sphere \cite{HaldaneHierarchy} in a straightforward way: $\mathcal{S} = 2 h_a$ \cite{Read2009}. It is the sum of the conformal dimensions of the vertex operator and of the field $\psi$ (noted $h_\psi$)
\begin{equation}
	h_a\, =  \, \frac{1/\nu}{2} + h_\psi \, .
\end{equation}
The field $a^\dagger(z)$ must carry a $U(1)$ charge which is opposite to the one of $a(z)$, and similarly in the statistics sector. For example in the case of the $\mathbb{Z}_k$ parafermions, the field $a(z) = e^{i \varphi(z)/\sqrt{\nu}} \otimes \psi_1(z)$ carries a $\mathbb{Z}_k$ charge $1$ (mod $k$), so $a^\dagger$ must carry a $\mathbb{Z}_k$ charge $k-1$ (mod $k$). We have then
\begin{equation}
	a^\dagger (z) \, = \, e^{- i \varphi(z)/\sqrt{\nu}} \times \psi^\dagger (z)
\end{equation}
where $\psi(z_1) \psi^\dagger(z_2) \sim 1/(z_1-z_2)^{2 h_\psi} + \dots$ The fields $a(z)$ and $a^\dagger(z)$ generate the chiral algebra $\mathcal{A}$ by the operator product expansion.

\subsubsection{Complex conjugation}
In this paper, we need to work with wavefunctions given by conformal correlators, such as ($\ref{eq:wfblock}$), but also with their complex conjugate. For this purpose, it is useful to introduce an anti-chiral copy of the CFT, and an anti-chiral field
\begin{equation}
	\bar{a}(\bar{z}) \, = \, e^{i \overline{\varphi}(\overline{z})/\sqrt{\nu}} \times \overline{\psi}(\overline{z})
\end{equation}
such that the complex conjugate of $\Psi(z_1,\dots,z_N)$ is given by the following correlator in the anti-chiral CFT
\begin{equation}
	[\Psi(z_1,\dots,z_N)]^* \, = \, \frac{1}{\sqrt{Z_N}}\overline{\left<N\right|} \prod_{i=1}^N \overline{a}(\overline{z_i}) \overline{\left|0\right>} \; .
\end{equation}
The field $\overline{a}(\overline{z})$, together with its conjugate $\overline{a}^\dagger(\overline{z})$, generate a copy of the same chiral algebra $\mathcal{A}$.

\subsection{Edge excitations}
\label{sec:edge}
In this section we construct wavefunctions that resemble the
``ground state'' wavefunction (\ref{eq:wfblock}), but which we interpret as the
``edge excitations''. Note that we do not address physical Hamiltonians in this paper.
Instead, the wavefunctions for the edge excitations are required to have the
same short-range properties (cluster properties) as the ground state (\ref{eq:wfblock}), but they have different global properties, such as angular momenta.

\begin{figure}
	\centering
	\includegraphics[width=0.46\textwidth]{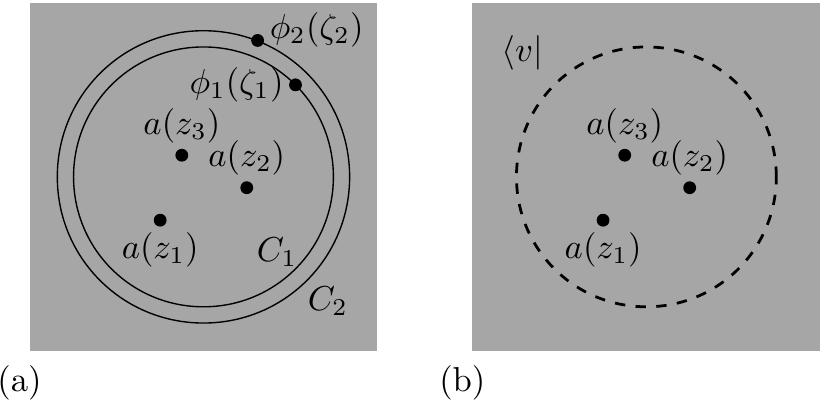}
	\caption{The edge state wavefunctions are obtained (a) by inserting contour integrals in the correlator as in formula (\ref{eq:contours}), or equivalently (b) by computing the matrix element (\ref{eq:edgestates1}) with an out state $\left< v \right|$ in radial quantization.}
	\label{fig:contour_integrals_edge}
\end{figure}

To construct these wavefunctions, we consider some set of fields $\phi_k(\zeta_k)$ ($k=1,\dots,p$) in the chiral algebra $\mathcal{A}$, and we compute the correlator
\begin{equation}
\label{eq:zetainsertion}
	\left< N'\right| \phi_p(\zeta_p) \dots \phi_1(\zeta_1) ~ \prod_{i=1}a(z_i) \left|0\right> 
\end{equation}
where the appropriate charged vacuum $\left< N' \right|$ is inserted (it must neutralize the total charge of the $a(z_i)$'s and the $\phi_k(\zeta_k)$'s). This correlator is a function of the $\zeta_k$'s and of the $z_i$'s. The short-distance properties as two or more of the $z_i$'s come close to each other are inherited from the underlying operator product expansions. Therefore, they must be the same as the ones of the ground state wavefunction (\ref{eq:wfblock}). This implies that the function (\ref{eq:zetainsertion}) is analytic in the $z_i$'s everywhere in the plane, except possibly at the points $\zeta_k$'s, where it can have some singularities. The function is always single valued, and the possible singularities at the $\zeta_k$'s are poles. In other words, the correlator (\ref{eq:zetainsertion}) is meromorphic, and it is not yet a valid wavefunction for particles in the LLL. However, let us consider instead the contour integrals
\begin{eqnarray}
\label{eq:contours}
\nonumber &&	\left<N' \right| \oint_{C_p} d\zeta_p ~\zeta_p^{m_p} \phi_p(\zeta_p) \dots \oint_{C_1} d\zeta_1 ~\zeta_1^{m_1} \phi_1(\zeta_1) ~ \prod_{i=1}a(z_i) \left|0\right> \\
&&
\end{eqnarray}
where the contours $C_1,\dots,C_p$ encircle all the $z_i$'s as shown in Fig.~\ref{fig:contour_integrals_edge}.a (and the contours are radially ordered: $C_1$ is encircled by $C_2$, and so on), and $m_k \in \mathbb{Z}$. In certain cases, the correlator (\ref{eq:contours}) can be zero (for instance, this may happen if some of the $m_k$'s are negative, when the correlator (\ref{eq:zetainsertion}) is analytic at $\zeta_k \rightarrow \infty$). The contours can now be deformed, without changing the value of expression (\ref{eq:contours}). In particular, they can be taken as circles with arbitrary large radii ({\it i.e.} they can be ``deformed around infinity''). Thus the expression (\ref{eq:contours}) is analytic in the $z_i$'s in the whole complex plane, just like the ground state (\ref{eq:wfblock}). If it is non-zero and normalizable for the norm (\ref{eq:normalizable}), it can be used as a FQHE trial wavefunction. Another advantage of using the contour integrals (\ref{eq:contours}) rather than the meromorphic functions (\ref{eq:zetainsertion}) is that one gets angular momentum eigenstates, which will be more convenient for the manipulations below.

If we introduce the Hilbert space of the CFT $\hilb_{\rm CFT}$, which is a (irreducible) module over the chiral algebra $\mathcal{A}$, then (\ref{eq:contours}) can be reformulated as follows. To each state $\left| v \right> \in \hilb_{\rm CFT}$, we associate its dual $\left< v \right| \in \hilb_{\rm CFT}^*$. Then we construct the correlator
\begin{equation}
	\label{eq:edgestates1}
	 \left<v \right| \prod_{i=1}^N a(z_i) \left|0 \right>
\end{equation}
which we use as a wavefunction when it is non-zero (these wavefunctions appeared previously in \cite{JacksonRS, FJM}). Thus, by definition, we have a linear mapping from the (dual) CFT Hilbert space $\hilb_{\rm CFT}^*$ to the space of edge states. This mapping is in general not injective.

We now consider two concrete examples, the Laughlin wavefunction and the MR (Pfaffian) wavefunction, to illustrate how this construction works in practice.

\subsubsection{First example: edge states for \\the Laughlin wavefunction}

For the Laughlin wavefunction, we have $a(z) = e^{i \varphi(z)/\sqrt{\nu}}$. There is no statistics sector. The chiral algebra $\mathcal{A}$ is generated by the vertex operators $e^{i \varphi(z)/\sqrt{\nu}}$ and $e^{-i \varphi(z)/\sqrt{\nu}}$ and by the operator product expansions. In particular, the $U(1)$ current $i\partial \varphi(z)$ is generated by $e^{i \varphi(z_1)/\sqrt{\nu}} e^{-i \varphi(z_2)/\sqrt{\nu}} \sim (z_1-z_2)^{-1/\nu} \left[1 + (z_1-z_2) i\partial \varphi(z_2)/\sqrt{\nu} +\dots \right]$. The modes of the $U(1)$ current $J_n = \frac{1}{2\pi i}\oint d\zeta ~\zeta^{n} i \partial \varphi(\zeta)$ satisfy the commutation relations
\begin{equation}
	\label{eq:U1KacMoody}
	\left[ J_n , J_m \right] = n~\delta_{n+m,0} \, .
\end{equation}
As claimed in section \ref{sec:blocks}, the ground state wavefunction (\ref{eq:wfblock}) is, up to the normalization factor,
\begin{equation}
	\left< N\right| \prod_{j=1}^N e^{i \varphi(z_j)/\sqrt{\nu}} \left|0\right> \, =\, \prod_{i<j} (z_i-z_j)^{1/\nu}  \; .
\end{equation}
The neutral edge states are obtained by exciting the out vacuum $\left<N \right|$ with the positive modes $J_n$ ($n\geq 0$):
\begin{eqnarray}
\nonumber 	&& \left< N \right| J_n ~ \prod_{j=1}^N e^{i \varphi(z_j)/\sqrt{\nu}} \left| 0 \right> \, = \, \frac{1}{\sqrt{\nu}} \sum_{k=1}^N z_k^n  \, \prod_{i<j} (z_i-z_j)^{1/\nu} \\  
\nonumber &&	\left< N \right| J_{n_2} J_{n_1} ~ \prod_{j=1}^N e^{i \varphi(z_j)/\sqrt{\nu}} \left| 0 \right> \, = \\ 
\nonumber && \qquad  \frac{1}{\sqrt{\nu}^2} \sum_{k=1}^N z_k^{n_1} \sum_{p=1}^N z_p^{n_2} \, \prod_{i<j} (z_i-z_j)^{1/\nu} \\ 
\nonumber &&	\left< N \right| J_{n_3} J_{n_2} J_{n_1} ~ \prod_{j=1}^N e^{i \varphi(z_j)/\sqrt{\nu}} \left| 0 \right> \, = \\ 
&& \qquad  \frac{1}{\sqrt{\nu}^3} \sum_{k=1}^N z_k^{n_1} \sum_{p=1}^N z_p^{n_2} \sum_{l=1}^N z_l^{n_3} \, \prod_{i<j} (z_i-z_j)^{1/\nu}  
\end{eqnarray}
and so on. The negative modes $J_n$ ($n<0$) do not need to be considered, since they annihilate the out vacuum $\left<N \right|$. In general, the positive mode $J_n$ of the $U(1)$ current $J(z)$ produces the corresponding sum of powers $\frac{1}{\sqrt{\nu}}\sum z_i^n$ \cite{JacksonRS,FJM}. These wavefunctions are the well-known (neutral) edge states for the Laughlin wavefunction \cite{WenEdge}. Equivalently, one could have obtained the same space of edge states (up to some change of basis) by acting with the modes $a_n = \frac{1}{2\pi i}\oint d\zeta ~ \zeta^{n-1+\frac{1}{2\nu}} e^{i \varphi(\zeta)/\sqrt{\nu}}$ and $(a^\dagger)_n=\frac{1}{2\pi i}\oint d\zeta~ \zeta^{n-1+\frac{1}{2\nu}} e^{-i \varphi(\zeta)/\sqrt{\nu}}$. The advantage of working with the modes of the $U(1)$ current here is that one gets a nice basis for the CFT space with the $U(1)$ charge corresponding to $N$ particles: $\left\{ \left|N \right>, J_{-1} \left|N\right>, J_{-2} \left|N\right>,J_{-1}^2 \left|N\right>,\dots\right\}$. In other words, the bosonic Fock space structure is transparent. It would be more painful to write such a basis in terms of states of the form $a_{-n_1}\dots a_{-n_p} (a^\dagger)_{-m_1} \dots (a^\dagger)_{-m_p}\left|N\right>$, although in principle nothing prevents us from doing that.

Finally, although the discussion in this section is about neutral excitations (the number of particles $N$ is identical to the one in the ground state), it is not difficult to extend it to the case of charged excitations. To get those, one needs to add some particles. For example, for a single particle added, the
correlator $\left< N+1 \right| \prod_{i=1}^{N+1} a(z_i)\left| 0\right>$ is simply the Laughlin wavefunction with $N+1$ particles.

\subsubsection{Second example: edge states for the Moore-Read (Pfaffian) state}
The second example we consider is the MR (Pfaffian) wavefunction, which corresponds to $a(z) = e^{i \varphi(z)/\sqrt{\nu}} \times \psi(z)$, where $\psi(z)$ is
a free (Majorana) fermion field with propagator $\left<\psi(z)\psi(w) \right> = 1/(z-w)$. For even particle number $N$, the ground state is then, up to normalization,
\begin{equation}
	\label{eq:MRPfaffian}
	\left< N\right| \prod_{i=1}^N a(z_i) \left|0 \right> \,= \, \prod_{i=1}^N \left( z_i-z_j\right)^{1/\nu} \, {\rm Pf} \left( \left\{ \frac{1}{z_k-z_l} \right\}_{1 \leq k,l\leq N} \right)  .
\end{equation}
In addition to the $J_n$'s appearing in the $U(1)$ sector as in the Laughlin case, the chiral algebra contains fermionic modes $\psi_n = \frac{1}{2\pi i}\oint d\zeta~\zeta^{n-\frac{1}{2}}~\psi(\zeta)$, where $n \in \mathbb{Z}+\frac{1}{2}$, with
\begin{equation}
	\{\psi_n , \psi_m \} \, = \, \delta_{n+m,0} \; .
\end{equation}
The neutral excitations obtained from the $U(1)$ sector generate edge states which are similar to the ones of the Laughlin wavefunction. In the Majorana sector we have instead (for $N$ even)
\begin{eqnarray}
	\label{eq:MRPfaffian_edge}
	&& \left< N \right| \psi_{n_1} \psi_{n_2} ~ \prod_{j=1}^N a(z_i) \left| 0 \right> \, = \,  \prod_{i<j} (z_i-z_j)^{1/\nu}  \; \times \\
\nonumber &&  {\rm Pf \,} \left(  \begin{array}{cccccc}
		0 & 0 & z_1^{n_1-\frac{1}{2}} & z_2^{n_1-\frac{1}{2}} & \dots & z_N^{n_1-\frac{1}{2}} \\
		0 & 0 & z_1^{n_2-\frac{1}{2}} & z_2^{n_2-\frac{1}{2}} & \dots & z_N^{n_2-\frac{1}{2}} \\
		-z_1^{n_1-\frac{1}{2}} & -z_1^{n_2-\frac{1}{2}} & 0 & \frac{1}{z_1-z_2} & \dots & \frac{1}{z_1-z_N} \\
		-z_2^{n_1-\frac{1}{2}} & -z_2^{n_2-\frac{1}{2}} &	\frac{-1}{z_1-z_2} & 0 & \ddots &  \\ 
		\vdots &\vdots & \vdots & \ddots & \ddots & \frac{1}{z_{N-1}-z_N} \\
		-z_N^{n_1-\frac{1}{2}}& -z_N^{n_2-\frac{1}{2}} & \frac{-1}{z_1-z_N} &  & \frac{-1}{z_{N-1}-z_N} & 0
	\end{array} \right) \\
\nonumber	&& \, = \,  \prod_{i<j} (z_i-z_j)^{1/\nu}  \; \times \\
\nonumber &&   \frac{1}{2^{\frac{N}{2}-1} (\frac{N}{2}-1)!} \sum_{\sigma \in S_N}\frac{{\rm sgn}\, \sigma  \, \times \,  z_{\sigma(1)}^{n_1-\frac{1}{2}} z_{\sigma(2)}^{n_2-\frac{1}{2}}}{(z_{\sigma(3)}-z_{\sigma(4)})\dots(z_{\sigma(N-1)}-z_{\sigma(N)})},  
\end{eqnarray}
where the Pfaffian of the $(N+2) \times (N+2)$ skew-symmatric matrix is the free-fermion correlator $\left<\psi_{n_1} \psi_{n_2} \psi(z_1) \dots \psi(z_N) \right>$ that must be evaluated using Wick's theorem. Of course, more fermion modes can be inserted in the correlator to generate other edge states, and one recovers the wavefunctions constructed in \cite{milr}. The MR wavefunction with odd particle number also fits naturally in that framework, by inserting one fermion mode $\psi_{1/2}$ in the out vacuum $\left< N \right|$, as discussed in section \ref{sec:blocks}.

\section{Screening}
\label{sec:screening}
It is well known that the normalization factor $Z_N$ of the Laughlin wavefunction is exactly equal to the partition function of a two-dimensional one-component plasma in a background potential $V(z,\overline{z})$,
\begin{equation}
	Z_N = \frac{1}{N!} \int_{\mathbb{C}} \prod_{i=1}^N d^2 z_i \, e^{\sum_j V(z_j,\bar{z}_j) + \nu^{-1} \sum_{k<l} \log |z_k-z_l|^2 } .
\end{equation}
This exact relation, usually refered to as ``plasma mapping'', has been exploited in various ways in the literature. A key point in the plasma mapping is the observation that for an inverse filling fraction $\nu^{-1}$ lower than about $70$, the plasma is in a screening phase. This property was highlighted already by Laughlin, who used it to derive the fractional charge of the quasi-particles. It was used later to show that the quasi-particles also obey (abelian) fractional statistics under adiabatic exchange \cite{SchriefferWilczek}. Wen used the screening property, coupled to an electrostatic argument (the method of images \cite{Jackson}) to construct the edge theory of the Laughlin states from a microscopic point of view \cite{WenEdge}.

In this part, we use a generalization of the screening property of Laughlin's plasma, discussed recently in full details in \cite{Read2009}. This ``generalized screening assumption''  is formulated as follows. The normalization factor of the wavefunction ($\ref{eq:wfblock}$) can be viewed as the partition function of a two-dimensional system of itinerant particles subject to some interactions and in a background electrostatic potential $V(z,\bar{z})$
\begin{equation}
	\label{eq:plasmaZN}
	Z_N \, =\, \frac{1}{N!} \int_{\mathbb{C}} \prod_{i=1}^N e^{V(z_i,\bar{z}_i)} d^2 z_i \, \left| \left< N\right| \prod_{j=1}^N a(z_j) \left|0\right> \right|^2  .
\end{equation}
Contrary to the case of the Laughlin wavefunction, in general this partition function is not the one of a Coulombic plasma ({\it i.e.} involving only Coulomb two-body interactions). Instead, it is argued in \cite{Read2009} (see also earlier ideas sketched in \cite{WilczekNayak}) that the partition function (\ref{eq:plasmaZN}) should in general be viewed as the one of a perturbed CFT (in a grand-canonical description, as we do in section \ref{sec:screening_bc}). Then, two situations may occur, depending on the IR fixed point towards which the perturbed CFT is sent under the RG flow: either (i) the IR theory is massive, that is, all the connected correlations of local fields decay exponentially, or (ii) the IR theory contains massless modes and therefore long-range (power-law decaying) connected correlations. We say that ``generalized screening'' holds if the situation (i) occurs. It generalizes the case of the screening phase for the Laughlin wavefunction, which contains only exponentially decaying connected correlations.

In general, there is no known analytical argument that allows to discriminate between the situations (i) and (ii). Instead, one usually has to rely on indirect numerical checks of some of the consequences of the generalized screening assumption (i). Recently, a plasma mapping has been succesfully constructed for the MR (Pfaffian) wavefunction \cite{GurarieNayak, BondersonGurarieNayak}, which opened the route to a direct numerical check of the screening hypothesis for this state \cite{checkMRplasma}. The property (i) is therefore strongly supported by numerical evidence for the MR (Pfaffian) wavefunction, and it is plausible that it holds also for other states like the $k=3$ RR state. Also, some general arguments have been given in \cite{Read2009} which show that generalized screening cannot hold in some cases, as it would lead to contradictions (in particular in the case of non-unitary CFTs).

In what follows, the generalized screening property (i), namely the property that bulk correlations are all short-range, is assumed to hold; our purpose it to explore some of its consequences. This part of our paper is devoted to reformulating the screening property in the language of boundary CFT, and to using it to make a precise statement of the long expected ``bulk/edge correspondence''. The arguments in this section may be viewed as the natural generalization of Wen's microscopic theory of the edge excitations, which in its original formulation was applicable only to the Laughlin state \cite{WenEdge}.

\begin{figure}[htbp]
	\centering
	(a)\includegraphics[width=0.20\textwidth]{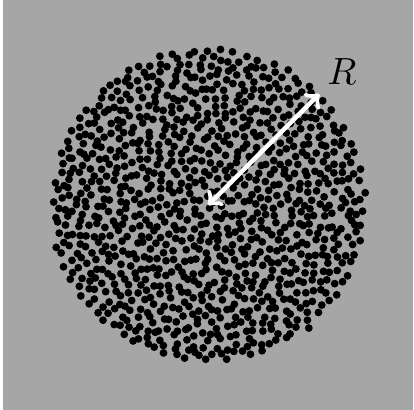}
	\; (b)\includegraphics[width=0.20\textwidth]{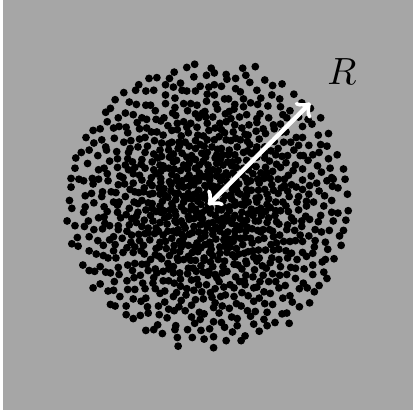}
	\caption{In the thermodynamic limit ($N\rightarrow \infty$), the particles fill a circular droplet of radius $R$. For the background potential (\ref{eq:potential}) corresponding to the plane, the density is uniform (a), for the sphere it is not uniform after the stereographic projection (b).}
	\label{fig:density}
\end{figure}

\subsection{The droplet}
Consider the distribution of particles corresponding to the partition function (\ref{eq:plasmaZN}) given by the normalization factor of the ground state wavefunction.
In the large $N$ limit, the particles fill some domain in the plane $\mathbb{C}$ (where the coordinates $z_i$'s are defined) called the ``droplet''. When the potential $V(z,\overline{z})$ is rotationally invariant, as in (\ref{eq:potential}), the droplet is circular and centered on the origin (see Fig. \ref{fig:density}). At large distances (that is, much larger than the mean particle spacing, which is of order of $\ell_B$ in the plane) the average density of particles is well approximated by a continuous function $ \sum_i \delta^{(2)}(z-z_i) \rightarrow \rho_0(z,\bar{z})$. It can be shown easily, for instance with a saddle-point approximation (a fully detailed derivation for the Laughlin case can be found in \cite{CappelliSaddlePoint}; it is easily extended to other cases), that the density of particles is fixed by the $U(1)$ charge sector only, and that it is equal (at scales $\gg \ell_B$) to the ``background charge density''
\begin{equation}
	\label{eq:background_charge}
	\rho_0(z,\bar{z}) \, = \, -\frac{\nu}{4\pi} \Delta V(z,\bar{z}), 
\end{equation}
where $\Delta = 4 \partial_z \partial_{\bar{z}}$ is the Laplacian. 
This background charge density is constant for the quadratic potential $V(z,\overline{z})$ corresponding to the plane in formula (\ref{eq:potential}). It is not constant for the sphere because the stereographic projection does not preserve the volume (Fig. \ref{fig:density}). Let us emphasize the fact that the relation (\ref{eq:background_charge}) has nothing to do with the screening property described in the previous section. In particular, it holds  in the crystallized phase for the Laughlin wavefunction, as well as in the screening phase, as long as the variations of the ``background charge'' $-\frac{\nu}{4\pi}\Delta V(z, \bar{z})$ occur on distances much larger than the mean particle spacing, such that a continuous description of the particle density is meaningful. In this paper, we will always be in the latter regime, where the the background potential $V(z,\bar{z})$ varies slowly. For the case of the plane in (\ref{eq:potential}), this is obviously true since the background charge does not vary at all, and for the sphere it varies on scales of order $\sim R_{S^2}$ (the radius of the sphere) while the mean particle spacing is of order $\sim R_{S^2}/\sqrt{N}$, so their ratio vanishes when $N \rightarrow \infty$.

Finally, for technical reasons, in the next sections we will often have to refer to the radius $R$ of the circular droplet for the rotationally invariant potentials (\ref{eq:potential}). The radius is fixed by the requirement that the number of particles inside the droplet is
\begin{equation}
	\label{eq:radius}
	N \, = \, \int_{|z|<R}  \rho_0(z,\bar{z}) \, d^2 z .
\end{equation}
In the case of the constant background charge for the plane (\ref{eq:potential}), this of course leads to the well-known radius $R=\sqrt{2 N/\nu} \ell_B$. On the sphere, the particles fill some spherical cap that is mapped onto the droplet by the stereographic projection, and the radius $R$ depends both on the radius of the sphere and on the size of the cap.

\subsection{Screening as a conformal boundary condition}
\label{sec:screening_bc}
In this section we assume that the generalized screening property holds, and we interpret its consequences at the edge of the droplet using the language of boundary conformal field theory (for a classical discussion of boundary CFT, see \cite{Cardy84,Cardy89}). We will have to make a certain number of technical choices in order to state our arguments. The technicalities, however, should not prevent the reader from catching the basic idea, which is simple. Let us summarize it here. We start from a non-chiral massless theory ${\rm CFT}\otimes \overline{{\rm CFT}}$ defined on a 2d surface. A subset of this surface, the ``droplet'', is filled with particles. In the field theory, these particles are equivalent to a perturbation of the form $\int a(z) \overline{a}(\overline{z}) d^2 z$ that is turned on inside the droplet (but not outside). This new term in the action drives the perturbed theory to a massive IR fixed point. That is the screening assumption. Now, outside the droplet, there is no perturbation, so the field theory ${\rm CFT} \otimes \overline{{\rm CFT}}$ is still massless. Inside te droplet, all the correlation functions decay exponentially, and the correlation length is zero at the IR fixed point. Thus, we are left with a non-chiral CFT {\it outside} the droplet, and the local fields in this theory must satisfy a local {\it boundary condition} along the boundary of the droplet. The purpose of this section is to find this boundary condition. Its consequences for the edge theory of the quantum Hall states will be discussed later. Now let us turn to a more detailed formulation of this argument.

\begin{figure}
	\centering
	\includegraphics[width=0.45\textwidth]{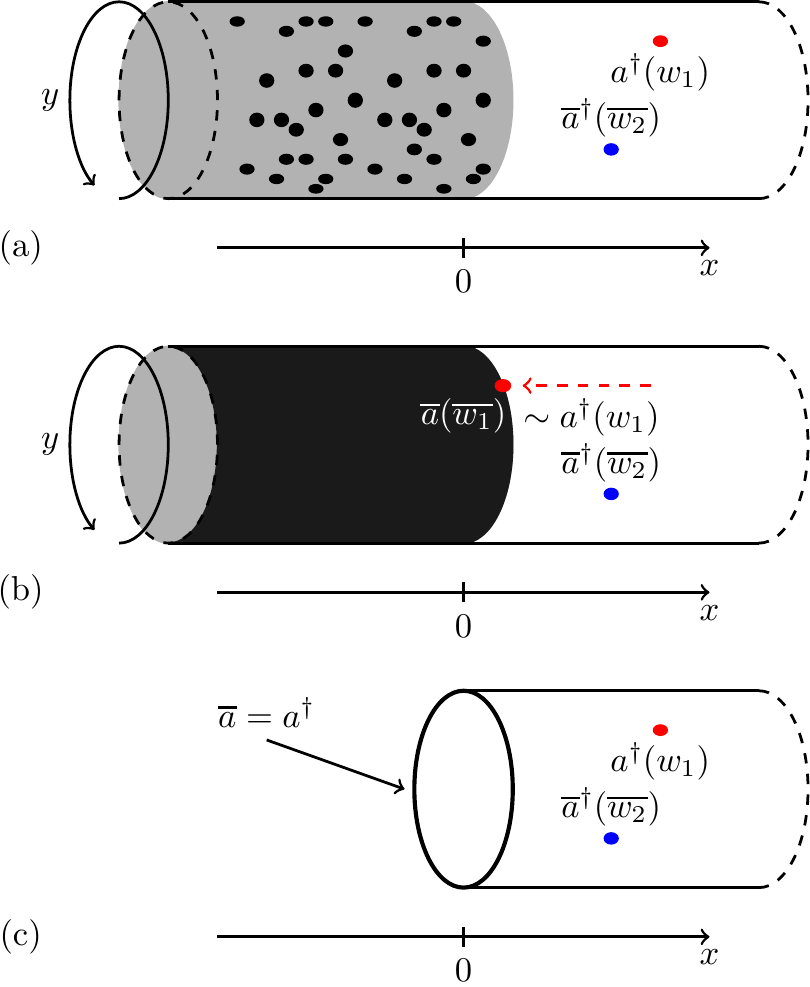}
	\caption{(color online) (a) Particles in a neutralizing backgroung on the left half-cylinder. We are interested in the correlation function of $a^\dagger(w_1)$ and $\overline{a}^\dagger(\overline{w_2})$ in the right part, in the presence of all the particles. (b) Assuming screening, and in the limit where the density of mobile charges in the left half-cylinder goes to infinity, an operator $a^\dagger(w_1)$ brought to the boundary from the outside is equivalent to an operator $\overline{a}(\overline{w_1})$. (c) In this limit, the correlation function of $a^\dagger(w_1)$ and $\overline{a}^\dagger(\overline{w_2})$ in the presence of all the mobile charges in the left is equivalent to the two-point correlation function in the right part, with the boundary condition $\overline{a}(x,y) = a^\dagger(x,y)$ at $x=0$.}
	\label{fig:cylinder}
\end{figure}

It is more convenient to work in the grand-canonical ensemble as in \cite{Read2009}. Also, to avoid phase factors and normalization constants that would obscure the argument, we work on the cylinder $\mathcal{C}$ (see Fig. \ref{fig:cylinder}) parametrized by $w = x+i y$ where $y$ is
identified with $y + L$. We imagine that the left half-cylinder $\mathcal{C}_l$ (${\rm Re\,}w < 0$) is filled by the particles in a uniform neutralizing background with, say, constant density $\rho_0$. The way to treat this background charge was discussed in \cite{MR}. Eventually, one can regularize this at $x \rightarrow -\infty$ by integrating over the region $-\Lambda <{\rm Re \,}w<0$, such that the total background charge is finite, and then take $\Lambda \rightarrow \infty$ in the end of the calculation. The partition function of this system is
\begin{equation}
	\label{eq:Zrho0}
	Z(\rho_0) \, = \, \left< e^{\lambda \int_{\mathcal{C}_l} d^2 w \, a(w) \overline{a}(\overline{w})} e^{-i \frac{\rho_0}{\sqrt{\nu}} \int_{\mathcal{C}_l} d^2 w' [\varphi(w')+ \overline{\varphi}(\overline{w'})]} \right>_{\mathcal{C}} ,
\end{equation}
where the first exponential generates integrals over $\mathcal{C}_l$ of pairs $a (w) \overline{a}(\overline{w})$. Following \cite{Read2009}, we look at this term as a perturbation of the action of the non-chiral theory ${\rm CFT}\otimes \overline{{\rm CFT}}$ by the local operator $a(w) \overline{a}(\overline{w})$ in the region $\mathcal{C}_l$. The coefficient $\lambda$ is included such that the term in the exponential is dimensionless. It can be tuned arbitrarily, giving different weights to the terms with different number of pairs $a(w)\overline{a}(\overline{w})$. This will be important later. For now, note that because of the charge neutrality of the correlator (\ref{eq:Zrho0}), only one term (the one with the $U(1)$ charge that is exactly opposite to the total background charge contained in region $\mathcal{C}_l$) in the expansion of the exponential actually contributes to the partition function $Z(\rho_0)$. The correlator is evaluated on the cylinder rather than in the plane, hence the subscript $\mathcal{C}$, and the propagator of the free chiral boson is $\left< \varphi(w_1)\varphi(w_2) \right>_{\mathcal{C}} = -\log \left[ \tanh \frac{\pi(w_1-w_2)}{L} \right]$. The argument will not depend on the details of the background charge though, the important point here is simply that there is a non-zero density of particles $\rho_0$ in the left half-cylinder $\mathcal{C}_l$.

Imagine that we want to compute the correlation function of two operators $a^\dagger(w_1)$, $\overline{a}^\dagger(\overline{w_2})$ in the right half-cylinder ${\rm Re \, }x > 0$, as shown in Fig. \ref{fig:cylinder},
\begin{equation}
	\label{eq:correl_screening}
	\frac{ \left< a^\dagger(w_1)\overline{a}^\dagger(\overline{w_2}) \,  e^{ \lambda \int_{\mathcal{C}_l} d^2 w \, a(w)\overline{a}(\overline{w})} e^{- i \frac{\rho_0}{\sqrt{\nu}} \int_{\mathcal{C}_l} d^2 w'[\varphi(w')+\overline{\varphi}(\overline{w'}) ]} \right>_{\mathcal{C}} }{Z(\rho_0)} .
\end{equation}
(The reason why such correlators of fields {\it outside} the droplet are interesting will become clear below.) 
The two operators $a^\dagger(w_1) = e^{-i \varphi(w_1)/\sqrt{\nu}}\times \psi^\dagger(w_1)$ and $\overline{a}^\dagger(\overline{w_2}) = e^{-i \overline{\varphi}(\overline{w_2})/\sqrt{\nu}} \times \overline{\psi}^\dagger(\overline{w_2})$ add a $U(1)$ charge to the correlator (\ref{eq:correl_screening}). The latter is still non-zero though, because there is again a term in the generating function of the integrals of pairs $a (w) \overline{a}(\overline{w})$ that has exactly the right $U(1)$ charge required to ensure the global neutrality of the correlator. Similarly, in the statistics sector, the global neutrality (under $\mathbb{Z}_k$ transformations for example, if the statistics sector is generated by a $\mathbb{Z}_k$-parafermionic current) is forced by the exponential that generates the pairs.

To get some insight, let us first consider the $U(1)$ charge sector only. We use an electrostatic language, assuming screening in the ``plasma'' that fills the left half-cylinder $\mathcal{C}_l$. If the left-right symmetric operators $e^{i \varphi(w)/\sqrt{\nu}} e^{i \overline{\varphi}(\overline{w})/\sqrt{\nu}}$ and $e^{-i \varphi(w)/\sqrt{\nu}} e^{-i \overline{\varphi}(\overline{w})/\sqrt{\nu}}$ are termed ``electric charges'', then the operators $e^{\pm i \varphi (w)/\sqrt{\nu}}$ and $e^{\pm i \overline{\varphi}(\overline{w})/\sqrt{\nu}}$ themselves contain both electric and magnetic charge. In the plasma, the magnetic charge is confined, that is correlators of operators carrying magnetic charge fall exponentially (and their expectations in an infinite system vanish, so there is no need to subtract their disconnected parts). The electric charge is screened in the plasma, so correlators of electric charges fall also exponentially, with a correlation length of order $\sim 1/\sqrt{\rho_0}$ set by the density. This then implies that, when we take the operator $a^\dagger(w_1)$ to the boundary of the plasma from outside $({\rm Re\,}w_1 \rightarrow 0)$, it has its electric charge neutralized, leaving its magnetic charge. Thus, at the boundary of the plasma (${\rm Re\,} w_1 =0$), when inserted in the correlator (\ref{eq:correl_screening}), the operator $e^{-i \varphi(w_1)/\sqrt{\nu}}$ can be replaced by $e^{i \overline{\varphi}(\overline{w_1})/\sqrt{\nu}}$ when the density $\rho_0$ goes to infinity. Although this replacement apparently violates charge neutrality because $e^{-i \varphi(w_1)/\sqrt{\nu}}$ and $e^{i \overline{\varphi}(\overline{w_1})/\sqrt{\nu}}$ have opposite electric charge (but the same magnetic charge), it is valid inside the correlator (\ref{eq:correl_screening}), once again thanks to the exponential generating integrals of pairs $a(w) \overline{a}(\overline{w})$ that ensures global charge neutrality. Thus, for the $U(1)$ sector, we have the boundary condition along the imaginary axis (${\rm Re\,} w =0$)
\begin{equation}
	\label{eq:bcU1}
	e^{- i \varphi(w)/\sqrt{\nu}} \, = \, e^{i \overline{\varphi}(\overline{w})/\sqrt{\nu}} .
\end{equation}

Now let us come back to the case of the full operator $a^\dagger(w_1) = e^{-i \varphi(w_1)/\sqrt{\nu}} \times \psi^\dagger(w_1)$. Again, one brings $a^\dagger(w_1)$ to the boundary (${\rm Re \,}w_1 \rightarrow 0$) from the outside. In the correlator (\ref{eq:correl_screening}), it can fuse with one of the fields $a(w)\overline{a} (\overline{w})$, leaving only the field $\overline{a} (\overline {w})$. Assuming exponentially decaying correlations inside $\mathcal{C}_l$, the ``pairing'' between $a^\dagger(w_1)$ and $a(w)$ can occur only on distances $\lesssim 1/\sqrt{\rho_0}$. Therefore, when the density $\rho_0$ goes to infinity, we are left with a CFT in the region ${\rm Re\,}w >0$, where the fields are constrained by the boundary condition
\begin{equation}
	\label{eq:bc}
	 a^\dagger(w) = \overline{a}(\overline{w})
\end{equation}
along the imaginary axis ${\rm Re}\, w =0$. This is a generalization of the boundary condition (\ref{eq:bcU1}) that includes the statistics sector. For example, when the statistics sector is generated by a $\mathbb{Z}_k$-parafermion current, the constraint (\ref{eq:bc}) is a boundary condition on the $\mathbb{Z}_k$-current that was discussed in \cite{MaldacenaMooreSeiberg,parafermionsLukyanov}. Again, the boundary condition (\ref{eq:bc}) apparently violates charge neutrality, but the correlator (\ref{eq:correl_screening}) is still globally neutral thanks to the generating function of integrals of pairs $a(w)\overline{a}(\overline{w})$. Strictly speaking, the replacement $a^\dagger(w_1)$ by $\overline{a}(\overline{w_1})$ close to the boundary is only correct up to a multiplicative constant, which depends on the value of $\lambda$ in (\ref{eq:correl_screening}). Such a multiplicative constant should also appear in the boundary condition (\ref{eq:bc}). However, the coefficient $\lambda$ can always be tuned such that the multiplicative constant is $1$, leading the simplest form of the boundary condition (\ref{eq:bc}).

The calculation of the correlator (\ref{eq:correl_screening}) then boils down to the one of the two-point correlator
\begin{equation}
\left< a^\dagger(w_1) \overline{a}^\dagger(\overline{w_2}) \right>_{\mathcal{C}_r}
\end{equation}
in the domain $\mathcal{C}_r$, with the boundary condition (\ref{eq:bc}). This is a considerable simplification of the problem. We will use this trick again in section \ref{sec:equality_bulk_edge} to compute equal-time correlators at the edge of a quantum Hall system.

The boundary condition (\ref{eq:bc}) for the non-chiral CFT outside the droplet is the main result of this section. It will play a crucial role in the rest of this paper. We obtained it from the specific form of the perturbation of the action $\int a(w) \overline{a}(\overline{w}) d^2w$ inside the droplet, and assuming that the generalized screening hypothesis holds. The boundary condition (\ref{eq:bc}) is a local constraint along the boundary. It is a conformally invariant boundary condition: it is invariant under conformal mappings $w \mapsto f(w)$ of the domain {\it outside the droplet} which preserve the shape of the boundary \cite{Cardy84}.

Finally, to conclude this section, we reformulate the boundary condition (\ref{eq:bc}) using the operator formalism in CFT. This is a purely technical step that will be used in the next section, when we analyze the consequences of (\ref{eq:bc}) for the edge theory of the FQHE. It is a standard procedure in boundary CFT \cite{Ishibashi, Cardy89}.
The fields $a(w)$ and $a^\dagger(w)$ can be expanded in Fourier modes on the cylinder,
\begin{eqnarray}
	a(w) &=& \left(\frac{2\pi}{L}\right)^{h_a} \sum_{n-h_a ~\in \mathbb{Z}} e^{\frac{2\pi w}{L} n} \, a_{-n}  \\
	a^\dagger(w) &=& \left(\frac{2\pi}{L} \right)^{h_a} \sum_{n-h_a ~\in \mathbb{Z}} e^{\frac{2\pi w}{L} n}  \, (a_n)^\dagger ,
\end{eqnarray}
with the hermiticity condition $(a_{-n})^\dagger = (a^\dagger)_n$ (in this Euclidean field theory, the hermitian conjugate must be taken {\it after} continuation to real time on the cylinder, namely $x \rightarrow i t$, so $w^\dagger = -w$). One has similar expansions for $\overline{a}(\overline{w})$ and $\overline{a}^\dagger(\overline{w})$. 
The boundary condition (\ref{eq:bc}) can be written in terms of the modes as $(a^\dagger)_{n} = \overline{a}_{-n}$ for any $n \in \mathbb{Z}+h_a$. More precisely, this identity must hold while acting on a {\it boundary state} $\left| B \right>$,
\begin{equation}
	\label{eq:Ishibashi}
	\left[ (a^\dagger)_{n} - \overline{a}_{-n} \right]  \left| B \right> \, = \, 0 \; .
\end{equation}
Such boundary states are known in the CFT literature as Ishibashi states \cite{Cardy89,Ishibashi}. It is convenient to think of the constraint (\ref{eq:Ishibashi}) as the expression of an intertwiner between the chiral and the anti-chiral representations of $\mathcal{A}$. Since the operator product expansions of $a(w)$ and $a^\dagger(w)$ generate the full chiral algebra $\mathcal{A}$, and because we are assuming that the representations of the chiral algebra $\mathcal{A}$ that appear in this paper are irreducible, Schur's lemma implies that the state $\left| B \right>$ is completely fixed, up to a global normalization constant.

Before we go ahead and analyze the consequences of these boundary CFT ideas, let us point out that other technical choices are possible for the analysis carried out here. We have adopted a grand-canonical formalism in order to be able to write the boundary condition (\ref{eq:bc}), which violates neutrality. The neutrality of correlators is restored thanks to the fact that the particle number is not fixed. One could have adopted other conventions. For instance, one possibility would be to work in the  canonical ensemble, and then focus on the {\it neutral subalgebra} of the chiral algebra $\mathcal{A}$, which is generated by all the neutral operators. For instance, in the $U(1)$ sector, the neutral operators are the generated by the operator product expansions of the current $i \partial \varphi(z)$ and its derivatives. In the statistics sector, the neutral subalgebra contains the stress-tensor, which modes generate a Virasoro algebra, and possibly some other local operators, which yield some extension of the Virasoro algebra (like a $\mathcal{W}_k$-algebra for $\mathbb{Z}_k$-parafermions, see \cite{FateevZamo2,FateevLukyanov,Wreview}). Then one would have found a boundary condition analogous to (\ref{eq:bc}) but for the neutral currents rather than for the operators $a(w)$ and $\overline{a}(\overline{w})$ themselves. Also, another appealing possibility to circumvent the problems caused by the violation of charge neutrality, while working in the canonical ensemble, would be to use ``shift operators'' that would map the CFT vacuum with $N$ charges, $\left| N\right>$, onto the one with $N \pm 1$ charges. These operators are not local. In terms of such a shift operator $\mathcal{S}$, one would obtain a boundary condition of the form $a(w) =  \mathcal{S} ~\overline{a}^\dagger(\overline{w}) ~\mathcal{S}$. The {\it locality} of the boundary condition would be somewhat hidden in this kind of expression. That is why, in this section, rather than dealing with these shift operators, we decided to use a grand-canonical formulation in order to reach the {\it local} boundary condition (\ref{eq:bc}), which should appear as more natural to the reader. Of course, although all these technical conventions need to be treated carefully for the global consistency of the argument, they will have no influence on our final results.

\subsection{Back to the droplet}
\label{sec:back_droplet}
In this section, we want to go back to the plane, where the particles fill a droplet of radius $R$. We want to understand how one should handle the state
\begin{equation}
	\label{eq:state_droplet}
	\frac{1}{\lambda^N ~Z_N} \exp \left[\lambda \int_{\mathbb{C}} e^{V(z,\overline{z})} d^2 z \, a(z)\overline{a}(\overline{z}) \right] \left| 0 \right> \overline{\left| 0 \right>}
\end{equation}
when it appears inside correlators of the form
\begin{eqnarray}
	\label{eq:correl_droplet}
&&	\overline{\left<N\right|} \left< N \right| \phi(\zeta_1) \dots \phi(\zeta_p) \\ 
\nonumber && \times  \frac{1}{\lambda^N~Z_N} \exp \left[\lambda \int_{\mathbb{C}} e^{V(z,\overline{z})} d^2 z \, a(z)\overline{a}(\overline{z}) \right]  \left| 0 \right> \overline{\left|0\right>}.
\end{eqnarray}
Each of the operators $\phi(\zeta_j)$ is one field $a(\zeta_j)$, $\overline{a}(\overline{\zeta_j})$, $a^\dagger(\zeta_j)$, or $\overline{a}^\dagger(\overline{\zeta_j})$. They are all lying outside the dropplet: $|\zeta_j|>R$. We are interested in these correlators in the ``scaling region'', which we now define. we first fix some number $M>0$, and consider the correlators of the form (\ref{eq:correl_droplet}) such that $p \leq M$. The non-zero contribution to the
correlator (\ref{eq:correl_droplet}) comes from the term generated by the exponential
that has exactly the right total charge. This charge is contained in the interval $\left[ N-\frac{M}{2} , N+\frac{M}{2} \right]$. Then we consider the limit $N \rightarrow \infty$, keeping $M$ fixed. In that process, the radius $R$ of the droplet grows, so one has to push the operators $\phi(\zeta_j)$ such that they stay out of the droplet (for example, when $R$ is changed to $R'$, one can rescale $\zeta_j \mapsto \zeta_j \times R'/R$).
In the scaling region, only terms with a number of particles within the range $\left[N-\frac{M}{2},  N+\frac{M}{2}\right]$ matter. Different particle numbers should in principle correspond to circular droplets with different radii $R + \delta R$, however we have defined the scaling region precisely such that $\delta R/R \rightarrow 0$ when the number of particle goes to infinity, so the variations of the size of the droplet become negligeable. Therefore, in what follows the radius of the droplet is always $R$, even if the number inside it can fluctuate around $N$.

\begin{figure}[htbp]
	\includegraphics[width=0.47\textwidth]{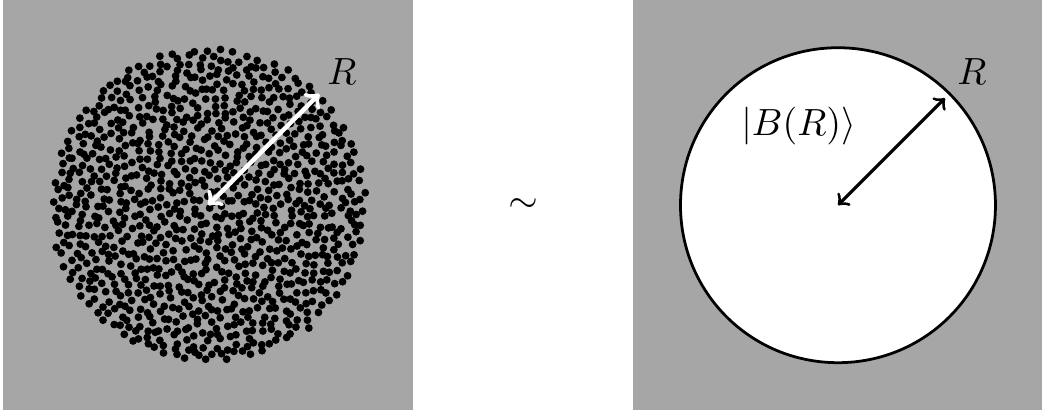}
	\caption{In the thermodynamic limit, assuming screening, the state (\ref{eq:state_droplet}) is equivalent to the conformal boundary state $\left|B(R) \right>$, in the so-called ``scaling region''.}
\end{figure}

Now we are ready to apply the ideas of the previous section. If the screening hypothesis holds inside the droplet, then when the number of particles goes to infinity, the droplet becomes equivalent to a boundary condition at $|z|=R$ for the non-chiral CFT that remains outside the droplet. The exterior of the droplet can be mapped onto the right half-cylinder by the conformal mapping $z \mapsto w= \frac{L}{2\pi} \log (z/R)$, and we know that the boundary condition on the cylinder is (\ref{eq:bc}). Again, the boundary condition (\ref{eq:bc}) requires some fine-tuning of the parameter $\lambda$, such that configurations with different particle numbers contribute with equal weights. Let us skip this detail for now. We reach the conclusion that, in radial quantization in the plane, the boundary condition is again encoded by the boundary state (\ref{eq:Ishibashi}) up to a scale transformation (in order for the boundary to be at $|z|=R$ rather than at $|z|=1$)
\begin{equation}
	\label{eq:droplet_BR}
	\left| B(R) \right> \, = \, R^{L_0+\overline{L_0}} \left| B \right>  .
\end{equation}
At this point, the reader might be worried by a technical aspect: a neutralizing background was explicitly included in the previous section, while here we have traded the neutralizing background for the factor $e^{V(z,\bar{z})}$ in the integration measure for the particles. However, since the boundary state $\left| B \right>$ is already a sum over all the possible charge sectors with equal weights, this does not affect the expression of $\left| B \right>$. Thus, when it is inserted in correlators, and in the scaling region, the state (\ref{eq:state_droplet}) can be safely replaced by the boundary state $\left|B(R) \right>$ in the limit $N \rightarrow \infty$ (up to a global normalization factor which still needs to be fixed).



\subsection{Consequence for the overlaps \\between the edge states}
\label{sec:edge_conjecture}
Now we come back to the edge states which we defined in section \ref{sec:edge}, and explore the consequences of screening for the overlaps between these wavefunctions. Following the formula (\ref{eq:edgestates1}), we define
\begin{equation}
	\label{eq:edge_def}
	\Psi_{\left< v \right|}(z_1,..,z_N) \, =\, \frac{1}{\sqrt{Z_N}} \left<v \right| \frac{1}{R^{\Delta L_0}} \prod_{i=1}^N a(z_i) \left|0 \right> ,
\end{equation}
where we have introduced the rescaling operator $1/R^{\Delta L_0}$. Here, $R$ is again the radius of the planar droplet defined by (\ref{eq:radius}), and $\Delta L_0 = L_0 - \left< N \right| L_0 \left| N \right>$ measures the conformal dimension relative to the one of the vacuum ($\left| N\right>$ is the vacuum with $N$ charges). For an angular momentum eigenstate ($L_0 \left| v\right> = h_v \left| v\right>$), one can easily check that the factor $1/R^{\Delta L_0}$ ensures that $\Psi_{\left< v \right|}(z_1,\dots,z_N)$ is dimensionally homogeneous to the ground state wavefunction $\Psi_{\left< N \right|}(z_1,\dots,z_N)$, $z_i$ and $R$ being two lengths. The normalization (\ref{eq:edge_def}) of the wavefunctions for the edge states will also allow us to express the result (\ref{eq:edge_conjecture}) in a particularly simple form.

Actually, the definition (\ref{eq:edge_def}) is valid for the neutral edge excitations only, as it is implicitly assumed that the out state $\left< v \right|$ is a state with charge $N$. To be able to express our final result (\ref{eq:edge_conjecture}) in a more general form, we also need to include charged excitations. Therefore, for a state $\left< v \right|$ with charge $N+n$ ($n$ can be positive or negative), we define the wavefunction for the excited state $\Psi_{\left<v\right|}$ as
\begin{eqnarray}
	\label{eq:edge_def_charge}
&&	\Psi_{\left<v\right|} (z_1,\dots,z_N,z_{N+1},\dots,z_{N+n})  \\
\nonumber && \qquad \qquad  = \,\frac{\lambda^{n/2}}{\sqrt{Z_N}} \left<v\right| \frac{1}{R^{\Delta L_0}} \prod_{i=1}^{N+n} a(z_i) \left|0 \right>
\end{eqnarray}
where the coefficient $\lambda$ is the same as in (\ref{eq:state_droplet}). Now that our conventions for the edge states are fixed, we can consider the overlap between two wavefunctions $\Psi_{\left< v_1 \right|}(z_1,\dots,z_N)$ and $\Psi_{\left< v_2 \right|}(z_1,\dots,z_N)$,
\begin{eqnarray}
	\label{eq:qm_innerprod}
	&& \left<\left< \, \Psi_{\bra{v_1}} \, \left|\, \Psi_{\bra{v_2}}\, \right>\right>\right. \, = \, \frac{1}{N!} \int_{\mathbb{C}} \prod_{i=1}^N  e^{V(z_i,\bar{z}_i)} d^2 z_i  \qquad  \\
\nonumber && \qquad \qquad \qquad \left[ \Psi_{\bra{v_1}}(z_1,\dots,z_N) \right]^* \, \Psi_{\bra{v_2}}(z_1,\dots,z_N) .
\end{eqnarray}
This is an overlap between two neutral edge excitations. More generally, between two charged excitations with some charge $N+n$, the overlap is defined as
\begin{eqnarray}
	\label{eq:qm_innerprod_Nn}
	&& \left<\left< \, \Psi_{\bra{v_1}} \, \left|\, \Psi_{\bra{v_2}}\, \right>\right>\right. \, = \, \frac{1}{(N+n)!} \int_{\mathbb{C}} \prod_{i=1}^{N+n}  e^{V(z_i,\bar{z}_i)} d^2 z_i  \qquad  \quad \\
\nonumber && \qquad \qquad \qquad \left[ \Psi_{\bra{v_1}}(z_1,\dots,z_{N+n}) \right]^* \, \Psi_{\bra{v_2}}(z_1,\dots,z_{N+n}) .
\end{eqnarray}
The overlap between two wavefunctions with different particle numbers is always zero.
Using our definition (\ref{eq:edge_def}-\ref{eq:edge_def_charge}), these overlaps are equal to
\begin{eqnarray}
&&	\left<\left< \, \Psi_{\bra{v_1}} \,\left|\, \Psi_{\bra{v_2}}\, \right>\right>\right. \, = \,\overline{\left<v_1 \right|}  \left< v_2 \right| R^{-\Delta L_0 - \overline{\Delta L_0}} \\
\nonumber && \qquad \qquad  \times \; \frac{1}{\lambda^N~Z_N} \exp \left[ \lambda \int_{\mathbb{C}}  e^{V(z,\bar{z})} d^2 z  \,a(z) \overline{a(z)}  \right] \left| 0\right> \overline{\left|0 \right>} \, .
\end{eqnarray}
According to the previous section, the whole expression in the second line can be replaced with the boundary state $ R^{L_0+\overline{L_0}} \left| B \right>$, at least as far as we are in the scaling region. This means that the charge of the states $\left<v_1 \right|$ and $\left< v_2\right|$ must be kept to some value $N+n$, where $|n| \leq M$, and $M$ is fixed while we send $N$ to infinity. This is precisely what we do here. Then, we have
\begin{equation}
	\left<\left< \, \Psi_{\bra{v_1}} \,|\, \Psi_{\bra{v_2}}\, \right>\right> \, \underset{N \rightarrow \infty}{\sim } \,\frac{R^{h_N +\overline{h_N}}}{Z_N}  \left[ \overline{\left<v_1 \right|} \left< v_2 \right|\right] \left|B \right> 
\end{equation}
with $h_N + \overline{h_N} = \left<N \right| L_0 \left| N \right> + \overline{\left<N \right| L_0 \left| N \right>}$. Using some basis of states $a_{n_1} a_{n_2}\dots (a^\dagger)_{m_1}(a^\dagger)_{m_2}\dots \left| 0 \right>$, one can easily show that the property (\ref{eq:Ishibashi}) implies that for any $\left|v_1 \right>$ and $\left| v_2\right>$
\begin{equation}
	\label{eq:inner_Ishibashi}
	 \overline{\left< v_1 \right|}\left< v_2\right| \left| B\right> \, = \, \left< v_2 \left| v_1 \right>\right. \, \times \, {\rm constant}
\end{equation}
where the constant does not depend on $\left| v_1\right>$ and $\left| v_2\right>$, and comes only from the global undetermination of the normalization of $\left| B\right>$. The property (\ref{eq:inner_Ishibashi}) is a very well-known property of Ishibashi states (see \cite{Ishibashi,Cardy89}). Actually, it could even be used as a definition of an Ishibashi state, instead of (\ref{eq:Ishibashi}).

The normalization of $\left| B\right>$ is fixed by the requirement that the ground state wavefunction $\Psi_{\left<N \right|}$ is normalized: $\left<\left< \Psi_{\left<N\right|}\left|\Psi_{\left<N\right|} \right>\right>\right.= 1$. Thus the constant in (\ref{eq:inner_Ishibashi}) must be equal to $Z_N/R^{h_N+\overline{h_N}}$. The final result is then
\begin{equation}
	\label{eq:edge_conjecture}
	\left<\left<\, \Psi_{\left< v_1\right|}\,\left| \,\Psi_{\left< v_2 \right|} \, \right>\right>\right. \, \underset{N \rightarrow \infty}{\longrightarrow} \, \left< v_2\left| v_1\right>\right. \, .
\end{equation}
This formula, which is a direct consequence of the generalized screening hypothesis, is the main result of this paper. It shows that the linear mapping (\ref{eq:edge_def}) from the (dual of the) Hilbert space of the CFT to the space of the physical edge states becomes an {\it isometric isomorphism} in the thermodynamic limit. This is a precise formulation of the long expected ``bulk/edge correspondence''. It implies, in particular, that the underlying CFT has a positive definite inner product, or in other words, that it is a {\it unitary} CFT. In the MR construction (section \ref{sec:blocks}), only the rationality of the CFT for the statistics sector was assumed. We see that, if generalized screening holds, then we arrive at (\ref{eq:edge_conjecture}), which implies that the CFT is unitary. This is one additional argument that shows that the use of a non-unitary CFT in the statistics sector cannot be consistent with the fact that all the connected correlations of local operators in the bulk are short-range, as it clearly leads to contradictions (for previous arguments, see Refs. \cite{Read2009,Nick_arXiv}). Therefore, for consistency, the FQHE trial states given by the MR construction should correspond to rational and unitary CFTs only.

The formula (\ref{eq:edge_conjecture}) will also play a key role when we study the entanglement spectrum in part \ref{sec:ES}. We will provide some direct numerical checks of this result in section \ref{sec:numerical}. In the next section we derive another result that is directly related to this bulk/edge correspondence.

To conclude this section, let us comment on the factor $\lambda$ that appears each time one has to deal with particle numbers that differ from $N$. We have for instance
\begin{equation}
	\label{eq:def_lambda}
	\left<\left< \Psi_{\left< N+1 \right|} \left| \Psi_{\left< N+1 \right|} \right>\right>\right.  \, \underset{N \rightarrow \infty}{\longrightarrow} \, 1,
\end{equation}
while by using the definitions (\ref{eq:plasmaZN}), (\ref{eq:edge_def_charge}) and (\ref{eq:qm_innerprod_Nn}),
\begin{equation}
	\left<\left< \Psi_{\left< N+1 \right|} \left| \Psi_{\left< N+1 \right|} \right>\right>\right. \, = \, \lambda ~\frac{Z_{N+1}}{R^{h_{N+1}+\overline{h_{N+1}}}}    \frac{R^{h_{N}+\overline{h_{N}}}}{Z_N}.
\end{equation}
In the previous sections, we explained that the coefficient $\lambda$ had to be tuned such that it gives rise to the result (\ref{eq:edge_conjecture}) for charged excitations (not only for the wavefunctions for neutral excitations). Thus, the formula (\ref{eq:def_lambda}) is rather a {\it definition} of the coefficient $\lambda$ than an actual result. Indeed, in general $\lambda$ may depend on the radius $R$, and therefore on the number of particles $N$. However, once $\lambda$ is fixed, there is no other free parameter in (\ref{eq:edge_conjecture}). For instance, by evaluating $\left<\left< \Psi_{\left< N+n\right|} \left| \Psi_{\left<N+n\right|} \right>\right>\right.$, one finds that $Z_{N+n}/Z_N \sim \lambda^{-n} R^{h_{N+n}+\overline{h_{N+n}}-h_N -\overline{h_N}}$, where the coefficient $\lambda$ is no longer a free parameter. As an exercise, one can check that this is consistent with the large $N$ behaviour of the partition function $Z_{N+n}$ in the Laughlin case, either with direct free-fermion calculations in the integer quantum Hall effect or with the results of the semi-classical expansion of Wiegmann and Zabrodin for the Laughlin wavefunction \cite{ZabrodinWiegmann3}.

\subsection{Equality of edge and bulk CFT correlators}
\label{sec:equality_bulk_edge}

In this section we show that screening, reformulated as the conformal boundary condition (\ref{eq:bc}), implies that the equal-time correlators measured along the edge of the quantum Hall system are equal to the correlators in the CFT that is used to construct the ground-state wavefunction (\ref{eq:wfblock}). For instance, we can compute the correlator of particle creation/annihilation operators $c/c^\dagger$ along the boundary of the droplet ($|z_j|=|z_j'|=R$),
\begin{equation}
	\label{eq:fermion_corr}
	\left<\left< \Psi \right.\right|  c^\dagger(z_1') \dots c^\dagger(z_n') c(z_1) \dots c(z_n) \left|\left. \Psi \right>\right> .
\end{equation}
The ground state $\left|\left. \Psi \right>\right>$ is the wavefunction (\ref{eq:wfblock}) with $N$ particles. In first quantization, the correlator (\ref{eq:fermion_corr}) is
\begin{eqnarray}
	\label{eq:fermion_corr_1q}
	&& \frac{A}{(N-n)!} \int_{\mathbb{C}} \prod_{i=n+1}^N e^{V(z_i,\overline{z_i})} d^2 z_i \\
\nonumber   && \qquad  ~ \left[ \Psi(z_1',\dots,z_n', z_{n+1}\dots,z_N) \right]^* \Psi(z_1, \dots,z_N) \\
\nonumber	&& = \,A~ \overline{\left< N \right|} \overline{a(z_1')\dots a(z_n')}  \left< N\right|a(z_1) \dots a(z_n) \,\times \\
\nonumber && \qquad \frac{1}{\lambda^{N-n} ~Z_N} \exp \left[ \lambda \int_\mathcal{C} e^{V(z,\bar{z})}d^2 z ~a(z) \overline{a(z)} \right] \left|0 \right> \overline{\left| 0\right>} \\
\nonumber && \underset{N \rightarrow \infty }{\longrightarrow} \,A~ \lambda^n~\overline{\left< N \right|} \overline{a(z_1')\dots a(z_n')}  \left< N\right|a(z_1) \dots a(z_n)   \left|B(R) \right> .
\end{eqnarray}
The factor $A$ is the product $\prod_{p=1}^n e^{V(z_p,\overline{z_p})/2} e^{V(z'_p,\overline{z'_p})/2}$. Since we assume that $V(z,\overline{z})$ is rotationally invariant and the $z_p$ and $z_p'$ are on the circle of radius $R$, one has $A= e^{n V(R)}$. 
In the last line we have replaced the state (\ref{eq:state_droplet}) by the boundary state (\ref{eq:droplet_BR}), as explained in section \ref{sec:back_droplet}. Finally, we use the fact that the boundary state $\left|B(R)\right>$ implements the conformal boundary condition $a^\dagger(w)=\overline{a}(\overline{w})$ on the cylinder, which can be mapped onto the plane by the conformal mapping $w \mapsto z = R \exp \frac{2\pi w}{L}$. This leads to the boundary condition along the circle of radius $R$ in the plane 
\begin{equation}
	\left( \frac{dz}{dw} \right)^{h_a} a^\dagger(z) \, = \, \left( \frac{d\overline{z}}{d\overline{w}} \right)^{\overline{h_a}} \overline{a}(\overline{z}) .
\end{equation}
The factors $\frac{dz}{dw},\frac{d\overline{z}}{d\overline{w}}$ appear because the operators $a$ and $\overline{a}$ transform covariantly under conformal transformations (see \cite{BYB,SaleurItzyksonZuber,Mussardo}). These two factors are equal to $\frac{2\pi z}{L}$ and $\frac{2\pi \overline{z}}{L}$ respectively. Thus, the boundary condition at $|z|=R$ is $\overline{a}(\overline{z}) = (z/\overline{z})^{h_a} a^\dagger(z)$ (recall that $h_a = \overline{h_a}$). In the end, the correlator (\ref{eq:fermion_corr}) converges to the following correlator in the chiral CFT in the plane
\begin{eqnarray}
\nonumber &&	\left(\lambda ~ e^{V(R)} \right)^n  \prod_{p=1}^n \left(\frac{z_p}{\overline{z_p}}\right)^{h_a}  \\ && \qquad \times \left< a^\dagger(z_1') \dots a^\dagger(z_n') a(z_1) \dots a(z_p)\right> \, .
\end{eqnarray}
In particular, the particle propagator along the edge is proportional to the two-point correlator in the chiral CFT, $\left< a^\dagger(z') a(z) \right>$
\begin{equation}
	\left<\left< \Psi \right.\right|   c^\dagger(z') c(z)   \left|\left. \Psi\right>\right> \, \propto \, \frac{1}{|z'-z|^{2 h_a}}
\end{equation}
in the thermodynamic limit. This shows that equal-time correlators evaluated along the edge are given by correlators in the CFT that is used to construct the trial wavefunction for the ground state (\ref{eq:wfblock}). This result has been assumed in many places in the literature, although no general argument has ever been given. It generalizes the one obtained by Wen for the Laughlin wavefunction in Ref. \cite{WenEdge}. For recent work on this topic, see also \cite{BBR2012}. Note that we have restricted the result to equal-time correlators because we do not address physical Hamiltonians in this paper (see, however, \cite{Nick_edge, Shankar_bulkedge}).

The discussion in this section can be extended to the case of equal-time correlation functions of quasi-particle and quasi-hole operators along the edge. In the thermodynamic limit and assuming short-range correlations in the bulk, the same calculation based on the boundary condition (\ref{eq:bc}) can be done, leading to the equality (up to normalization and phase factors) between these correlation functions and the correlators in the bulk chiral CFT that underlies the trial wavefunction.

\subsection{Numerical checks}
\label{sec:numerical}

The formula (\ref{eq:edge_conjecture}) can be checked numerically. In this section we present numerical evidence that shows that it holds for the Laughlin wavefunction and for the MR (Pfaffian) wavefunction. In both cases, we do a Monte-Carlo (MC) simulation for a system of $N$ particles, which is tractable both for the Laughlin and Pfaffian states. The MC calculation allows us to estimate numerically the ground state expectation value of any observable $\mathcal{O}(\{z_i\})$ that depends on the positions $z_i$'s
\begin{equation}
	\left<\left< \Psi \right.\right| \mathcal{O}(\{z_i\}) \left|\left.\Psi \right>\right> .
\end{equation}

\subsubsection{Laughlin at $\nu =1/3$}
The edge states for the Laughlin wavefunction are given explicitly in section \ref{sec:edge}. We have
\begin{eqnarray}
	\label{eq:MCexplain}
\nonumber	&& \left<\left< \Psi_{\left< N\right|J_{k_1} \dots J_{k_p}} \left| \Psi_{\left< N\right|J_{k'_1} \dots J_{k'_q}} \right>\right>\right. \\ && \, = \, \left<\left< \Psi \right.\right|  S^*_{k_1}\dots S^*_{k_p} S_{k'_1}\dots S_{k'_q}   \left|\left.\Psi \right>\right>
\end{eqnarray}
where 
\begin{equation}
	S_k = \frac{1}{\sqrt{\nu}} \sum_{i=1}^N \frac{z_i^k}{R^k} \,,
\end{equation}
and $S_k^*$ is the complex conjugate of $S_k$. It is the right-hand side of (\ref{eq:MCexplain}) that we measure numerically with MC techniques. The result predicted by (\ref{eq:edge_conjecture}) when $N\rightarrow \infty$ is
\begin{eqnarray}
\nonumber	&& \left<\left< \Psi \right.\right| S^*_{k_1}\dots S^*_{k_p} S_{k'_1}\dots S_{k'_q} \left|\left.\Psi \right>\right> \\  && \, = \, \left< J_{k'_1} \dots J_{k'_q} J_{-k_p} \dots J_{-k_1} \right> .
\end{eqnarray}
For $\nu =1/3$, $N=100$, with $10^8$ MC steps and with the quadratic potential (\ref{eq:potential}) corresponding to the plane, we find the following results for the first exited states. They are in very good agreement with our analytic prediction in the $N\rightarrow \infty$ limit.
\begin{center}
\begin{tabular}{|c|c|c|}
\hline &  MC & analytic   \\ \hline \hline 
$\left< \left< \Psi \right.\right| S_1^* S_1 \left|\left. \Psi \right>\right>$ & 1.007 & $\left<J_1 J_{-1} \right>=$ 1\\
\hline
$\left< \left< \Psi \right.\right| S_2^* S_2 \left|\left. \Psi \right>\right>$ & 2.017 & $\left<J_2 J_{-2} \right>=$ 2\\
\hline
$\left< \left< \Psi \right.\right| (S_1^*)^2 S_2 \left|\left. \Psi \right>\right>$ & 0.003 & $\left<J_2 J_{-1}^2 \right>=$ 0 \\
\hline
$\left< \left< \Psi \right.\right| S_2^* S_1^* S_1 S_2 \left|\left. \Psi \right>\right>$ & 2.034 & $\left<J_2 J_1 J_{-1} J_{-2} \right>=$ 2 \\
\hline
$\left< \left< \Psi \right.\right| (S_2^*)^2 (S_2)^2 \left|\left. \Psi \right>\right>$ & 8.048 & $\left<J_2^2 J^2_{-2} \right>=$ 8 \\
\hline
\end{tabular}
\end{center} 

\vspace{0.5cm}

\subsubsection{Moore-Read (Pfaffian) at $\nu =1/2$}
For the Pfaffian state, there are two types of edge excitations: the excitations in the $U(1)$ charge sector, which are similar to the ones of the Laughlin state, and the excitations in the Majorana sector. For the $U(1)$ excitations, we find for $\nu =1/2$, $N=100$ and with $10^8$ MC steps:
$$
\begin{tabular}{|c|c|c|}
\hline &  MC & analytic   \\ \hline \hline 
$\left< \left< \Psi \right.\right| S_1^* S_1 \left|\left. \Psi \right>\right>$ & 1.002 & $\left<J_1 J_{-1} \right>=$ 1\\
\hline
$\left< \left< \Psi \right.\right| S_2^* S_2 \left|\left. \Psi \right>\right>$ & 1.993 & $\left<J_2 J_{-2} \right>=$ 2\\
\hline
$\left< \left< \Psi \right.\right| (S_1^*)^2 S_2 \left|\left. \Psi \right>\right>$ & 0.005 & $\left<J_2 J_{-1}^2 \right>=$ 0 \\
\hline
$\left< \left< \Psi \right.\right| (S_1^*)^2 (S_1)^* \left|\left. \Psi \right>\right>$ & 1.995 & $\left< J_1^2 J_{-1}^2 \right>=$ 2 \\
\hline
$\left< \left< \Psi \right.\right| (S_2^*)^2 (S_2)^2 \left|\left. \Psi \right>\right>$ & 7.934 & $\left<J_2^2 J^2_{-2} \right>=$ 8 \\
\hline
\end{tabular}
$$
In the Majorana sector, the excitations are of the form (\ref{eq:MRPfaffian_edge}). For instance, with two excited fermion modes only, we get the overlaps
\begin{eqnarray}
	\label{eq:MCexplain2}
\nonumber	&& \left<\left< \Psi_{\left< N\right|\psi_{n_1} \psi_{n_2}} \left| \Psi_{\left< N\right|\psi_{n'_1} \psi_{n'_2}} \right>\right>\right. \\ && \, = \, \left<\left< \Psi \right.\right| F^*_{n_1,n_2} F_{n'_1,n'_2} \left|\left.\Psi \right>\right>
\end{eqnarray}
where $F_{n_1,n_2}$ is the following ratio:

\begin{widetext}
\begin{equation}
	\frac{ {\rm Pf \,} \left(  \begin{array}{cccccc}
		0 & 0 & z_1^{n_1-\frac{1}{2}} & z_2^{n_1-\frac{1}{2}} & \dots & z_N^{n_1-\frac{1}{2}} \\
		0 & 0 & z_1^{n_2-\frac{1}{2}} & z_2^{n_2-\frac{1}{2}} & \dots & z_N^{n_2-\frac{1}{2}} \\
		-z_1^{n_1-\frac{1}{2}} & -z_1^{n_2-\frac{1}{2}} & 0 & \frac{1}{z_1-z_2} & \dots & \frac{1}{z_1-z_N} \\
		-z_2^{n_1-\frac{1}{2}} & -z_2^{n_2-\frac{1}{2}} &	\frac{-1}{z_1-z_2} & 0 & \ddots &  \\ 
		\vdots &\vdots & \vdots & \ddots & \ddots & \frac{1}{z_{N-1}-z_N} \\
		-z_N^{n_1-\frac{1}{2}}& -z_N^{n_2-\frac{1}{2}} & \frac{-1}{z_1-z_N} &  & \frac{-1}{z_{N-1}-z_N} & 0
	\end{array} \right) }{R^{n_1+n_2} \;\times \; {\rm Pf} \left( \left\{ \frac{1}{z_i-z_j} \right\}_{1\leq i,j\leq N} \right)}.
\end{equation}
Similar formulas hold for more fermionic excitations, for example for $F_{n_1,n_2,n_3,n_4}$, etc. 
Again, the right-hand side of (\ref{eq:MCexplain2}) can be measured numerically in a MC simulation. When $N\rightarrow \infty$, we expect
\begin{eqnarray}
 \left<\left< \Psi \right.\right| F^*_{n_1,n_2} F_{n'_1,n'_2}  \left|\left.\Psi \right>\right>  &  = & \left< \psi_{n'_1} \psi_{n'_2} \psi_{-n_2} \psi_{-n_1} \right> .
\end{eqnarray}
We have checked this for a few matrix elements for sizes $N=10,20,30$ and $100$ (each of them with $10^8$ MC steps). The results are in good agreement with our analytic prediction, although the convergence is slower than in the $U(1)$ sector.
$$
\begin{tabular}{|c|c|c|c|c|c|}
\hline &  $N=10$ & $N=20$ & $N=30$ & $N=100$ & analytic ($N \rightarrow \infty$)  \\ \hline \hline 
$\left< \left< \Psi \right.\right| F^*_{\frac{1}{2},\frac{3}{2}} F_{\frac{1}{2},\frac{3}{2}} \left|\left. \Psi \right>\right>$ & 1.371 & 1.288 & 1.231 & 1.124 & $\left<\psi_{\frac{1}{2}} \psi_{\frac{3}{2}} \psi_{-\frac{3}{2}} \psi_{-\frac{1}{2}} \right>=$ 1\\
\hline
$\left< \left< \Psi \right.\right| F^*_{\frac{1}{2},\frac{5}{2}} F_{\frac{1}{2},\frac{5}{2}} \left|\left. \Psi \right>\right>$ & 1.460 & 1.323 & 1.268 & 1.151 & $\left<\psi_{\frac{1}{2}} \psi_{\frac{5}{2}} \psi_{-\frac{5}{2}} \psi_{-\frac{1}{2}} \right>=$ 1\\
\hline
$\left< \left< \Psi \right.\right| F^*_{\frac{3}{2},\frac{5}{2}} F_{\frac{3}{2},\frac{5}{2}} \left|\left. \Psi \right>\right>$ & 1.432 & 1.393 & 1.331 & 1.189 & $\left<\psi_{\frac{3}{2}} \psi_{\frac{5}{2}} \psi_{-\frac{5}{2}} \psi_{-\frac{3}{2}} \right>=$ 1\\
\hline
$\left< \left< \Psi \right.\right| F^*_{\frac{1}{2},\frac{7}{2}} F_{\frac{3}{2},\frac{5}{2}} \left|\left. \Psi \right>\right>$ & -0.033 & -0.004 & -0.003 & -0.001 & $\left<\psi_{\frac{3}{2}} \psi_{\frac{5}{2}} \psi_{-\frac{7}{2}} \psi_{-\frac{1}{2}} \right>=$ 0\\
\hline
$\left< \left< \Psi \right.\right| F^*_{\frac{1}{2},\frac{3}{2},\frac{5}{2},\frac{7}{2}} F_{\frac{1}{2},\frac{3}{2},\frac{5}{2},\frac{7}{2}} \left|\left. \Psi \right>\right>$ & 1.617 & 1.764 & 1.700 & 1.381 & $\left<\psi_{\frac{1}{2}} \psi_{\frac{3}{2}} \psi_{\frac{5}{2}} \psi_{\frac{7}{2}}  \psi_{-\frac{7}{2}} \psi_{-\frac{5}{2}} \psi_{-\frac{3}{2}} \psi_{-\frac{1}{2}} \right>=$ 1\\
\hline
\end{tabular}
$$
\end{widetext}

\subsection{Corrections to scaling}
\label{sec:corrections_overlaps}
So far we have shown that, assuming short-range bulk correlations only---the generalized screening assumption---, the universal formula (\ref{eq:edge_conjecture}) gives the inner products between the edge states in the thermodynamic limit $N \rightarrow \infty$. In this section we show how the corrections to scaling can be tackled.

We will use ideas that come from the field of surface critical phenomena (for a review, see Ref. \cite{Diehl}). Let us sketch some of the main points here. Like bulk critical phenomena, surface critical phenomena can be understood within the framework of the renormalization group (RG). Let us consider a classical statistical system which is critical in the bulk, such as a critical Ising model in $d$ dimensions. At the surface (which is $d-1$ dimensional), let us imagine that the spins are free. Imagine also that one can turn on a magnetic field at the surface. The spins at the surface tend to align with the magnetic field. Thus, the surface of the system undergoes a transition, although the bulk is still critical. Under the RG flow, the surface of the system flows towards a fixed boundary condition where all the spins are aligned. This is called a boundary RG flow; it brings the system from one unstable boundary condition in the UV to a more stable one in the IR. A boundary RG fixed point is a scale-invariant boundary condition. For most systems, scale-invariance extends to conformal invariance, and we get a conformal boundary condition (which means a {\it conformally invariant} boundary condition). In the vicinity of a boundary RG fixed point, the scaling behaviour can be understood in terms of perturbing operators along the boundary of the system. In our example of the Ising model, the operator which is coupled to the magnetic field at the surface is the one corresponding to the local magnetization. This operator, which we note $\phi_a({\bf x})$, is located at the surface, therefore it is called a boundary operator. Like in the more familiar case of bulk critical phenomena, one can classify the boundary operators as relevant, irrelevant, or marginal, depending on their scaling dimension $h_a$. When such an operator appears as a perturbation at the boundary, it adds a term of the form $\lambda_a \int d^{d-1} {\bf x} ~\phi_a({\bf x})$ to the action of the theory. The coupling $\lambda_a$ scales like $\ell^{h_a+1-d}$, where $\ell$ is some UV cutoff, such as the lattice spacing if our system is a statistical model on a lattice. Under the RG flow, $\phi_a$ is said to be relevant when $h_a < d-1$, irrelevant if $h_a>d-1$, and marginal if $h_a = d-1$. Generically, all the operators which respect the symmetries of the system are expected to appear as boundary perturbations. For more information on surface critical phenomena and boundary RG flows, we refer the reader to \cite{Diehl,CardyBook}. Now, let us use this framework to analyze the corrections to scaling for the overlaps between the edge states.

\subsubsection{Locality of the boundary perturbation}
As explained in the previous sections, our result (\ref{eq:edge_conjecture}) relies on the fact that, in the thermodynamic limit and assuming generalized screening, we are left with a non-chiral CFT that lives outside the droplet, constrained by a boundary condition along the edge of the droplet. The interior of the droplet decouples from the exterior thanks to screening. In this framework, it is natural to include boundary perturbations that modify the conformal boundary condition. The action of the field theory, $S_{\rm CFT}$, is then modified along the circle $|z|=R$ by boundary perturbations $S_{\rm CFT} \rightarrow S_{\rm CFT} + S_b$. The latter are of the form
\begin{equation}
	\label{eq:perturbation_2}
	S_{b} \, =  \, \sum_a \lambda_a \int_{|z|=R}|dz| \, \phi_a(z)
\end{equation}
where the $\phi_a$'s are some local boundary operators with scaling dimensions $h_a$, and with coupling constants $\lambda_a$. The boundary condition (\ref{eq:bc}) should be stable under the RG flow, which means that there can be no relevant perturbation, namely all the scaling dimensions satisfy $h_a \geq 1$. The boundary perturbation (\ref{eq:perturbation_2}) might look completely generic, however one should emphasize that it implicitly assumes locality, in the sense that it is a sum of local operators. This holds thanks to the locality of the action of the CFT outside the droplet, and thanks to the screening assumption inside it. Without screening, the degrees of freedom along the edge might be coupled at long distances through the bulk, which would typically lead to non-local perturbations of the action, $S_{\rm CFT}$. Since we are assuming screening, this cannot happen, and the perturbation (\ref{eq:perturbation_2}) has the most generic form. The coupling $\lambda_a$ scales with some power of the UV cutoff, which is of the order of the mean particle spacing close to the edge, or equivalently $\sqrt{\rho_0}^{-1}$ where $\rho_0$ is the mean particle density close to the edge. Thus, $\lambda_a \sim (\sqrt{\rho_0}^{-1})^{h_a-1}$. Note that the perturbation (\ref{eq:perturbation_2}) must be real, so $S_b$ is hermitian: $S_b = S_b^\dagger$.

The perturbation (\ref{eq:perturbation_2}) modifies our formula for the overlaps between the edge states in the scaling region (which implies $\sqrt{\rho_0}^{-1}/R \rightarrow 0$):
\begin{equation}
	\label{eq:perturb_conj1_2}
	\left<\left<\, \Psi_{\left< v_1\right|} \,\left|\, \Psi_{\left< v_2\right|} \,\right>\right>\right. \, = \, \frac{\left< v_2 \right| e^{-S_b} \left| v_1 \right>}{\left< e^{-S_b} \right>} \, .
\end{equation}
Of course, the leading order in this formula is nothing but the universal result (\ref{eq:edge_conjecture}), but this refined expression generates the corrections to scaling we are interested in. The first correction comes from the least irrelevant operator ({\it i.e.} with the smaller $h_a$) and leads to a term of order $(\sqrt{\rho_0}^{-1}/R)^{h_a-1} \sim (1/\sqrt{N})^{h_a-1}$. The denominator $\left< e^{-S_b} \right> = \left< N \right| e^{-S_b} \left| N \right>$ is fixed by the requirement that the ground state $\left|\left. \Psi_{\left<N\right|} \right>\right>$ is normalized: $\left<\left< \, \Psi_{\left< N \right|} \, \left| \, \Psi_{\left< N \right|} \, \right>\right>\right. = 1$. By redefining $S_b \rightarrow S_b + {\rm constant}$, one can absorb this denominator in the definition of $S_b$ itself. This is what we do in the following, and we have thus the following formula for the inner products:
\begin{equation}
	\label{eq:perturb_conj_2}
	\left<\left<\, \Psi_{\left< v_1\right|} \,\left|\, \Psi_{\left< v_2\right|} \,\right>\right>\right. \, = \, \left< v_2 \right| e^{-S_b} \left| v_1 \right> \, .
\end{equation}

\subsubsection{RG analysis of the corrections to scaling: \\ the example of the Laughlin state}
\label{sec:corrections_Laughlin}
So far, we have just expressed the fact that, if generalized screening holds, then the corrections to scaling for the overlaps can be understood in terms of {\it local} boundary perturbations to the {\it local} boundary condition (\ref{eq:bc}). The next step is to discuss what local terms are allowed in $S_b$, namely what are the least irrelevant terms that are compatible with the symmetries of the system. This requires a case by case analysis. Let us do this for the Laughlin state in some more details now.

First, we note that, for different particle number $N \neq N'$, two states $\left| \left. \psi_{\left<N,k\right|} \right>\right>$
and $\left| \left. \psi_{\left<N',k'\right|} \right>\right>$ always have a zero overlap. This rules out the possibility of having the operator $e^{i \varphi(z)/\sqrt{\nu}}$ or $e^{-i \varphi(z)/\sqrt{\nu}}$ in the boundary perturbation $S_b$, or any other vertex operator, as it would allow non-zero overlaps between states with different particle numbers. In other words, we must have $[J_0,S_b]=0$, where $J_0$ is the zero mode of the $U(1)$-current, which is the number operator. Similarly, because of rotational invariance, two edge states with different angular momenta have zero overlap, which can be expressed as the contraint $[L_0,S_b]=0$. The most generic local perturbation along the boundary takes the form of a sum of polynomials in the (derivatives of the) $U(1)$ current $i\partial \varphi(z)$. To avoid technical issues caused by the extensive $U(1)$ charge of the droplet, it is more convenient to work with the {\it shifted} chiral bosonic field $\tilde{\varphi}(z) = \varphi(z) + i \frac{N}{\sqrt{\nu}} \log z$. It is defined such that $i \partial \tilde{\varphi}(z) = \frac{1}{z}\tilde{J}_0 + \sum_{n\neq 0} z^{n-1} J_{-n}$, where $J_n = \frac{1}{2\pi i} \oint d\zeta ~\zeta^n i \partial \varphi(\zeta)$ is a Fourier mode of the original ({\it i.e.} not shifted) $U(1)$ current $i \partial \varphi(z)$ (see also section \ref{sec:edge}.1), and only the zero mode is shifted: $\tilde{J}_0 = J_0 - N/\sqrt{\nu}$. This ensures that, when it acts on the CFT vacuum with $N + \Delta N$ charges, the eigenvalue of $\tilde{J}_0$ is $\tilde{J}_0 \left| N +\Delta N \right> = \Delta N/\sqrt{\nu} \left| N +\Delta N \right>$, which is of order $O(1)$ in the scaling region, while the eigenvalue of $J_0$ would rather be of order $O(N)$. Leaving aside these technicalities for now, the most generic boundary perturbation has the form
\begin{eqnarray}
	\label{eq:SbA_Laughlin_2}
	&& S_b \, =   \\
	\nonumber &&  \sum_{\{k\}} \lambda_{\{k \}} \oint \frac{dz}{2\pi i} \left(\frac{z}{R}\right)^{k_1+\dots+k_p-1} ~ (i\partial_z^{k_1} \tilde{\varphi})\dots (i\partial_z^{k_p} \tilde{\varphi})(z),
\end{eqnarray}
where the sum runs over the finite sets $\{ k\} = \{ k_1,\dots,k_p\}$ with $k_1,\dots,k_p \geq 1$. The polynomials in $i\partial \tilde{\varphi}(z)$ are normal-ordered, and the factor $\frac{z}{R}$ can be viewed as the jacobian (up to factors $2\pi$) of the conformal mapping from the cylinder to the plane. It ensures, in particular, that $S_b$ does not break rotational invariance: $[L_0,S_b]=0$. The coupling $\lambda_{\{k\}}$ is of order $(\sqrt{\rho_0}^{-1})^{k_1+\dots+k_p-1}$; it leads to a correction of order $(\sqrt{\rho_0}^{-1}/R)^{k_1 +\dots +k_p -1} \sim  (1/\sqrt{N})^{k_1 +\dots +k_p -1}$.

The least irrelevant operator is actually marginal: it is the $U(1)$-current $i \partial \tilde{\varphi}(z)$ itself, which has a scaling dimension $1$. Its zero mode is nothing but the number operator $\tilde{J}_0$, so it plays a role only when sectors with different particle numbers are involved. The weights of these different sectors may eventually be fixed such that the inner product do not depend on the number operator at the leading order, and thus the associated coefficient $\lambda_{\{1\}}$ is zero. Actually, this is exactly what we did in the previous sections, when we explained that the somewhat mysterious coefficient $\lambda$ in (\ref{eq:Zrho0}) and the subsequent equations could be chosen arbitrarily, and we tuned it such that the isometry (\ref{eq:edge_conjecture}) holds not only for neutral excitations, but also for charged excitations. We see now that the fact that this coefficient required some fine-tuning was simply due to the presence of a marginal operator. That being said, we assume as previously that the weights of the sectors with different particle numbers are tuned such that the $U(1)$ current does not appear at this order, so we can safely turn to the next boundary perturbations, which are all irrelevant for the Laughlin state.

The next possible contribution corresponds to the stress-tensor---more precisely, a ``shifted'' stress-tensor, which involves the shifted zero mode $\tilde{J}_0$ rather than $J_0$--- $\tilde{T}(z) = \frac{1}{2}(i\partial \tilde{\varphi})^2(z)$, with scaling dimension $2$. It turns out that this term cannot appear in $S_b$, at least for the inner products of the edge states in the plane or in the sphere that we are considering in this part (however, it will appear later, in a modified version of these inner products associated with the real-space entanglement spectrum). The reason why the stress tensor $\tilde{T}(z)$ cannot appear in $S_b$ is essentially translational invariance in the plane (or rotational invariance on the sphere); we come back to that in more details in the next part (section \ref{sec:ES_PP}). For now, let us focus on the other possible operators. There is another candidate with scaling dimension $2$: $i\partial^2 \tilde{\varphi}(z)$, which is nothing but the derivative of the $U(1)$-current, so one can again dismiss it. Thus, there are actually no perturbating operators of scaling dimension $2$ in $S_b$.

There are three operators with scaling dimension $3$, namely $(i\partial \tilde{\varphi})^3(z)$, $(i\partial \tilde{\varphi})(i\partial^2 \tilde{\varphi})(z)$, and $(i\partial^3 \tilde{\varphi})(z)$. One can show (for instance using translation invariance, as below), that the leading contribution to $S_b$ for the Laughlin state comes from $(i \partial \tilde{\varphi})^3 (z)$, and the two other operators---which are total derivatives---don't contribute. Then, at the next order there are five possible operators with dimension $4$, and so on.

In conclusion, for the Laughlin state, the overlaps in the plane and on the sphere are given by the formula (\ref{eq:perturb_conj_2}) where the leading contribution to $S_b$ is of order $O(1/N)$, and is given by
\begin{equation}
	\label{eq:Laughlin_W3}
	S_b \, = \, {\rm const.} \times \frac{1}{N} \oint \frac{d z}{2\pi i} ~z^{2} (i\partial \tilde{\varphi})^3(z) \;+\; O(N^{-3/2})  \, .
\end{equation}
The proportionality constant in (\ref{eq:Laughlin_W3}) is some non-universal number. In general, such coefficients cannot be determined simply by symmetry arguments, like those we are giving here. The constraint derived in the next paragraph for inner products in the plane may help fixing a few of these coefficients; in full generality, however, there will always remain some undetermined coefficients that cannot be calculated by the methods we use in this paper.

\subsubsection{Translation invariance constraint on the boundary perturbation $S_b$ for the overlaps in the plane}
\label{sec:loop_equation}
In this paragraph, we derive a constraint on the boundary perturbation $S_b$ that generates the corrections to scaling in the formula (\ref{eq:perturb_conj_2}), which holds for any quantum Hall states given by the MR construction (not only the Laughlin state). The constraint is valid in the plane, namely when one uses the potential $V(z,\overline{z}) = -\frac{|z|^2}{2\ell_B^2}$ (see (\ref{eq:potential})) in the integration measure. It essentially expresses translation invariance in the plane, and it is written as:
\begin{equation}
	\label{eq:loop_equation}
	[e^{-S_b},J_{-1}] \, = \, \frac{\sqrt{\nu}}{N}  \tilde{L}_{-1} ~e^{-S_b} ,
\end{equation}
where $J_{-1}$ is the first Fourier mode of the $U(1)$ current, and $\tilde{L}_{-1}$ is the Fourier mode of the {\it total} stress-tensor of the CFT, which involves both the $U(1)$ sector and the statistics sector: $\tilde{T}^{U(1)}(z) + T^\psi(z)$. The reason why we use the notation $\tilde{T}(z)$ is the same as above: in the $U(1)$ sector, to avoid technical problems due to the extensive $U(1)$ charge inside the droplet, we use the {\it shifted} current $i \partial \tilde{\varphi}(z)$ to construct the {\it shifted} stress-tensor in the $U(1)$ sector $\tilde{T}^{U(1)}(z) = \frac{1}{2}(i \partial \tilde{\varphi})^2(z)$. This shift does not affect the statistics sector at all. The Fourier mode $\tilde{L}_{-1}$ is simply defined as $\oint \frac{d\zeta}{2\pi i}  [\tilde{T}^{U(1)}(\zeta)+T^{\psi}(\zeta)]$, or, in other words, by
\begin{equation}
	\label{eq:shiftedL1}
	\tilde{L}_1 \, = \, \sum_{n>0} J_{-n-1} J_{n} + J_{-1}\tilde{J}_0  \; + \; L_{-1}^{\psi} \, .
\end{equation}
Before we explain how to derive the constraint (\ref{eq:loop_equation}) in the plane, let us explain why, when it is associated to the locality of $S_b$ expressed in (\ref{eq:perturbation_2}), it becomes a useful equation. For the Laughlin state $\tilde{T}(z) = \tilde{T}^{U(1)}(z)$, and as explained in the previous paragraph, $S_b$ could in principle include a leading contribution from the stress-tensor, which has scaling dimension $2$. Then we would have $S_b = \frac{\alpha}{\sqrt{N}} \tilde{L}_0 + O(N^{-1})$, for some undetermined coefficient $\alpha$ of order $O(1)$. When we plug this into the constraint (\ref{eq:loop_equation}), and expand $e^{-S_b} = 1 - \frac{\alpha}{\sqrt{N}} \tilde{L}_0 + O(N^{-1})$, we see that we get $\frac{\alpha}{\sqrt{N}} [\tilde{L}_0,J_{-1}] = O(N^{-1})$, which implies $\alpha = 0$. Thus, the absence of the $U(1)$ part of the stress-tensor in $S_b$ follows from the constraint (\ref{eq:loop_equation}). However, it is important to note that the constraint (\ref{eq:loop_equation}) does not prevent the statistics part of the stress-tensor, $T^\psi(z)$, to appear. Therefore, in general, one should expect to have a term proportional to $\frac{1}{\sqrt{N}} L^\psi_0$ in $S_b$. In particular, this means that the first correction to the universal formula for the inner products generated by the formula (\ref{eq:perturb_conj_2}) is usually of order $1/\sqrt{N}$ (at least) for the edge excitations in the statistics sector, while in the $U(1)$ sector the corrections are of order $1/N$. This nicely agrees whith our numerical results for the MR (Pfaffian) state in section \ref{sec:numerical}: the convergence towards our universal formula is much faster for the excitations in the $U(1)$ sector (because the corrections are of order $1/N$) than for those in the statistics sector (corrections of order $1/\sqrt{N}$).

More systematically, once one has postulated the local form of the perturbation $S_b$ (\ref{eq:perturbation_2}), one can use the constraint (\ref{eq:loop_equation}) recursively to analyze new terms in $S_b$, that are more and more irrelevant. This leads to a large-$N$ expansion for the inner products when the resulting formula for $S_b$ is plugged into (\ref{eq:perturb_conj_2}). Of course, not all possible terms are constrained by (\ref{eq:loop_equation}), as we have seen for instance for $L_0^\psi$. For the Laughlin state, since there is no statistics sector, the constraint is slightly more powerful. For example, it allows to compute the proportionality constant in (\ref{eq:Laughlin_W3}). At the next order $O(1/N^{3/2})$, however, the operators of scaling dimension $4$ appear, and these include $(i\partial^2 \tilde{\varphi})^2$, which commutes with $J_1$ (this operator will appear again in the entanglement spectrum). Thus the coefficient of this operator is not fixed by the constraint (\ref{eq:loop_equation}). As the scaling dimension increases, there are more and more operators that are not constrained by this relation.

Finally, let us show how to derive the constraint (\ref{eq:loop_equation}). For any pair of CFT states $\ket{v_1}$, $\ket{v_2}$ (with the same charge corresponding to $N$ particles), consider the overlap between the corresponding edge states $\Psi_{\left<v_1 \right|}$ and $\Psi_{\left< v_2 \right|}$ given by (\ref{eq:qm_innerprod}). Since all the coordinates $(z_i,\overline{z}_i)$'s are integrated over the entire complex plane $\mathbb{C}$, the integral is invariant under a reparametrization $(z_i,\overline{z}_i) \rightarrow (z_i + \epsilon,\overline{z}_i)$ corresponding to a translation of the entire droplet along some axis. For an infinitesimal $\epsilon$, this implies:
\begin{eqnarray}
\nonumber	0 &=& \epsilon \, \frac{1}{N!} \int_{\mathbb{C}} \prod_{k=1}^N d^2 z_k \sum_{i=1}^N \partial_{z_i} \left[ e^{-\frac{\sum_j|z_j|^2}{2\ell^2_B}} \right. \\
\nonumber	&&  \left. \left[ \Psi_{\left<v_1\right|}(z_1,\dots,z_N) \right]^* \Psi_{\left< v_2\right|}(x_1,\dots,z_N)  \right] \,.
\end{eqnarray}
Since the $\partial_{z_i}$ derivatives don't act on the $\overline{z}_i$ variables, we find
\begin{eqnarray}
\nonumber 0&=&  -\frac{1}{N!} \int_{\mathbb{C}} \prod_{k=1}^N d^2 z_k \left[e^{-\frac{\sum_j|z_j|^2}{2\ell^2_B}} \right. \\
\nonumber	&&  \left. [  \sum_i \frac{z_i}{2 \ell_B^2} \Psi_{\left<v_1\right|}(z_1,\dots,z_N) ]^* \Psi_{\left< v_2\right|}(x_1,\dots,z_N)  \right] \\
\nonumber &&+ \,  \frac{1}{N!} \int_{\mathbb{C}} \prod_{k=1}^N d^2 z_k \left[ e^{-\frac{\sum_j|z_j|^2}{2\ell^2_B}} \right. \\
\nonumber	&&  \left. \left[ \Psi_{\left<v_1\right|}(z_1,\dots,z_N) \right]^* \sum_i \partial_{z_i} \Psi_{\left< v_2\right|}(x_1,\dots,z_N)  \right] \, .
\end{eqnarray}
In other words, we just performed an integration by part, and used the fact that there are no contact terms since we integrate over the entire complex plane. Now, using the notations of this paper, we can rewrite this as
\begin{eqnarray}
\nonumber 0&=&  - \frac{R\sqrt{\nu}}{2\ell_B^2} \left<\left<\left. \, \Psi_{\left<v_1\right|J_1} \,\right| \, \Psi_{\left< v_2\right|}\,  \right>\right>  \\
\nonumber &&+ \, \frac{1}{R}  \left<\left< \left.\, \Psi_{\left<v_1\right|} \, \right| \, \Psi_{\left< v_2\right|L_{-1}} \, \right>\right> \, .
\end{eqnarray}
The factors $R$ come from the normalization of the edge states (\ref{eq:edge_def}). We have used the fact that, in conformal field theory, $L_{-1}$ is the generator of translations: it acts as $\sum_i \partial_{z_i}$ on the conformal block in $\Psi_{\left< v_2\right|}(z_1,\dots,z_N)$. Plugging this into (\ref{eq:perturb_conj_2}), and using the radius (\ref{eq:radius}) of the droplet in the plane, $R = \sqrt{2N/\nu}~\ell_B$, we find:
\begin{equation}
\label{eq:loop_deriv1}
\frac{N}{\sqrt{\nu}} \left<v_2 \right| e^{-S_b}  J_{-1} \left| v_1\right> \, = \,  \left< v_2 \right| L_{-1} ~ e^{-S_b} \left| v_1\right> \, .
\end{equation}
Here, the operator $L_{-1}$ is the Fourier mode of the full stress-tensor $T^{U(1)}(z)+T^\psi(z)$, and it is {\it not shifted}. The shifted mode $\tilde{L}_{-1} = (\tilde{L}_1)^\dagger$, given by (\ref{eq:shiftedL1}), is related to $L_{-1}$ by
\begin{equation}
	 \tilde{L}_{-1} \, =\, L_{-1} - \frac{N}{\sqrt{\nu}} J_{-1} \, .
\end{equation}
The advantage of using $\tilde{L}_{-1}$ rather than $L_{-1}$ is, once again, that it leads to terms of order $O(1)$, rather than terms of order $O(N)$ (or higher powers of $N$) when it is used in the formulas (\ref{eq:edge_conjecture}) or (\ref{eq:perturb_conj_2}). For instance, the overlaps $\left<\left< \Psi_{\left<v_1 \right|\tilde{L}_1} \left| \Psi_{\left< v_2\right|\tilde{L}_1}\right>\right>\right.$ and $\left<\left< \Psi_{\left<v_1 \right|L_1} \left| \Psi_{\left< v_2\right|L_1}\right>\right>\right.$ are {\it not} equal: the former is of order $O(1)$, while the latter is of order $O(N^2)$ when $N\rightarrow \infty$. Substituting $\tilde{L}_{-1}$ in 
(\ref{eq:loop_deriv1}), we find the constraint (\ref{eq:loop_equation}).

Let us finally emphasize that our results are intimately related to a series of papers by Zabrodin and Wiegmann, who have studied the behaviour of the Laughlin droplet as the potential $V(z,\bar{z})$ is varied, using the so-called ``loop equation'' \cite{ZabrodinWiegmann1,ZabrodinWiegmann2,ZabrodinWiegmann3}. Our constraint (\ref{eq:loop_equation}) is actually nothing but a reformulation, in the particular case of a quadratic potential $V(z,\overline{z})$, of their ``loop equation''.
The details of the calculations that relate the results of Zabrodin and Wiegmann to ours are beyond the scope of this paper, but we believe that our point of view, which emphasizes the crucial role of the screening assumption and its main consequences, which are the conformal boundary condition (\ref{eq:bc}) and the locality of the boundary perturbations (\ref{eq:perturbation_2}), is slightly different from theirs, and sheds some new light on their results.

\section{Entanglement spectrum}
\label{sec:ES}
We are ready to apply the formalism developed in parts \ref{sec:part_edge} and \ref{sec:screening} to the analysis of the entanglement spectrum of the ground state wavefunction. In particular, the universal result (\ref{eq:edge_conjecture}), which relates the overlaps between the edge states to the inner product in the CFT space, will play a key role in our analysis. Our main focus is on real-space partition (RSP, see section \ref{sec:intro_entanglement}). However, the techniques we use can also be applied to the case of the particle partition (PP) in the so-called ``scaling region'', as explained below (section \ref{sec:ES_PP}).

In this part it is more convenient to use a second quantized formalism. By definition, in second quantization, the wavefunction (\ref{eq:wfblock}) becomes
\begin{eqnarray}
	\label{eq:statepsi}
	\left|\left. \Psi \right>\right> & = & \frac{1}{N!} \int_{\mathbb{C}^N} \prod_{i=1}^N e^{V(z_i,\overline{z_i})/2} d^2 z_i \\
\nonumber	&& \qquad \quad  \Psi(z_1,\dots,z_N) \, c^\dagger(z_1) \dots c^\dagger(z_N)
\left|\left. 0 \right>\right> ,
\end{eqnarray}
where the $c^\dagger(z)$'s create fermions/bosons at position $z$ and $\left| \left.0 \right>\right>$ is the fermionic/bosonic vacuum annihilated by all the $c(z)$'s. The modes obey the canonical (anti-)commutation relations $\{c(z),c^\dagger(z') \} = \delta^{(2)}(z-z')$ (fermions) or $[ c(z),c(z') ] = \delta^{(2)}(z-z')$ (bosons). To define the RSP, we fix $R>0$, and we define $A$ as the disc of radius $R$, $A= \{ z\in \mathbb{C};\, |z|<R \}$, and $B$ as the complementary subset $B = \mathbb{C} \setminus A$. Our goal is to compute the Schmidt decomposition
\begin{equation}
	\left|\left. \Psi \right> \right> \, = \, \sum_{N_A=0}^N \sum_{k} e^{-\xi(N_A,k)/2} \left|\left. \Psi^A_{N_A,k} \right>\right>  \left|\left. \Psi^B_{N_B,k} \right>\right>
\end{equation}
where $\{ \left|\right. \Psi^A_{N_A,k} \left>\right> \}$ (as well as $\{ \left|\right. \Psi^B_{N_B,k} \left>\right> \}$) is a set of orthogonal states with $N_A$ particles in part $A$ (and $N_B=N-N_A$ particles in part $B$). The set of pseudoenergies $\xi(N_A,k)$ is the entanglement spectrum \cite{HaldaneLi}.

Since the bipartition $A\cup B$ is rotationally invariant, the angular momentum in part $A$, $L_{\hat{z}}^A$, is a good quantum number. The Schmidt eigenvalues/eigenvectors can be classified according to $L_{\hat{z}}^A$. As claimed in section \ref{sec:intro_entanglement}, in general, there is a non-degenerate lowest pseudoenergy $\xi$ at some values $N_{A0}$ and $L_{\hat{z}0}^A$. These values depend on the radius $R$. In general, the radius $R$ is the one of a droplet with $N_{A0}$ particles given by (\ref{eq:radius}), and $L_{\hat{z}0}^A$ is the angular momentum of the ground state (\ref{eq:wfblock}) with $N_{A0}$ particles. We define $\Delta \xi$, $\Delta N_A$ and $\Delta L_{\hat{z}}^A$ by subtracting off these values.

\subsection{Decomposition of the ground state from \\ a completeness relation in the CFT space}
\label{sec:completeness}

\begin{figure}
	\includegraphics[width=0.48\textwidth]{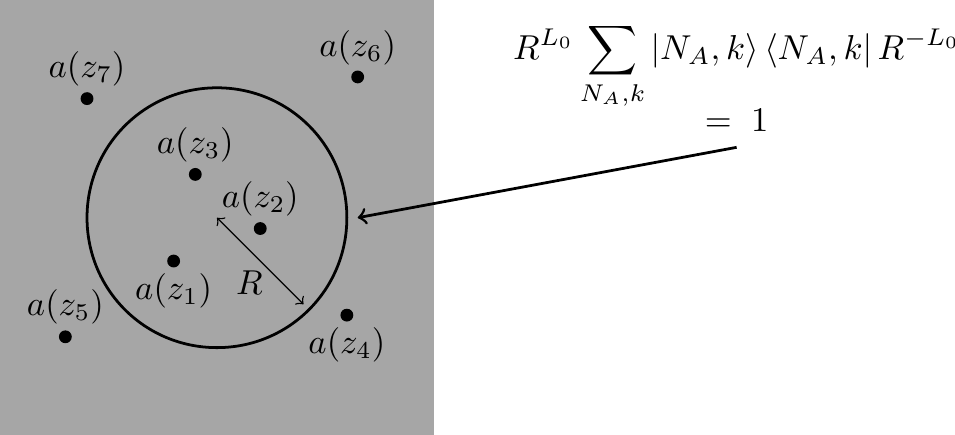}
	\caption{We use the completeness relation (\ref{eq:completeness_R}) to decompose the conformal correlator $\left< \prod_{i} a(z_i) \right>$ as a sum of products of one correlator which involve only particles in part $A$ ($|z|<R$) with another one which involves particles in part $B$ ($|z|>R$).}
	\label{fig:completeness}
\end{figure}

In this section, we use a completeness relation to decompose the ground state as a sum of products of terms of the form $\left|\left.\Psi^A \right>\right> \left|\left. \Psi^B \right>\right>$. In the end, these terms will turn out to be orthogonal only in the thermodynamic limit, when $N_A,N_B \rightarrow \infty$ (and assuming screening), so this will be a true Schmidt decomposition only in that limit. However, it will not be difficult, once our procedure has been exposed, to adapt it to take into account corrections to scaling. This will be explained later in section \ref{sec:finitesize}. For now, we initiate the process of calculating the decomposition in a naive way, relying only on the fact that the ground state wavefunction is a conformal block.

Like in section \ref{sec:edge}, we think of the CFT in radial quantization, and we use a complete set of states in the CFT Hilbert space to write the identity operator as
\begin{equation}
	\label{eq:completeness}
	1 \, = \, \sum_{N_A,k} \left| N_A, k\right> \left< N_A,k \right| \, .
\end{equation}
In this expression, $N_A$ is the number of particles, and $k$ stands for all the excited states in a given $N_A$ sector (called the {\it descendants} in the CFT literature). 
Strictly speaking, in radial quantization, the operator (\ref{eq:completeness}) actually acts on the CFT states corresponding to the field configurations on a circle at radius $|z|=1$, so we need to rescale this operator such that it acts at $|z|=R$ (see also Fig. \ref{fig:completeness})
\begin{equation}
	\label{eq:completeness_R}
	1 \, = \, \sum_{N_A,k} R^{ L_0} \left| N_A, k\right> \left< N_A,k \right| R^{-L_0} \, .
\end{equation}
Now we write the state (\ref{eq:statepsi}) as
\begin{eqnarray}
	\label{eq:statepsi_exp}
	&& \left| \left. \psi \right>\right> \, = \, \frac{1}{\lambda^{\frac{N}{2}} ~\sqrt{Z_N}}  \times \\
\nonumber	&& \left< N \right| \mathcal{R} \exp \left[\lambda^{\frac{1}{2}} \int_{\mathbb{C}} e^{V(z,\overline{z})/2} d^2 z ~ a(z) \otimes c^\dagger(z)  \right] \left| 0\right> \left|\left. 0 \right>\right> \, .
\end{eqnarray}
Note that $\left| 0\right>$ is the CFT vacuum, while $\left|\left. 0 \right>\right>$ is the physical fermionic/bosonic vacuum (annihilated by the $c(z)$'s). The symbol $\mathcal{R}$ denotes radial ordering (see \cite{BYB,SaleurItzyksonZuber,Mussardo}). The coefficient $\lambda$ will be useful later, as a tunable parameter associated with the terms with different particle numbers in $A$ and $B$. When one expands the exponential, only the term with $N$ particles in total remains, because of the projection onto the vacuum with $N$ charges $\left| N\right>$. This expression is then exactly equal to (\ref{eq:statepsi}). The integration over $\mathbb{C} = A \cup B$ can be split into an integration of $A$ and another over $B$, so the exponential is equal to
\begin{eqnarray}
	\label{eq:two_exp}
\nonumber	&& \mathcal{R} \exp \left[ \lambda^{\frac{1}{2}} \int_{B} e^{V(z,\overline{z})/2} d^2 z ~ a(z) \otimes c^\dagger(z)  \right] \\ 
	&& \qquad \times \quad 
\mathcal{R} \exp \left[\lambda^{\frac{1}{2}} \int_{A} e^{V(z,\overline{z})/2} d^2 z ~ a(z) \otimes c^\dagger(z)  \right] . \quad 
\end{eqnarray}
Then we insert the completeness relation (\ref{eq:completeness_R}) between these two exponentials (Fig. \ref{fig:completeness}). The result can be written as
\begin{equation}
	\label{eq:semi_decomp}
	\left|\left. \Psi \right>\right> \, = \,  \sqrt{\frac{Z_{N_{A0}}\, Z_{N_{B0}}}{Z_N}} \sum_{ N_A=0}^N \sum_k   \left|\left. \,\Psi^A_{\left< N_A,k \right| }\, \right>\right>  \, \left|\left. \,\Psi_{\left| N_A,k \right>}^B\, \right>\right> 
\end{equation}
where $N_{B0}=N-N_{A0}$, and
\begin{subequations}
\label{eq:decomp_states}
\begin{eqnarray}
	&& \left|\left. \, \Psi^A_{\left< N_A,k \right|}\, \right>\right>  \, = \, \frac{1}{\lambda^{\frac{N_{A0}}{2}}~ \sqrt{Z_{N_{A0}}}} \left< N_A,k \right|\frac{1}{R^{\Delta L_0}}  \; \times  \\
\nonumber	&&  \qquad \mathcal{R}  \exp \left[\lambda^{\frac{1}{2}} \int_A e^{V(z,\overline{z})/2} d^2z ~ a(z) \otimes c^\dagger(z) \right] \left|0 \right>\, \left|\left. 0 \right>\right>    \\
	&& \left|\left. \, \Psi^B_{\left|N_A,k\right>} \, \right>\right>  \, = \, \frac{1}{\lambda^{\frac{N_{B0}}{2}}~\sqrt{Z_{N_{B0}}}} \left< N \right| \; \times \\
\nonumber   &&  \mathcal{R} \exp \left[\lambda^{\frac{1}{2}} \int_B e^{V(z,\overline{z})/2} d^2z ~ a(z) \otimes c^\dagger(z) \right] R^{\Delta L_0} \left|N_A,k \right>\, \left|\left. 0 \right>\right> .
\end{eqnarray}
\end{subequations}
We have used the notation $\Delta L_0 = L_0 - \left< N_{A0}\right| L_0 \left| N_{A0}\right>$. The normalizing factors $Z_{N_{A0}}$ and $Z_{N_{B0}}$ are fixed such that the
states $\left|\left. \, \Psi_{\left< N_{A0} \right|}^A \, \right>\right>$ and $\left|\left. \, \Psi_{\left| N_{A0} \right>}^B \, \right>\right>$ corresponding to the vacuum $\left|N_{A0} \right>$ are both normalized,
\begin{subequations}
\label{eq:decomp_ZN}
\begin{eqnarray}
 \nonumber Z_{N_{A0}} & = & \frac{1}{N_{A0}!}\int_{A} \prod_{i=1}^{N_{A0}} e^{V(z_i,\overline{z_i})} d^2 z_i\; |\left< N_{A0} \right| \prod_{i=1}^{N_{A0}} a(z_i) \left| 0 \right> |^2  \\   \\
\nonumber Z_{N_{B0}} & = & \frac{1}{N_{B0}!}\int_{B} \prod_{i=1}^{N_{B0}} e^{V(z_i,\overline{z_i})} d^2 z_i\; |\left< N \right| \prod_{i=1}^{N_{B0}} a(z_i) \left| N_{A0} \right> |^2 . \\
\end{eqnarray}
\end{subequations}
The states $\left|\left.\Psi^A_{\left< N_A,k \right|} \right>\right>$ appearing in the decomposition (\ref{eq:semi_decomp}) should be viewed as the edge excited states that we constructed in part \ref{sec:part_edge}. Similarly, the states $\left| \left. \Psi^B_{\left|N_A,k \right>} \right>\right>$ can be viewed as the edge excited states for part $B$ (although, strictly speaking, we have not constructed these states in section \ref{sec:edge}, the extension of our formalism to the part $B$ is straightforward). One difference with the edge states introduced previously is the integration domain: here the positions of the particles are restricted to the domain $A$, so the $\left|\left. \Psi^A_{\left< N_A,k \right|} \right>\right>$'s are projections onto the domain $A$ of the edge states of part \ref{sec:part_edge}. With that minor distinction in mind, we see that the formula (\ref{eq:semi_decomp}) is a natural decomposition of the ground state as a sum of products of edge excited states for part $A$ and part $B$. It is quite close to the Schmidt decomposition we are looking for, except that the states appearing in the right-hand side of (\ref{eq:semi_decomp}) are not necessarily orthogonal. In the next section we argue that, in the ``scaling region'' (to be defined below), the states are othogonal if the screening assumption holds, such that we can use our result for the overlaps between the edge states (\ref{eq:edge_conjecture}).

It is worth emphasizing that, although the decomposition (\ref{eq:semi_decomp}) may look quite formal, it becomes very explicit when one considers concrete examples. In the case of the Laughlin wave function, for instance, the decomposition of the ground state in terms of the edge state wavefunctions becomes very elegant; it can be found explicitly in \cite{DRR}. For the MR (Pfaffian) wavefunction (\ref{eq:MRPfaffian}), it is an instructive exercise to write down the decomposition of the Pfaffian in terms of the edge excited states (\ref{eq:MRPfaffian_edge}).

\subsection{Pseudo-energies at the leading order\\ in the scaling region}
\label{sec:scaling_region}

Let us start by defining what we call the {\it scaling region}. It is the extension of the concept of ``scaling region'' that we used previously in part \ref{sec:screening}, which needs to be adapted to the bipartition $A \cup B$.  Roughly speaking, it is the set
of Schmidt eigenvalues and eigenvectors for which $\Delta N_A$ is small compared to $N$, and $\Delta L_{\hat{z}}^A$ is small compared to $L_{\hat{z}0}^A$. More precisely, we imagine that the radius $R$ of the disc $A$ depends on $N$ such that $N_{A0}/N$ is fixed. This is the situation that is usually considered in numerical studies, where $A$ and $B$ correspond to the two hemispheres of the sphere after stereographic projection, with a ratio $N_{A0}/N = 1/2$ (for an even number of particles on the sphere). Then we fix some positive number $M$, and keep all the eigenvalues and eigenvectors such that $|\Delta N_A|, |\Delta L_{\hat{z}}^A| \leq M$. Finally, we take the thermodynamic limit $N \rightarrow \infty$. This scaling region is the one where the entanglement spectrum is conjectured to possess the ``scaling property'' \cite{DRR}. Note that, by definition, in the scaling region, both $N_A$ and $N_B$ go to infinity. In this section we analyze the limiting behavior of the pseudoenergies in this range of quantum numbers $\Delta N_A$, $\Delta L_{\hat{z}}^A$. We do this only at the leading order; subleading corrections will be tackled in the next section.

First, we focus on the prefactor appearing in the right-hand side of (\ref{eq:semi_decomp}). This is a ratio of partition functions,
\begin{equation}
	\label{eq:ratioZ}
	\sqrt{\frac{Z_{N_{A0}} Z_{N_{B0}}}{Z_N}} = \exp\left[{-(f_{N_{A0}} +f_{N_{B0}} - f_N)/2} \right],
\end{equation}
where $f_{N_{A0}} = - \log Z_{N_{A0}}$, etc. Each of these three terms in the exponential is the free energy of a statistical system of itinerant particles in part $A$, part $B$, or in $\mathbb{C}=A \cup B$. To discuss them, it is natural to adopt a statistical mechanics point of view as in part \ref{sec:screening}. The following argument is standard in that context; it has been used in several works (see for instance \cite{CalabreseCardy}) to support the area law for the entanglement entropy \cite{arealaw} or other similar quantities such as the (logarithm of the) ratio (\ref{eq:ratioZ}) \cite{DS1}. The leading contribution to the free energy is a bulk part, which in general could be expressed as a functional of the local density of particles inside the domain. This bulk part is usually extensive for statistical systems with short-range interactions, or super-extensive if the interactions are long-range, like Coulomb interactions in the plasma mapping for the Laughlin wavefunction. Another contribution, which is subleading, comes from the boundary of the statistical system. It usually scales linearly with the length of the boundary, and may be interpreted as a surface tension. The free energies $f_{N_{A0}}$ and $f_{N_{B0}}$ should both contain such a surface tension term, proportional to the length of the cut between $A$ and $B$ (that is $2\pi R$). The bulk parts of $f_{N_{A0}}$ and $f_{N_{B0}}$ cancel the bulk part of $f_{N}$, leaving only the surface tension term at the boundary between $A$ and $B$. Therefore, we are left with
\begin{equation}
	\label{eq:shiftES}
	-\log \left[ \sqrt{\frac{Z_{N_{A0}} Z_{N_{B0}}}{Z_N}} \right] \, \sim \, \alpha ~ 2\pi R  -  \gamma  +  o(1)
\end{equation}
when $R$ is large compared to mean particle spacing close to the cut (which plays the role of the UV cutoff). The coefficient $\alpha$ is non-universal and depends on the UV cutoff. We have also included a possible $O(1)$ term, noted $-\gamma$. This term is dimensionless, and it is universal: it is the topological entanglement entropy \cite{KitaevPreskill,LevinWen}. The presence of this term here can be shown by adapting the original argument of KP; this is explained in more details in the appendix \ref{appendix:topoterm}.

Second, assuming screening, in the limit $N_A \rightarrow \infty$, the overlaps between the states $\left|\left.\, \Psi^A_{\left< N_A,k \right|} \, \right>\right>$ are given by our result (\ref{eq:edge_conjecture}):
\begin{equation}
	\label{eq:limitA}
	\left<\left< \, \Psi^A_{\left< N_A ,k\right|} \, \left| \, \Psi^A_{\left< N_A,k'\right|} \, \right> \right>\right. \, \underset{N_A\rightarrow \infty}{\longrightarrow}  \, \left< N_A,k' \left| N_A,k \right>\right. \, = \, \delta_{k,k'} .
\end{equation}
Here we have used the fact that the basis of states $\{ \left| N_A,k \right> \}$ in (\ref{eq:completeness}) is orthonormal. 
States with different particle number $N_A' \neq N_A$ are always orthogonal. 
The same formula can be obtained for the overlaps between the states in part $B$ by a straightforward extension of the arguments developed in part \ref{sec:screening},
\begin{equation}
	\label{eq:limitB}
	\left<\left< \, \Psi^B_{\left| N_A ,k\right>} \, \left| \, \Psi^B_{\left| N_A,k'\right>} \, \right> \right>\right. \, \underset{N_B\rightarrow \infty}{\longrightarrow} \, \left< N_A,k \left| N_A,k' \right>\right. \, = \, \delta_{k,k'} .
\end{equation}
One might be worried about the integration domains: here the positions of the particles $z_i$ are integrated only over part $A$ (or $B$), while the result (\ref{eq:edge_conjecture}) was claimed for overlaps computed by integrating over the whole plane $\mathbb{C}$. This does not make any difference though, as long as we focus on the states $\left| \left. \Psi^A_{\left|N_A,k \right>} \right>\right>$ or $\left| \left. \Psi^B_{\left<N_A,k \right|} \right>\right>$ that are in the scaling region. For these states, the contribution of the integration over the whole plane is exponentially suppressed by the factor $e^{V(z,\bar{z})}$ outside the droplet of radius $R$, so the difference between the overlaps computed by integrating over $A$ ($B$) or over $\mathbb{C}$ is exponentially small. The choice of the non-universal coefficient $\lambda$ in (\ref{eq:statepsi_exp}), which must be tuned in order to give the correct weight to the terms with different particle numbers, might be sensitive to the fact that one is integrating over $A$ or $B$ rather than the whole complex plane $\mathbb{C}$. However, as discussed at the end of sction \ref{sec:edge_conjecture}, there is always one choice of $\lambda$ that leads to the conformal boundary condition (\ref{eq:bc}) and to the universal result (\ref{eq:edge_conjecture}). Therefore the limits (\ref{eq:limitA}) and (\ref{eq:limitB}) are correct, and the subtleties about the integration domains will become relevant only in the next section, when we deal with subleading corrections to these formulas.

Let us come back to the decomposition (\ref{eq:semi_decomp}). We have reached the conclusion that, in the scaling region, the two sets of vectors in the right-hand side are orthogonal at the leading order. Thus, in that limit, (\ref{eq:semi_decomp}) is really a Schmidt decomposition. In particular, this implies that in each $L^A_{\hat{z}}$ subsector, the Schmidt rank matches exactly the dimension of the corresponding subspace of angular momentum eigenstates in the conformal field theory. Also, the Schmidt eigenvectors turn out to be exactly the edge states constructed in section \ref{sec:part_edge}, up to the subtleties involving the integration domain. The pseudoenergies are all equal to
\begin{equation}
	\label{eq:shiftES2}
	\xi(\Delta N_A,k) \, = \, \alpha ~2\pi R  - \gamma + o(1)
\end{equation}
in the scaling region, independently from $\Delta N_A$ and from the descendant labelled by $k$. 
The splitting between the pseudoenergies appears only at higher order, when one tackles subleading corrections to the formulas (\ref{eq:limitA}-\ref{eq:limitB}), as we do in the next section.

\subsection{Corrections to scaling, RG analysis, and locality of the pseudo-Hamiltonian}
\label{sec:finitesize}

As in section \ref{sec:corrections_overlaps}, we adopt a statistical mechanics point of view, and we use ideas from the theory of surface critical phenomena \cite{Diehl,CardyBook}. Let us start by summarizing the main points discussed in section \ref{sec:corrections_overlaps}. Our main result (\ref{eq:edge_conjecture}) and its variations (\ref{eq:limitA}-\ref{eq:limitB}), which are useful for the calculation of the entanglement spectrum, rely on the fact that, in the thermodynamic limit and assuming generalized screening, we are left with a non-chiral CFT that lives outside the droplet, constrained by a conformal boundary condition along the edge of the droplet. The interior of the droplet decouples from the exterior thanks to screening. To tackle the corrections to scaling, one has to include boundary perturbations. The action of the field theory, $S_{\rm CFT}$, is modified along the circle $|z|=R$ by these boundary perturbations $S_{\rm CFT} \rightarrow S_{\rm CFT} + S_b$, where $S_b$ has the generic form
\begin{equation}
	\label{eq:perturbation}
	S_{b} \, =  \, \sum_a \lambda_a \int_{|z|=R}|dz| \, \phi_a(z) \, .
\end{equation}
The $\phi_a$'s are local boundary operators with scaling dimensions $h_a$, and with coupling constants $\lambda_a \sim (\sqrt{\rho_0}^{-1})^{h_a-1}$, where $\rho_0$ is the mean particle density close to the edge. The perturbation (\ref{eq:perturbation}) modifies our formula for the overlaps between the edge states in the scaling region, as in formula (\ref{eq:perturb_conj_2}). Here, since we are dealing with overlaps in part $A$, this formula becomes
\begin{equation}
	\label{eq:perturb_conj}
	\left<\left<\, \Psi^A_{\left< v_1\right|} \,\left|\, \Psi^A_{\left< v_2\right|} \,\right>\right>\right. \, = \, \left< v_2 \right| e^{-S_b(A)} \left| v_1 \right> ,
\end{equation}
and the same formula holds for part $B$. However, since the boundary perturbation $S_b$ is not necessarily the same for $A$ and $B$, in what follows will label it by $S_b(A)$ or $S_b(B)$. The formula (\ref{eq:perturb_conj}), for parts $A$ and $B$, generate the subleading corrections to the universal formula for the overlaps, and determine the ES, as we show next.

\subsubsection{Locality of the pseudo-Hamiltonian}
The physical space of edge states is isometric to the CFT Hilbert space, up to corrections that are {\it local} along the edge, and that are encoded in the operators $S_b(A)$ and $S_b(B)$. Now we show that this property implies that the ES is the spectrum of a pseudo-Hamiltonian that is {\it local} along the cut between $A$ and $B$. The ES is related to $S_b(A)$ and $S_b(B)$ as follows. Let us consider the operator $S_{\rm ES}$ defined by
\begin{equation}
	\label{eq:eSBeSA}
	e^{-\frac{S_{\rm ES}}{2}} \, =\, e^{-\frac{S_b(B)}{2}} e^{-\frac{S_b(A)}{2}},
\end{equation}
which acts in the Hilbert space of the chiral CFT. This operator is nothing but the ``pseudo-Hamiltonian'' we are looking for: its spectrum {\it is} the ES, up to the global additive constant (\ref{eq:shiftES2}). To see that, consider the singular value decomposition (SVD) of $e^{-\frac{S_{\rm ES}}{2}}$,
\begin{equation}
	e^{-\frac{S_{\rm ES}}{2}} \, = \, \sum_{N_A,k} e^{- \frac{\Delta \xi(N_A, k)}{2}} \left| u_{N_A,k} \right> \left< v_{N_A,k} \right|  .
\end{equation}
By definition of the SVD, $\{ \left| u_{N_A,k} \right> \}$ and $\{ \left| v_{N_A,k} \right> \}$ are two sets of orthogonal vectors in the CFT Hilbert space. Here the $\Delta \xi(N_A,k)$'s are simply the eigenvalues of $S_{\rm ES}$. Next, we write the identity operator as
\begin{eqnarray}
	&& 1  \; =    \\  \nonumber &&  R^{L_0}~ e^{\frac{S_b(B)}{2}} \left( \sum_{N_A,k} e^{- \frac{\Delta \xi(N_A, k)}{2}} \left| u_{N_A,k} \right> \left< v_{N_A,k} \right| \right) e^{\frac{S_b(A)}{2}} ~R^{-L_0}, 
\end{eqnarray}
and we insert it in (\ref{eq:two_exp}), like we did previously in section \ref{sec:completeness}. This time, however, we get
\begin{eqnarray}
	\label{eq:decomp}
	\nonumber \left|\left. \Psi \right>\right> & = & \sqrt{\frac{Z_{N_{A0}}\, Z_{N_{B0}}}{Z_N}} \sum_{ N_A} \sum_k   e^{-\frac{\Delta \xi(N_A, k)}{2}}   \\
&&  \left|\left. \,\Psi^A_{\left< v_{N_A,k} \right| e^{\frac{S_b(A)}{2}}}\, \right>\right>  \, \left|\left. \,\Psi_{e^{\frac{S_b(B)}{2}} \left| u_{N_A,k} \right>}^B\, \right>\right> . \qquad \qquad 
\end{eqnarray}
The difference between this last identity and (\ref{eq:semi_decomp}) is that the states in the right-hand side are now orthogonal, because of (\ref{eq:perturb_conj}). We also use the fact that the operators $S_b(A)$ and $S_b(B)$ are self-adjoint; again, this reflects the fact that they represent boundary perturbations of the action of the field theory, which must be real. Then, (\ref{eq:decomp}) is truly a Schmidt decomposition. The pseudoenergies are directly given by the eigenvalues of $S_{\rm ES}$ (noted $\Delta \xi(N_A,k)$), up to the global additive constant (\ref{eq:shiftES2}),
\begin{equation}
	\xi(N_A,k) \, = \, \alpha~2 \pi  R - \gamma + \Delta \xi (N_A,k) \, .
\end{equation}

We have thus found the ES in terms of $S_b(A)$ and $S_b(B)$. The key point is now that, thanks to the particular relation (\ref{eq:eSBeSA}) between $S_{\rm ES}$, $S_b(A)$ and $S_b(B)$, the pseudo-Hamiltonian $S_{\rm ES}$ inherits the locality property of $S_b(A)$ and $S_b(B)$. This follows from the Baker-Campbell-Hausdorff formula: $S_{\rm ES}$ can be expanded in terms of the commutators $[S_b(A),S_b(B)]$, $[S_b(A),[S_b(A),S_b(B)]]$, etc. Since $S_b(A)$ and $S_b(B)$ are both integrals of local operators, each of these commutators is also an integral of a local operator. Note that this is a completely general observation, which remains valid beyond the framework of conformal field theory. Finally, the locality of $S_{\rm ES}$ itself implies that it must also have the generic form (\ref{eq:perturbation}), just like $S_b(A)$ and $S_b(B)$ (but with different coupling constants $\lambda_a$, and possibly different operators $\phi_a$). The calculation of the ES in the scaling region has thus boiled down to a standard renormalization group (RG) discussion: one needs to determine which {\it local} operators $\phi_a(z)$ (which, again, can be relevant, irrelevant or marginal, depending on their scaling dimension $h_a$) do or do not contribute to the expression of $S_{\rm ES}$ relying on symmetry arguments.

This completes our proof of the ``scaling property'' conjectured in \cite{DRR}, which states that the entanglement spectrum in the scaling region is the spectrum of a sum of local operators along the cut between $A$ and $B$. Our line of arguments relies heavily on the short-rangeness of bulk correlations---the generalized screening hypothesis---, and makes use of standard RG arguments and field-theoretic tools.

\subsubsection{Pseudo-Hamiltonian, Hamiltonian of the $1+1d$ CFT, and role of the stress-tensor}

\label{sec:ES_RSP}
So far, we have shown that, if generalized screening holds, then the pseudo-Hamiltonian $S_{\rm ES}$ is the integral of a local operator in the $1+1d$ CFT that underlies the wavefunction. We know that it must have the generic form (\ref{eq:perturbation}), but we have not yet determined this operator. We are now going to argue that, in most situations, and for real-space partition (RSP), this operator is the Hamiltonian of the CFT, namely $\frac{v}{R} L_0$ for some ``velocity'' $v$, with irrelevant or marginal perturbations. In other words, the real-space ES of the trial wavefunctions is the spectrum of a perturbed CFT.

To see this, we need to discuss what terms may appear in $S_{\rm ES}$, which has the form (\ref{eq:perturbation}). In the case of our FQHE states, there is always an operator with scaling dimension $1$: the $U(1)$ current $i \partial \varphi(z)$. In most cases, this is the most relevant operator (more precisely, it is marginal as a boundary perturbation, see section \ref{sec:corrections_overlaps}). The integral of the current along the cut is the number operator $J_0$, so this term plays a role only when sectors with different particle numbers are involved. When the system is symmetric under the exchange of $A$ and $B$ (as is a sphere cut along the equator, with hemispheres $A$ and $B$), the ES must be invariant under the exchange $\Delta N_A \rightarrow -\Delta N_A$, and the number operator is forbidden by this symmetry. In other cases, as explained in part \ref{sec:screening}, the effect of the number operator is not difficult to comrehend, so we can safely turn to the next perturbations, which usually have scaling dimensions $h_a>1$. In particular, the stress-tensor $T(z)$ appears generically as a perturbation. The coupling $\lambda_T$ has the dimension of a length, and is of order $\lambda_T \sim \sqrt{\rho_0}^{-1} \sim \ell_B$. In statistical mechanics, the length $2\pi\lambda_T$ is known as the extrapolation length and is ubiquitous in the study of surface critical phenomena \cite{Diehl}. Although it has the dimension of a length, we use the notation $v = 4\pi \lambda_T$, because $v$ is going to be the velocity that appears in the ES. Generically, there are of course other perturbations, but for the sake of simplicity, let us assume first that this one is the least irrelevant, and that it is the only operator at this order. Then, one is left with
\begin{equation}
	\label{eq:linear_disp}
S_{\rm ES} = \frac{ v}{R} L_0 + \dots
\end{equation}
This means that the entanglement spectrum is the spectrum of $\frac{v}{R}L_0$ in the chiral CFT that underlies the trial wavefunction, with a linear dispersion relation. Our results thus support the claim made in \cite{KitaevPreskill,QKL}.

More generally, however, other boundary perturbations are present in (\ref{eq:perturbation}), and may consequently appear in the ES. The latter is thus given by the spectrum of the Hamiltonian of a {\it perturbed} CFT. In particular, for most states in the MR construction, the stress-tensor of the full theory actually breaks down into two pieces: one for the charge sector $T^{U(1)}(z)$ and another one for the statistics sector $T^{\psi}(z)$. These two operators do not, in general, appear with the same coefficient $\lambda_T$. Thus, for the MR (Pfaffian) states or for the RR states, or any state in the MR construction but the Laughlin state, one should generically expect two branches (rather than one) of gapless excitations in the ES, with different velocities. This is of course a well-known feature of the energy spectrum of a quantum Hall system with a physical edge.

 All the perturbations that preserve the symmetries should generically appear in the ES, and this needs to be discussed case by case. There are perturbating operators in the $U(1)$-charge sector, in the statistics sector, and possibly also mixed terms between the charge and statistics sectors. For most trial states, we expect the perturbing operators $\phi_a(z)$ (other than the stress-tensor) appearing in $S_{\rm ES}$ to be more irrelevant than the stress-tensor, namely they should have scaling dimensions $h_a > h_T =2$. When this is true, the ES in the scaling region collapses onto a branch (or two branches) of excitations with linear dispersion relation as $N \rightarrow \infty$. The corrections coming from the more irrelevant operators introduce some splitting between the pseudoenergies at the next order.

We would like to warn the reader about the fact that, although the stress-tensor is itself {\it irrelevant} (as a boundary perturbation) in the RG sense ($h_T =2>1$), it is crucial to take it into account to understand the structure of the ES, at least beyond the zeroth order where all the pseudoenergies are degenerate (\ref{eq:shiftES2}). When discussing the influence of the different perturbing operators in $S_{\rm ES}$, most of them turn out to be {\it irrelevant} (as boundary perturbations) in the sense that $h_a>1$, but they can still have spectacular effects on the ES if they are less or equally irrelevant to $T(z)$ (that is, $h_a \leq 2$). The standard RG terminology (relevant/irrelevant/marginal) must thus be used cautiously here, as we refer to the relevance of the {\it boundary perturbations} at the boundary of a $2d$ theory, which differs from the relevance of {\it bulk perturbations} of the $1+1d$ theory that gives the ES.

Finally, since we have shown that the real-space ES is the spectrum of a perturbed CFT, most of the usual results on the edge spectrum can be safely used for the ES as well. This of course includes the case when there is an operator that is less or equally irrelevant to the stress-tensor in the spectrum (for a recent discussion of the possible perturbations in the physical edge spectrum, see \cite{Nick_edge}). As an illustrative example that is not exactly a FQHE state, but that is closely related, one can look at the case of $\ell \pm i\ell$-paired superfluids (spinless when $\ell$ is odd). For those, the edge theory is known to correspond to $\ell$ chiral Majorana modes at the boundary \cite{RG}. One generically expects to have a bilinear term that couples the different Majorana modes. Such a term in the {\it bulk} of the $1+1d$ theory is relevant; it has scaling dimension $h=1 < 2$ (this term is directly a bulk perturbation in the ES, we do not view it as a boundary perturbation along the edge of the sample). It leads to a splitting of the energy levels of Majorana modes \cite{Nick_edge}. As should be expected from the generic arguments presented here, such a splitting should also appear in the real-space ES of these chiral superfluids. This was indeed found in \cite{DR}.

\subsubsection{RG analysis of the ES: the example of the Laughlin state}

Let us discuss the case of the Laughlin wavefunction in some more details now. Following the arguments in section \ref{sec:corrections_Laughlin}, we arrive at the fact that $S_{\rm ES}$ must have the form
\begin{eqnarray}
	\label{eq:SbA_Laughlin}
	&& S_{\rm ES} \, = \,  \\
	\nonumber &&  \sum_{\{k\}} \lambda_{\{k \}} \oint \frac{dz}{2\pi i} \left(\frac{z}{R}\right)^{k_1+\dots+k_p-1} ~ (i\partial_z^{k_1} \tilde{\varphi})\dots (i\partial_z^{k_p} \tilde{\varphi})(z),
\end{eqnarray}
where the sum runs over the finite sets $\{ k\} = \{ k_1,\dots,k_p\}$ with $k_1,\dots,k_p \geq 1$. Here, the shifted chiral bosonic field is defined as $\tilde{\varphi}(z) = \varphi(z) + i \frac{N_{A0}}{\sqrt{\nu}} \log z$, such that, when acting on the CFT vacuum with $N_{A0} + \Delta N_{A0}$ charges, the eigenvalue of $\tilde{J}_0$ is $\tilde{J}_0 \left| N_{A0} +\Delta N_{A0} \right> = \Delta N_{A0}/\sqrt{\nu} \left| N_{A0} +\Delta N_{A0} \right>$, which is of order $O(1)$ in the scaling region, while the eigenvalue of $J_0$ would rather be of order $O(N_{A0})$.

Further restrictions can be put on the operators that can appear $S_{{\rm ES}}$. Usually, in the literature, the cut between $A$ and $B$ is chosen such that the bipartition is symmetric with respect to the exchange of $A$ and $B$. This is the case in particular for the sphere, when $A$ and $B$ are chosen as the northern and southern hemispheres. For the Laughlin wavefunction, the symmetry under the exchange of $A$ and $B$ implies a symmetry the transformation of the $U(1)$ current $i\partial \tilde{\varphi}(z) \mapsto -i \partial \tilde{\varphi}(z)$, which prevents all the terms that contain an odd number of factors $(i \partial^{k_j} \tilde{\varphi}(z))$, such as $i\partial \tilde{\varphi}(z)$, $i\partial^2 \tilde{\varphi}(z)$ or $(i\partial \tilde{\varphi})^3(z)$, from appearing in the operator $S_{\rm ES}$. Then the least irrelevant operator is indeed the stress-tensor, leading to the linear dispersion relation (\ref{eq:linear_disp}). The next leading corrections are due to higher order terms, such as $(i\partial^2 \tilde{\varphi})^2(z)$, or $(i\partial \tilde{\varphi})^4(z)$, both with scaling dimension $4$, leading to additional splitting between the pseudoenergies at order $(\sqrt{\rho_0}^{-1}/R)^3 =1/N_{A0}^{3/2}$. There is no new operator leading to a splitting of order $1/N_{A0}$, because the only possible operator of scaling dimension $3$ that is even under $i\partial \tilde{\varphi}(z) \mapsto - i \partial \tilde{\varphi}(z)$ is $(i\partial \tilde{\varphi})( i\partial^2 \tilde{\varphi})(z)$, which is the derivative of the stress-tensor.

It turns out that very similar RG arguments can be given for the ES obtained from particle partition (PP), so we now turn to this discussion. We will come back later to the real-space ES of the Laughlin state (section \ref{sec:Laughlin_RSP_num}), and we will provide numerical comparison between the spectrum that can be obtained from our scaling analysis and the actual ES computed numerically for the Laughlin state.

\subsection{Particle partition entanglement spectrum in the scaling region}
\label{sec:ES_PP}

As mentioned in the introduction of this paper (section \ref{sec:intro_entanglement}), the particle partition (PP) has also been considered in the literature. Mathematically, this bipartition is very similar to the RSP, which allows us to treat the particle ES (namely the ES for PP) in a way that is almost identical to the real-space ES \cite{DRR, RSPBernevig}. In particular, our definition of the scaling region (section \ref{sec:scaling_region}) can be extended straightforwardly to PP. On the sphere, using the $SO(3)$ rotational invariance, the Schmidt eigenvalues and eigenstates for PP can be classified according to their total angular momentum $\Delta L^A$ (rather than the $\Delta L_{\hat{z}}^A$ component as we did for RSP). We also fix some finite ratio $N_{A0}/N$ (usually $N_{A0} = N/2$ for $N$ even). The scaling region then corresponds to the set of all Schmidt eigenvalues/eigenstates which correspond to values of $\Delta N_A$ and $\Delta L^A$ that remain bounded ($\leq M$ for some fixed $M$) when one sends $N$ to infinity. 

Although we just emphasized the fact that the total angular momentum $L^A$ (rather than $L^A_{\hat{z}}$) is the most natural quantity to use to classify the Schmidt eigenvectors/eigenvalues, it is still worth looking at what happens at large $L^A_{\hat{z}}$. Indeed, the Schmidt eigenstate corresponding to the largest value of $L^A_{\hat{z}}$ is nothing but a state where $N_A=N/2$ particles with ``pseudospin'' $A$ fill the northern hemisphere, while the ones with ``pseudospin'' $B$ fill the southern hemisphere. Therefore, PP at large values of $L^A_{\hat{z}}$ is very similar to RSP with a cut along the equator. We will use this below to analyze the scaling behaviour of the ES. After stereographic projection, the equator corresponds to a circle in the plane where the coordinates $z$ live. We note $R$ the radius of this circle.

\subsubsection{Completeness relation, corrections to scaling and entanglement spectrum}

The decomposition (\ref{eq:semi_decomp}), obtained from a completeness relation in the CFT space, is still valid for PP. More precisely, for any fixed $N_A$, the groundstate wavefunction $\Psi(z_1,\dots,z_N)$ can be decomposed as
\begin{eqnarray}
\nonumber	\Psi(z_1,\dots,z_N) & = & \frac{1}{\sqrt{Z_N}} \sum_k \left< N \right| \prod_{i=N_A+1}^N a(z_i) \left| N_A,k\right>  \\ &&\times \, \left< N_A,k\right| \prod_{j=1}^{N_A} a(z_j) \left| 0\right>
\end{eqnarray}
where we have used the completeness relation $1_{N_A} = \sum_k \left| N_A,k\right>\left<N_A,k\right|$ in the $N_A$ subsector. This equation can be derived using the procedure of section \ref{sec:completeness}, restricting the positions of the operators inside/outside a disc of some radius as in Fig. \ref{fig:completeness}. Since the correlators $\left<N_a,k \right| \prod_j a(z_j) \left| 0\right>$ and $\left<N \right| \prod_i a(z_i) \left| N_A,k\right>$ in the right-hand side are all analytic, the identity remains true for any positions of the points $z_i$'s and $z_j$'s, without the requirement that they lie inside/outside a disc of some radius. Thus, for PP, we have a decomposition identical to (\ref{eq:semi_decomp}), where the states $\left|\left. \, \Psi^A_{\left< N_A,k \right|}\, \right>\right>$ and $\left|\left. \, \Psi^B_{\left|N_A,k\right>} \, \right>\right> $ are replaced by
\begin{subequations}
\label{eq:decomp_states_PP}
\begin{eqnarray}
	&& \left|\left. \, \Psi^A_{\left< N_A,k \right|}\, \right>\right>_{\rm PP}  \, = \, \frac{1}{\lambda^{\frac{N_{A0}}{2}}~ \sqrt{Z^{\rm PP}_{N_{A0}}}} \left< N_A,k \right|\frac{1}{R^{\Delta L_0}}  \; \times  \\
\nonumber	&&  \qquad \mathcal{R}  \exp \left[\lambda^{\frac{1}{2}} \int_{\mathbb{C}} e^{V(z,\overline{z})/2} d^2z ~ a(z) \otimes c^\dagger(z) \right] \left|0 \right>\, \left|\left. 0 \right>\right>    \\
	&& \left|\left. \, \Psi^B_{\left|N_A,k\right>} \, \right>\right>_{\rm PP}  \, = \, \frac{1}{\lambda^{\frac{N_{B0}}{2}}~\sqrt{Z^{\rm PP}_{N_{B0}}}} \left< N \right| \; \times \\
\nonumber   &&  \mathcal{R} \exp \left[\lambda^{\frac{1}{2}} \int_{\mathbb{C}} e^{V(z,\overline{z})/2} d^2z ~ a(z) \otimes c^\dagger(z) \right] R^{\Delta L_0} \left|N_A,k \right>\, \left|\left. 0 \right>\right> .
\end{eqnarray}
\end{subequations}
Note that the integration domain in the exponential is the entire complex plane $\mathbb{C}$, rather than some subsets as for RSP. Again, the coefficient $\lambda$ is non-universal, and must be tuned like for RSP, to give the correct weight to the charged edge states. It usually depends on $N_{A0}$ (or $R$). The normalization factors for PP are:
\begin{subequations}
\label{eq:decomp_ZN}
\begin{eqnarray}
 \nonumber Z^{\rm PP}_{N_{A0}} & = & \frac{1}{N_{A0}!}\int_{\mathbb{C}} \prod_{i=1}^{N_{A0}} e^{V(z_i,\overline{z_i})} d^2 z_i\; |\left< N_{A0} \right| \prod_{i=1}^{N_{A0}} a(z_i) \left| 0 \right> |^2  \\   \\
\nonumber Z^{\rm PP}_{N_{B0}} & = & \frac{1}{N_{B0}!}\int_{\mathbb{C}} \prod_{i=1}^{N_{B0}} e^{V(z_i,\overline{z_i})} d^2 z_i\; |\left< N \right| \prod_{i=1}^{N_{B0}} a(z_i) \left| N_{A0} \right> |^2 . \\
\end{eqnarray}
\end{subequations}
Again, note that the integration domain is $\mathbb{C}$ rather than a subset $A$ or $B$, as was the case for RSP.

The states $\left|\left. \, \Psi^A_{\left< N_A,k \right|}\, \right>\right>_{\rm PP}$ and $\left|\left. \, \Psi^B_{\left|N_A,k\right>} \, \right>\right>_{\rm PP} $ are exactly the edge states constructed in section \ref{sec:edge}. In the scaling region, they all correspond to edge excitations of the droplet of radius $R$. Then the entire discussion of sections \ref{sec:scaling_region} and \ref{sec:finitesize} applies as well for PP in the scaling region. The conclusion is that the particle ES is also the spectrum of a local operator along the ``cut'', namely the circle of radius $R$. Up to a global constant shift, the pseudoenergies are the eigenvalues of $S_{\rm ES}^{\rm PP}$, related to the boundary perturbation of the action of the CFT $S_{b}^{\rm PP}(A)$ and $S_{b}^{\rm PP}(B)$ by the same formula as (\ref{eq:eSBeSA}). Then, the determination of the particle ES boils down to a discussion of the operators that can/cannot appear in the operator $S_{\rm ES}^{\rm PP}$, exactly like in the case of RSP. Once again, this needs to be done case by case.

\subsubsection{Absence of the stress-tensor in the $U(1)$ sector\\ for particle partition}

The most striking difference between PP and RSP is the following. For RSP we argued that the ES usually obeys a linear dispersion relation because of the presence of the stress-tensor of the chiral CFT $T(z)$ (more precisely $T^{U(1)}(z)$ and $T^{\psi}(z)$ separately) in the set of perturbing operators along the cut. This is expected to be a very generic boundary perturbation, and as we highlighted, it is related to the extrapolation length $2\pi \lambda_T$ that appears very often in the field of surface critical phenomena \cite{Diehl}. However, the stress-tensor $T^{U(1)}$ {\it cannot} appear in the particle ES. We already gave an explanation for why it cannot appear in $S_b(A)$ (or $S_b(B)$) in section \ref{sec:loop_equation}, based on the constraint (\ref{eq:loop_equation}) derived from translation invariance in the plane. We give now an alternative argument, which goes as follows. On the sphere, the $SO(3)$ rotational invariance implies that the pseudoenergies depend not directly on $\Delta L^A_{\hat{z}}$ (which would be related to the Virasoro generator $L_0 = L^{U(1)}_0 + L^\psi_0$, or more precisely the shifted mode $\tilde{L}_0 = \tilde{L}^{U(1)}_0 + L_0^\psi$), but rather on the total angular momentum $\Delta L^A$. This means that the combination $L^{U(1)}_0 + L^\psi_0$ cannot appear in the pseudo-Hamiltonian for PP; $L^\psi_0$, however could appear independently, as was pointed out in section (\ref{sec:loop_equation}).

\subsubsection{The example of the Laughlin state}

In the case of the Laughlin wavefunction, the analysis can be pushed further. As for RSP, the operators $S_{b}^{\rm PP}(A)$, $S_{b}^{\rm PP}(B)$, and $S_{\rm ES}^{\rm PP}$ have the generic form (\ref{eq:SbA_Laughlin}). Again, some additional restrictions are imposed: the operators in $S_{\rm ES}^{\rm PP}$ have to be even under $i \partial \tilde{\varphi}(z) \mapsto -i \partial \tilde{\varphi}(z)$. Since the stress-tensor and its derivatives are not here either, the least irrelevant allowed operator has scaling dimension $4$, and it leads to a splitting of the pseudoenergies of order $(\sqrt{\rho_0}^{-1}/R)^3 \sim 1/N_{A0}^{3/2}$ when $N,N_{A0} \rightarrow \infty$. Hence, this operator is $(i\partial^2 \tilde{\varphi})^2(z)$. Before we analyze the ES that follows from this operator, let us give a slightly more detailed argument which shows how this operator appears.

For large $L_{\hat{z}}^A$, the $N_A = N/2 +\Delta N_A$ particles fill the northern hemisphere, while the $N_B = N/2 - \Delta N_A$ particles fill the southern hemisphere. The free energies $- \log Z_{N_{A0}}^{\rm PP}$ and $-\log Z_{N_{B0}}^{\rm PP}$ are the ones of a one-component plasma (in its screening phase as long as $1/\nu$ is not too large) on the sphere, with a boundary along the equator. Generically, these two free energies should both contain a surface tension contribution, proportional to the perimeter of the cap that they fill. More precisely, such a term does not appear at $\nu =1$ (this is of course consistent with the fact that all the pseudoenergies are degenerate in that case), but when $\nu >1$ it is known that the surface tension is not zero \cite{WiegmannNote,Shakirov}. Now we imagine that the droplet is slightly deformed, namely that its boundary is a curve parametrized by $\theta = \frac{\pi}{2} + \delta \theta(\phi)$, for a small displacement $\delta \theta$. Here $(\theta,\phi)$ are the polar coordinates on the unit sphere. Then the length of the interface is $\int_0^{2\pi} d\phi \sqrt{\sin^2 \theta + (d\theta/d\phi)^2} \simeq 2\pi + \frac{1}{2} \int d \phi \left((d \delta \theta/d\phi)^2 - (\delta \theta)^2 \right)$. The displacement of the interface is proportional to the current $i\partial_\phi \tilde{\varphi} (\phi)$ (see \cite{WenEdge}). Thus, we find that the surface tension term is given by the integral along the equator of the operator $(i\partial^2_\phi \tilde{\varphi})^2(\phi)-(i\partial_\phi \tilde{\varphi})^2(\phi)$. One can check that this is the same operator as $(i\partial^2 \tilde{\varphi})^2(z)$ when one maps the sphere back to the plane, taking into account the Jacobian of this transformation carefully. Looking at the surface tension term on the sphere rather than on the plane has the advantage of making the $SO(3)$ rotational invariance more transparent. This term appears both in $S^{\rm PP}_b(A)$ and $S^{\rm PP}_b(B)$, with no obvious cancellation between the contributions coming from $A$ and $B$, therefore it must also appear as well in $S_{\rm ES}^{\rm PP}$. We analyze the spectrum of the operator that we just found next.

When it is written in terms of the modes of the (shifted) $U(1)$ current $i \partial \tilde{\varphi} = \frac{1}{z}\tilde{J}_0 + \sum_{n \neq 0} z^{n-1} J_{-n}$, the zero Fourier mode of $(i \partial^2 \tilde{\varphi})^2(z)$ leads to
\begin{equation}
	\label{eq:PPES_Laughlin}
	S_{\rm ES}^{\rm PP} \, = \, -\frac{C}{N_{A0}^{3/2}} \left[ \frac{1}{2} \tilde{J}_0^2 - \sum_{k\geq 1} (k^2-1) J_{-k}J_k \right] \, + \, O(1/N_{A0}^{2})
\end{equation}
where $C$ is some undetermined positive constant of order $O(1)$. This constant vanishes for the integer quantum Hall effect (Laughlin with $\nu =1$), but it is non-zero as soon as $\nu <1$. As explained previously, the particle ES is nothing but the spectrum of $S_{\rm ES}^{\rm PP}$, up to a global additive shift that ensures the normalization of the reduced density matrix ${\rm tr \,}\rho_A =1$. The spectrum of the operator (\ref{eq:PPES_Laughlin}) has several interesting features. The eigenstates of $S_{\rm ES}$ at the order $1/N_{A0}^{3/2}$ are all the states $\prod_{k \geq 1} J_{-k}^{n_k}\left| N_{A0} +\Delta N_A  \right>$, for any finite set of positive occupation numbers $n_k$'s. Then, for each finite set of non-zero occupation numbers, the splitting between the corresponding pseudoenergy and the ground state is, at the leading order
\begin{equation}
	\label{eq:PPES_Laughlin_occ}
	\Delta \xi \, = \, -\frac{C}{N_{A0}^{3/2}} \left[ \frac{(\Delta N_A)^2}{2\nu} - \sum_{k>0} n_{k}(k^2-1) k \right] .
\end{equation}
This spectrum possesses a few noticeable features. First, looking at the lowest pseudoenergy in each $\Delta N_A$ sector, which always corresponds to no non-zero occupation number $n_k$, one gets the {\it inverted parabola} observed numerically in \cite{DRR}: $\Delta \xi = - \frac{C}{N_{A0}^{3/2}} \frac{(\Delta N_A)^2}{2\nu}$. Second, for fixed $\Delta N_A$, the spectrum of $S_{\rm ES}$ is in general highly degenerate. This is due to the fact that $\Delta \xi$ does not depend on $n_1$. Thus, in the Schmidt decomposition, all the states $\left|\left. \Psi^A_{\left< N \right|} \right>\right>_{\rm PP}$, $\left|\left. \Psi^A_{\left< N \right| J_{1}} \right>\right>_{\rm PP}$, $\left|\left. \Psi^A_{\left< N \right| J_1^2} \right>\right>_{\rm PP}$, $\left|\left. \Psi^A_{\left< N \right| J_1^3} \right>\right>_{\rm PP}$,$\dots$, contribute with the same pseudoenergy (at least at this order), leading to a large degeneracy in the ES. Such a large degeneracy is to be expected, since the $SO(3)$ rotational implies that the pseudoenergies appear in multiplets.

Once again, we emphasize the fact that our results can be related to those of Zabrodin and Wiegmann. 
The calculations that lead to the operator $(i \partial^2 \tilde{\varphi})^2$ in their approach are beyond the scope of our paper. Here we just want to point out that one should be careful when trying to relate their results to ours. Indeed, the fact that the surface tension for the one-component plasma vanishes at $\nu =1$ has been the source of some confusion. In particular, the term corresponding to the surface tension, which is non-zero when $\nu<1$, is actually missing in the result stated by Zabrodin and Wiegmann in \cite{ZabrodinWiegmann3}, although they do consider it in their calculations. This term appears as a zero mode of the ``loop equation'' in their paper, which of course is reflected in our formalism by the fact that the operator in formula (\ref{eq:PPES_Laughlin}) commutes with the mode $J_1$, and therefore is a zero mode of the constraint (\ref{eq:loop_equation}). This term, however, should be there, as pointed out for instance in \cite{WiegmannNote,Shakirov}. We note that, in the FQHE literature, the possibility of a surface tension term was also quickly discussed in the appendix of Ref. \cite{RRviscosity}.

In Fig. {\ref{fig:Ed_plot}}, we compare the formula (\ref{eq:PPES_Laughlin_occ}) with numerical results for the particle ES of the Laughlin state with $\nu=1/3$. We plot low-lying pseudoenergies $\xi$ versus $L_{\hat{z}}^A$ for the scaling region near $L_{\hat{z}0}^A$, for $N_A=N/2$ and $N=12$ particles. The range of pseudoenergies included is a little larger than in Ref. \cite{DRR}, to show the structure. It is interesting to look at the upper envelope of the levels, namely the maximum $\xi$ for $L_{\hat{z}}^A=L_{\hat{z}0}^A-k$ for each $k=1$, $2$, \ldots. For $k\leq 6=N_A$, those levels fall on a curve, but the trend ends at this $k$; the next one $k=7$ is degenerate with that at $k=6$. These levels (with $k\leq6$) should agree with setting $n_k=1$, and others zero, for $k=1$, $2$, \ldots, $6$, in the above formula, and so lie on the cubic curve $\Delta\xi=(C/N_{A0}^{3/2})(k^3-k)$ plotted in the Figure. We find a reasonable agreement, as shown, for $C=0.5144$ at this size. At the same time, the {\em lowest} pseudoenergies for each $N_A$ should lie on an inverted parabola, and the coefficient should be the same according to the above formula (\ref{eq:PPES_Laughlin_occ}). A plot is shown in the inset on the Figure, with the same parameter value, and the agreement is reasonable. If the two curves are fitted to these forms independently, with one parameter for each, then the best parameter values are found to be within about $10\%$ of each other. Further features of the spectra also agree with the above form. Let us note first that the cutoff at $k=N_A$ noticed above makes sense as the standard cutoff from finite size effects in the edge spectra; the sums of powers $\sum_{i=1}^{N_A} z_i^k$, which create these elementary excitations if trial wavefunctions are used, become algebraically dependent on the lower ones if $k>N_A$. Levels in the spectra other than the extreme ones fitted above can be roughly explained as sums of multiple excitations, in rough agreement with the formula. In particular, the banded structure that is apparent in the Figure emanating from each of those extreme levels can be understood in each case as excitations with lower $k'<k$ added to the $k$th extreme one. Similar structures and trends are seen in smaller sizes, and for $N_A<N/2$ (not shown).

\begin{figure}[ht]
	\includegraphics[width=0.5\textwidth]{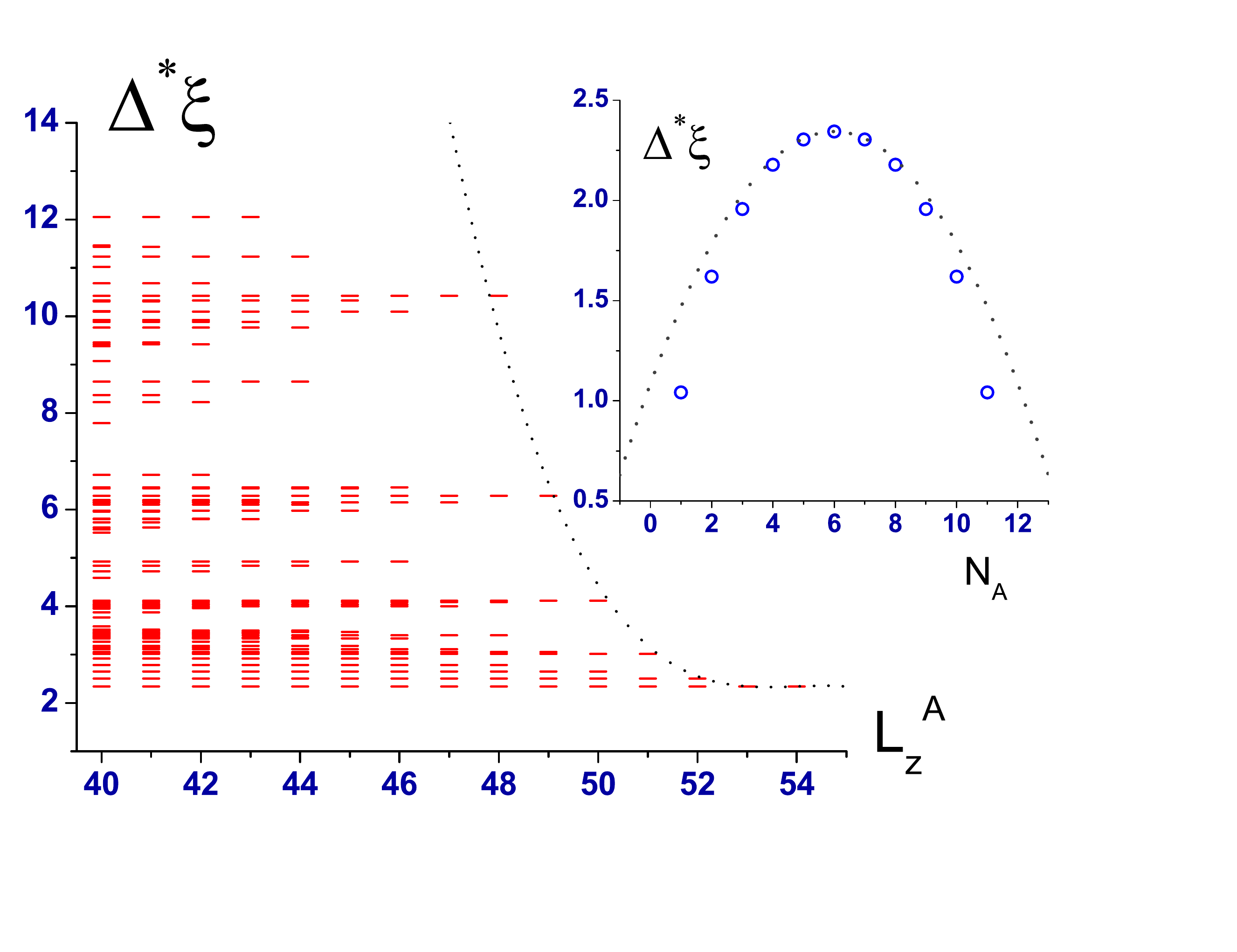}

\caption{Plots of PP pseudoenergies $\xi$ versus $L_{\hat{z}}^A$ for $N_A=N/2$ and versus $N_A$ for the $N=12$, $\nu=1/3$ Laughlin state. Here $\Delta^\ast\xi=\xi-N\ln 2$. Main figure: levels in the scaling region versus $L_{\hat{z}}^A$. The dotted curve is a one-parameter fit of a cubic to seven levels, explained in the text. Inset: lowest pseudoenergy for each $N_A$; the values for $N_A=0$, $12$, which are $\xi=N\ln 2$, are omitted. The curve is an inverted parabola, with the same parameter value as the main figure, as explained in the text.}
\label{fig:Ed_plot}

\end{figure}

\subsection{Numerical comparisons for the real-space ES of the Laughlin state at $\nu=1/3$}
\label{sec:Laughlin_RSP_num}

In this section we come back to RSP, and present some numerical results for the Laughlin state. As pointed out previously, the ``pseudo-Hamiltonian'' $S_{\rm ES}$ must be of the form (\ref{eq:SbA_Laughlin}).

\begin{figure}[h]
\includegraphics[width=0.35\textwidth]{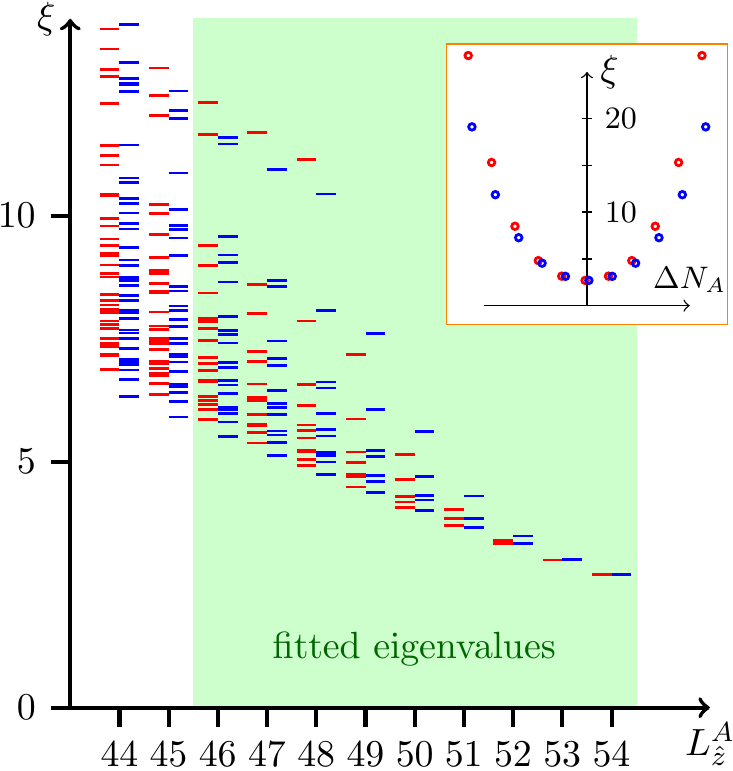}
\centering
\caption{(color online) Red: real-space ES of the Laughlin state at $\nu =1/3$ on the sphere, with a cut along the equator (for $N=12$, $N_{A0}=6$). Blue: spectrum of the operator (\ref{eq:S_ES_spec3}). Both spectra are plotted in the $\Delta N_A =0$ sector. The inset shows the lowest eigenvalue for the other values of $\Delta N_A$. The parameters $\alpha,\beta,\gamma$ are obtained from a least square fit that includes all the eigenvalues in the shaded (green) area.} 
\label{fig:RSP_ES_Laughlin}	
\end{figure}

We work on the sphere, with a cut along the equator, and with $N_{A0} = N/2$. Then the ES must be invariant under the exchange $\Delta N_A \mapsto - \Delta N_A$, which implies that $S_{\rm ES}$ is invariant under $i \partial \varphi(z) \mapsto - i \partial \varphi(z)$. The first operator that can appear in $S_{ES}$ is then the zero mode of the (shifted) stress-tensor $\tilde{L}_0 = \tilde{L}_0^{U(1)}$, which has scaling dimension $2$, and the mode expansion
\begin{equation}
	\tilde{L}_0 \, = \, \frac{1}{2} \tilde{J}_0^2 + \sum_{k\geq 1} J_k J_{-k} \, .
\end{equation}
The next linearly independent operators that are allowed in $S_{\rm ES}$ have scaling dimension $4$: it is the operator $(i\partial^2 \tilde{\varphi})^2(z)$ that we already encountered for PP, and the operator $(i \partial \tilde{\varphi})^4$. Their zero Fourier modes $O^1_0 = \oint dz~ z^3 (i \partial^2 \tilde{\varphi})^2(z)/2$ and $O_0^2=\oint dz~ z^3 (i \partial \tilde{\varphi})^4(z)/4!$ are:
\begin{eqnarray}
\nonumber	O^1_0 &=& \frac{1}{2} \tilde{J}^2_0 - \sum_{k \geq 1} (k^2-1) J_{-k} J_{k} \\
	O^2_0 &=& \frac{1}{4!} \sum_{k_1,k_2,k_3} :J_{k_1} J_{k_2}J_{k_3}J_{-k_1-k_2-k_3}: \, .
\end{eqnarray}
We then expect a pseudo-Hamiltonian of the form
\begin{equation}
	\label{eq:S_ES_spec3}
	S_{\rm ES} \, = \, \frac{\alpha}{N_{A0}^{1/2}} \tilde{L}_0 \, + \, \frac{\beta}{N_{A0}^{3/2}} O^1_0 \, + \, \frac{\gamma}{N_{A0}^{3/2}} O^2_0 \, + \, O(N_{A0}^{-2}) \, .
\end{equation}
In this formula, the coefficients $\alpha, \beta$ and $\gamma$ are of order O(1), however they need not be completely independent of $N_{A0}$: they can be expanded as $\alpha = \alpha_0 + \alpha_1/N_{A0}^{1/2} + \alpha_2/N_{A0} + \dots$, where $\alpha_0,\alpha_1,\alpha_2$, etc. do not depend on $N_{A0}$. These subleading pieces of the coefficients come from the operators that are total derivatives. For instance, $\alpha_1$ comes from the derivative of the stress-tensor $\partial(\partial\tilde{\varphi})^2/2$, which has scaling dimension $3$, has a zero Fourier mode $\oint dz \, z^2 \partial(\partial\tilde{\varphi})^2/2 = - 2 \tilde{L}_0$. In what follows, we keep $N_{A0}$ fixed, so we work only with the three real parameters $\alpha,\beta,\gamma$, and we do not focus on their dependence on $N_{A0}$.

We now compare our formula (\ref{eq:S_ES_spec3}) to the ES computed numerically for the Laughlin state at $\nu =1/3$ for $N =12$ particles in total ($N_{A0} =6$). We have fitted the three parameters $\alpha, \beta, \gamma$ to minimize the square of the difference between the two sets of eigenvalues. We impose some cutoff, such that the fit is done only on eigenvalues that are low enough, and that correspond to $|\Delta L^{A}_{\hat{z}}| \leq K$. We find that the ES is approximated by the one of $S_{\rm ES}$ with $\alpha =0.7603$, $\beta= -0.4108$ and $\gamma = 0.3653$. The comparison between the two spectra is shown in Fig. \ref{fig:RSP_ES_Laughlin}.

\begin{figure}[h]
\includegraphics[width=0.35\textwidth]{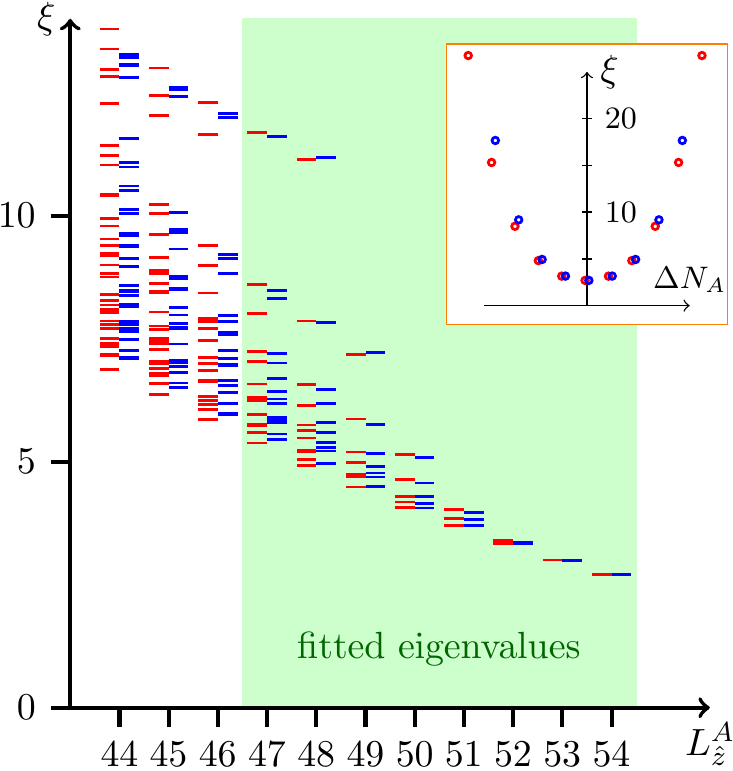}
\centering
\caption{(color online) Same as Fig.~\ref{fig:RSP_ES_Laughlin}, but with six linearly independent operators in $S_{\rm ES}$, instead of three, leading to a six-parameter fit on the eigenvalues contained in the shaded (green) area.} 
\label{fig:RSP_ES_Laughlin2}	
\end{figure}

For completeness, we also pushed this analysis to the next order. The next linearly independent operators that can appear in the spectrum are $(\partial^2 \tilde{\varphi})^2$, $\partial^3 \tilde{\varphi} (\partial \tilde{\varphi})^3$, and $(\partial \tilde{\varphi})^6$, of scaling dimension $6$. This leads to a fit with six real parameters. The result is shown in Fig.~\ref{fig:RSP_ES_Laughlin2}.

\section{Discussion: entanglement, trial wavefunctions, conformal blocks \\and Tensor Product States}
\label{sec:MPS}

Since the seminal work of Laughlin \cite{Laughlin}, it is known that an efficient approach to the theory of the FQHE consists in making various proposals for trial wavefunctions, which are not directly related to the eigenstates of the original Hamiltonian, but which still accurately capture some of its physical features. This approach is of course not specific to the FQHE effect, and other strongly correlated quantum systems can be studied with variational wavefunctions. Some powerful techniques have been developed for one-dimensional systems, especially for spin chains, which focus on some class of trial wavefunctions called Matrix Product States (MPS) \cite{MPS1}. As a simple example of these, let us consider an open spin-$1/2$ chain with $N$ sites. An MPS for this system can be constructed as
\begin{eqnarray}
	\label{eq:MPS}
	&& \left| \left. \Psi \right>\right> \, = \, \\ 
\nonumber && \sum_{s_1,\dots,s_N = \uparrow,\downarrow} \left<v_{\rm out} \right| A(s_N) \dots A(s_2) A(s_1) \left| v_{\rm in}\right> \, \left|\left. s_1 s_2 \dots s_N \right>\right>
\end{eqnarray}
where the states $\left|\left.s_1 s_2 \dots s_N \right>\right>$ with $s_i = \uparrow,\downarrow$ span the $2^N$-dimensional Hilbert space $\hilb = \{\uparrow,\downarrow\}^{\otimes N}$, and $A(\uparrow)$,$A(\downarrow)$ are both $D \times D$ matrices acting in an {\it auxiliary space} $V_D$. The two states $\left|v_{\rm in}\right>$ and $\left|v_{\rm out} \right>$ belong to the space $V_D$. The choice of these two states in the auxiliary space, and of the $D\times D$ matrix $A$, fixes the physical state $\left|\left. \Psi \right>\right>$.

An MPS possesses peculiar entanglement properties: dividing the system into two subchains of length $N_A$ and $N_B$ (with Hilbert spaces $\hilb_A = \{\uparrow,\downarrow \}^{\otimes N_A}$ and $\hilb_B = \{\uparrow,\downarrow \}^{\otimes N_B}$), one can easily check that the Schmidt decomposition corresponding to the bipartition $\hilb = \hilb_A \otimes \hilb_B$
\begin{equation}
	\label{eq:Schmidt}
	\left| \left. \Psi \right>\right> \, = \, \sum_{k} e^{-\xi_i/2} \left|\left. \Psi_{A,i} \right>\right> \otimes \left| \left. \Psi_{B,i} \right> \right>
\end{equation}
is of rank $D$ at most. This simply follows from inserting a completeness relation in the auxiliary space $V_D$ (namely $1_{V_D} = \sum_{k=1}^D \left| v_k \right>\left< v_k \right|$ with $\left< v_k | v_{k'} \right> = \delta_{k,k'}$) between the matrices $A(s_{N_A+1})$ an $A(s_{N_{A}})$.
This is a striking feature of an MPS, as it is radically different from generic states in $\hilb$ which would typically have Schmidt decompositions of rank ${\rm min}({\rm dim} \hilb_A, {\rm dim} \hilb_B)$. In that sense, the set of MPS in $\hilb$ for some fixed $D$ can be thought of as being a submanifold (of $\hilb$, viewed as a manifold) of minimally entangled states. In particular, when $D=1$, the space of MPS is nothing but the submanifold of separable states.

Starting from a Hamiltonian $\hat{H}$ for our spin-$1/2$ chain, one should minimize the expectation value $\left<\left< \Psi \right| \right. \hat{H} \left.\left| \Psi \right>\right> / \left<\left<\Psi \left| \Psi \right>\right>\right.$ within the set of $D$-dimensional MPS, which would give some approximation to the true ground state of $\hat{H}$. By increasing the dimension $D$ of the auxiliary space $V_D$, one can approximate the true ground state of the system with arbitrary accuracy. MPS are at the heart of numerical methods such as the Density Matrix Renormalization Group (DMRG) \cite{WhiteDMRG}, and have also been used recently as a key theoretical tool in the successful efforts made by several groups to classify one-dimensional symmetry-protected topological phases of matter \cite{PTBO,KitaevFidkowski,Classif1d_Wen,Classif1d_Cirac}. Their higher-dimensional analogs, the TPS 
(also called ``PEPS'' for Projected Entangled-Pair States \cite{PEPS}) are currently being used in attempts to classify higher-dimensional topological phases (see for instance \cite{GuLevinWen}). The basic idea of such a program is that any gapped phase of matter is believed to be accurately represented by one (or more) trial state, like an MPS for $D$ large enough, or some higher-dimensional generalization (see for instance \cite{Hastings1}). Such a trial state is not the ground state of the actual physical Hamiltonian, but it is the ground state of some local Hamiltonian \cite{HaldaneHierarchy, AKLT, PEPS_Hamiltonian} which is believed to be adiabatically connected to the physical one. Then the topological properties of the phase, namely the properties which are robust as one varies the different physical parameters within the phase are captured by the trial state, and the trial state has the enormous advantage of having some nice mathematical structure, which makes analytic and/or numerical calculations tractable, even for relatively large system sizes.

We would like to point out that the states given by the MR construction have a structure that is very similar to the one of an MPS or a TPS. Essentially, it can be viewed either as an MPS with $D = \infty$, or equivalently as a continuous version of a TPS. Indeed, we note that the conformal block in (\ref{eq:wfblock}) is nothing but
\begin{equation}
	\left<v_{\rm out} \right| a(z_N) \dots a(z_1) \left| v_{\rm in} \right> \, .
\end{equation}
where $\left| v_{\rm in} \right> = \left| 0 \right>$ is the vacuum of the CFT,
and $\left| v_{\rm out} \right>=\left|N \right>$ is the vacuum with $N$ charges. 
The structure of this correlator is clearly analogous to the MPS (\ref{eq:MPS}), with a space $V_D$ which is now the (infinite-dimensional) Hilbert space of the CFT. 
The use of conformal blocks as infinite-dimensional MPS has recently been advocated in Ref. \cite{CiracSierra}. In particular, in this work, conformal blocks are used as trial wavefunctions for critical (gapless) spin chains. We note that, in the context of the FQHE, such wavefunctions are nothing but the ones given by the MR construction (section \ref{sec:blocks}). The fact that $V_D$ is infinite-dimensional does not mean than the particular entanglement properties emphasized above completely breaks down, since in each given angular momentum subsector (that is, in each $L_0$ eigenspace) the auxiliary space is finite, and thus the Schmidt rank in that sector is finite. This is exactly the property that was observed by LH, namely that particular trial wavefunctions for the FQHE have an infinite entanglement gap. We emphasized the connection between the so-called ``level-counting'' in each angular momentum sector, the conformal block structure, and the completeness relation in the auxiliary space in section \ref{sec:completeness}.

Another way of looking at the trial wavefunctions for the FQHE is to view them as continuous TPS in 2d. Let us imagine that the ground state in a two-dimensional gapped phase of matter can be approximated by some TPS. For a continuous system of identical itinerant particles, a natural way of searching for a TPS approximation of the ground state would be to discretize the space, namely to put the particles on a lattice. For simplicity, let us assume that the particles are identical hard-core bosons. Then on each site $i$ of the lattice, either there is a particle, or there is none. There is one tensor associated to each of these two situations, $B^i_{e_{1},\dots,{e_p}}$ (one particle) and $A^{i}_{e_{1},\dots,{e_p}}$ (no particle), with $p$ indices corresponding to the $p$ edges that end at the site $i$. Each of the $e_1$ can take values from $1$ to $D$, where $D$ is the dimension of the auxiliary space $V_D$. Then, in the ground state, the amplitude of the configuration $c_{i_1}^\dagger \dots c_{i_N}^\dagger \left|\left. 0 \right>\right>$ with $N$ bosons on the sites $i_1,\dots , i_N$ (for $S$ sites in total) is given by the contraction of all the edge indices of a product of tensors $A$ ($S-N$ times) and $B$ ($N$ times). Then we sum over all the possible numbers of particles $n$, and we get the trial wavefunction:
\begin{equation}
	\left|\left. \Psi \right>\right> \, = \, {\rm tr }_{V_D} \,\left[ \prod_{{\rm vertex} \, i} \left( A^{i} \otimes 1 + B^i \otimes c^\dagger_{i} \right) \right] \left|\left.0 \right>\right>
\end{equation}
where the trace is taken over the auxiliary space $V_D$ for each edge of the lattice, and we have omitted all the edge indices. If we fix one configuration of edge indices $\{e_1,e_2,\dots\}$, $A^{1} A^{2} \dots A^N$ can be viewed as an amplitude for this configuration. The same observation holds with insertions of tensors $B^i$'s in some places on the lattice, giving different amplitudes for the same configuration $\{e_1,e_2,\dots \}$.
It is now clear what the natural extension of a TPS should be when we replace the lattice by a continuous space (here the two-dimensional plane, or sphere). The set of edge indices $\{e_1,e_2,\dots \}$ should be replaced by a set of local degrees of freedom, which we note $h({\bf x})$. A fixed configuration of degrees of freedom comes with an amplitude $\prod_i A^i$. In the continuum, the product on each point ``$\prod_{\bf x}$'' is more conveniently written as the exponential of a sum, which becomes an integral. Thus, to a given configuration of variables $\{h({\bf x}) \}$, we associate the amplitude 
$e^{\int d^2 {\bf x} F[h({\bf x})]}$. The continuous analog of the tensor $B^i$, which modifies the amplitude $A^1\dots A^i \dots A^S$ to $A^1\dots B^i \dots A^S$, is an operator $\mathcal{O}[h({\bf x})]$. Then the continuous TPS takes the familiar form
\begin{equation}
	\left|\left. \Psi \right>\right> \, = \, \int [d h] \, e^{\int d^2 {\bf x}\,  F[h(\bf x)]}\, e^{\int d^2 {\bf x}'\mathcal{O}[h({\bf x'})] \otimes c^\dagger({\bf x'})} \left|\left.0 \right>\right>,
\end{equation}
up to normalization. In other words, the amplitude of the configuration $c^\dagger({\bf x}_1)\dots c^\dagger({\bf x}_N) \left|\left.0 \right>\right>$ is $\int [dh] e^{\int d^2{\bf x} F[h(\bf x)]} \mathcal{O} [h({\bf x}_1)] \dots  \mathcal{O} [h({\bf x}_N)]$. This, of course, is nothing but the correlator of a set of {\it local operators} in a {\it local field theory}.

We thus conclude that, searching for a TPS-like Ansatz for a continuous system of mobile identical particles, one is naturally lead to considering local field theories, and correlators of local observables in those. This remark, of course, can be extended straightforwardly to the case of particles with additional degrees of freedom (such as spin), by using one operator $\mathcal{O}$ per degree of freedom (for instance $\mathcal{O}_\uparrow$ and $\mathcal{O}_\downarrow$, with an appropriate $SU(2)$ symmetry, for spin 1/2). Although lattice TPS may appear as more useful in numerical simulations, working with their continuous analogs can be helpful in a few cases. In the FQHE in particular, and more generally in phases of matter with breaking of parity and time-reversal symmetry, one may encounter various issues by trying to search for lattice TPS. One reason for that is the following. For fermionic systems, the field theory/lattice model that one uses to contruct the trial wavefunction must be fermionic as well. If one wants a trial state which breaks parity and time-reversal symmetry, one is naturally lead to consider chiral fermions. It is a notoriously difficult problem to construct lattice models which give rise to chiral fermions \cite{doublingproblem}. Working directly in the continuum allows to circumvent such problems. There are plenty of well-known examples of field theories of chiral fermions. Such examples include, of course, some chiral CFTs. The use of conformal blocks as trial wavefunctions is thus natural in that context. The trial states for the FQHE that are given by conformal blocks may all be viewed as continuous TPS.

To conclude this section, we note that the results given in this paper may be reformulated in light of the connection between the wavefunctions given by conformal blocks and the larger class of (lattice or continuous) TPS. For all of these TPS, one can construct some wavefunctions for the ``edge excitations'' in a way similar to ours. These ``edge states'' are exactly the ones that appear in the Schmidt decomposition of the ground state. In general, there are two important questions that are intimately related: (i) whether the space of ``edge states'' is---in the thermodynamic limit---isometric to the auxiliary space, up to corrections that are local along the edge (and up to distinctions of a finite number of topological sectors, which by definition are always non local); (ii) whether the entanglement Hamiltonian is isospectral to an operator that is local along the cut (up to distinctions of topological sectors). As explained in this paper, if the property (i) holds, then (ii) follows from a simple use of the Baker-Campbell-Hausdorff formula. A natural conjecture is that the properties (i) and (ii) hold as long as the connected correlations of local operators are all short-range in the bulk. This is exactly what we showed in the present paper, for the FQHE trial wavefunctions that are given by conformal blocks. In the lattice TPS context, we note the existence of a recent work along these lines, where the conjectured property (ii) is stated explicitely (Ref. \cite{PoilblancCirac}), and where numerical evidence is provided for the locality of the pseudo-Hamiltonian in a number of specific two-dimensional lattice models.

\section{Conclusion}
\label{sec:conclusion}
In summary, we have studied the trial states for the FQHE that are given by conformal blocks. We have constructed the wavefunctions for to the gapless edge excitations that are associated to those. We define those directly, relying on the conformal block structure: to each CFT state $\left| v\right>$, we have attached a physical states $\left|\left. \Psi_{\left< v\right|} \right>\right>$. 
We did not address (physical) special Hamiltonians, in contrast with most previous approaches. In all known cases, however, and in particular for Laughlin and the MR Pfaffian state, these edge states agree with the ones constructed previously, for instance in \cite{WenEdge,milr}.

Then, we have studied the overlaps between these edge states, relying heavily on the assumption that all the connected correlations of local observables are short-range in the bulk, a property dubbed {\it generalized screening} after \cite{Read2009}. The techniques involved in the study of the overlaps were the ones of {\it boundary} CFT \cite{Cardy84,Cardy89}. We used a generalized plasma analogy to compute these overlaps. The basic idea can be summarized as follows: the chiral CFT that underlies the trial wavefunctions lives in the complex plane $\mathbb{C}$. The complex conjugate of the wavefunctions are obtained from an anti-chiral copy $\overline{\rm CFT}$. In the thermodynamic limit, there is a simply connected region $\Omega \subset \mathcal{C}$ filled by the $N$ particles: the droplet. Inside the droplet, generalized screening means that the non-chiral theory ${\rm CFT} \otimes \overline{\rm CFT}$ is perturbed by a local operator $a(z) \otimes \overline{a}(\overline{z})$, and flows towards a massive fixed point under RG. Then we are left with a massive field theory inside the region $\Omega$, and a massless theory ${\rm CFT} \otimes \overline{\rm CFT}$ in $\mathbb{C} \setminus \Omega$. At the boundary of $\Omega$, the massless theory obeys a {\it conformal boundary condition}. We have determined this conformal boundary condition ($\overline{a}(\overline{z}) \propto a^\dagger(z)$ at the boundary), and shown how it implies that the (antilinear) mapping from the CFT space to the physical space $\left|v \right> \mapsto \left|\left. \Psi_{\left< v\right|} \right>\right>$ becomes an {\it isometric isomorphism} in the thermodynamic limit $N\rightarrow \infty$. This is a precise bulk/boundary correspondence, for which we provided some elementary numerical checks (Monte-Carlo) in the case of the Laughlin and the Moore-Read (Pfaffian) states. In particular, one consequence of this isometric isomorphism between the CFT space and the physical space of edge excitations is that the CFT must be unitary. We have also shown that another aspect of this bulk/boundary correspondence holds as a simple consequence of this boundary CFT formalism: the equal-time correlators (evaluated in the ground state) of local operators along the edge are {\it equal to} the analogous correlators in the chiral CFT.

We then tackled the entanglement spectrum of the ground state for real-space partition. A key point was the use of a completeness relation in the CFT space, that leads to a natural decomposition of the ground state $\left|\left. \Psi \right>\right>$ as a sum of terms of the form $\left|\left. \Psi^A_{\left< v \right|} \right>\right> \otimes \left|\left. \Psi^B_{\left| v \right>} \right>\right>$. In the limit $N_A, N_B \rightarrow$, because of the isometric isomorphism, all these states are orthogonal, and therefore this natural decomposition is really a Schmidt decomposition. All the pseudo-energies are degenerate in that limit. More seriously, most of the interesting features of the ES are hidden in the subleading corrections. We sketched how such corrections to scaling can be naturally apprehended with RG arguments: in reality, the massless theory ${\rm CFT} \otimes \overline{\rm CFT}$ that lives in $\mathbb{C} \setminus \Omega$ is subject to a {\it perturbed} conformal boundary condition along the edge. Therefore, one needs to discuss what perturbations are allowed by the different symmetries to be present at the boundary. A careful analysis of these perturbations leads to a deformation of our isometric isomorphism, due to the corrections to scaling. Of course, the latter need to be taken into account in the orthogonalization of the states that appear in the Schmidt decomposition. These corrections to scaling are thus reflected in the entanglement spectrum. Our most significant result here is that the ES, as a consequence of generalized screening, is the spectrum of a sum of (integrals of) local operators along the cut between $A$ and $B$. We also showed that the ES for particle partition (PP) can be analyzed in a similar fashion.

Finally, we have emphasized that our results suggest relations between the formalism of wavefunctions given by conformal blocks and Matrix Product States (in 1d) or Tensor Product States (in higher dimension), noted MPS or TPS. We have argued that the continuous analog of a TPS (defined on a lattice) is nothing but the {\it correlator of local operators in a local field theory}. Working in the continuum rather on the lattice has a number of advantages, in particular when one is dealing with the ground state of a phase of matter with broken inversion/time-reversal symmetry, such as the FQHE. In that case, assuming that good trial wavefunctions can be built as correlators in a local field theory, one is naturally lead to look for {\it chiral field theories}, and correlators of local fields in those. A good illustration of this is the case of complex paired superfluids, for which the BCS ground state can always be viewed as a correlator in some chiral free fermion theory. Looking for a {\it lattice} TPS approximation of the ground state is then equivalent to constructing a lattice discretization of a chiral field theory. Then one runs into trouble, as various problems arise, in particular the fermion doubling problem \cite{doublingproblem}. Starting from the continuum, namely from a field theory, to construct a trial wavefunction, is a natural way of circumventing such issues. This justifies, at least heuristically, the approach pioneered by MR, namely the use of conformal blocks as trial wavefunctions for the FQHE.




\acknowledgments
We thank A. Bernevig, P. Bonderson, B. Estienne, A. Gaynutdinov, V. Pasquier, N. Regnault, and J.-M. St\'ephan for useful discussions. We thank P. Wiegmann for sharing his unpublished work and for pointing out the existence of the missing term in \cite{ZabrodinWiegmann3}. JD and NR thank the Institut Henri Poincar\'e in Paris, where part of this work was done,
for kind hospitality during the workshop ``Advanced conformal field theory and applications''. JD also thanks the Galileo Galilei Institute in Florence and the INFN for support during the workshop ``New quantum states of matter in and out of equilibrium'', where this work was completed.

This work was supported by a Yale Postdoctoral Prize Fellowship (JD), by NSF grant no.\ DMR-1005895 (NR), and by DOE grant no.\ DE-SC0002140 (EHR).

\vspace{0.2cm}
{\it Note added}: shortly after the submission of this paper, the recent work \cite{ZaletelMong} appeared, which is based on a point of view that is identical to the one discussed in part \ref{sec:MPS}, namely that wavefunctions given by the MR construction can be regarded as infinite-dimensional MPS. In this work, a numerical scheme is set up, based on a truncation of the CFT Hilbert space, that allows to compute some observables in the MR (Pfaffian) and Laughlin states, and could possibly be extended to other states.


\appendix

\section{The topological term $\gamma$ appears in all the pseudo-energies}
\label{appendix:topoterm}
In this appendix we give one argument to support the fact that the topological entanglement entropy $\gamma$ appears in all the pseudo-energies in the scaling region (see section \ref{sec:scaling_region}). It is adapted from the argument sketched by Kitaev and Preskill at the end of Ref. \cite{KitaevPreskill}.

We have shown in this paper that the ES (for RSP) is, up to a constant shift, given by the spectrum of $\frac{v}{R} L_0 + o(\sqrt{\rho_0}^{-1}/R)$. The constant shift is fixed by the requirement that ${\rm tr\,}\rho_A = 1$. To compute this shift, let us start by dropping the terms of order $o(\sqrt{\rho_0}^{-1}/R)$. Within this approximation, the reduced density matrix $\rho_A$ is isospectral to
\begin{equation}
	\frac{1}{\mathcal{Z}} e^{- \frac{v}{R} L_0}
\end{equation}
acting on a module $V$ over the chiral algebra $\mathcal{A}$; usually the module of the identity. The normalization factor is
\begin{equation}
	\mathcal{Z} =  q^{\frac{c}{24}} \times {\rm tr}_V \,  q^{L_0- \frac{c}{24}}
\end{equation}
with $q= \exp \left({-2\pi \frac{v}{2\pi R}}\right)$. The central charge of the CFT is noted $c$. The object in the right-hand side is called a {\it character} in the CFT literature \cite{BYB,SaleurItzyksonZuber,Mussardo}.

The constant shift we are looking for is the order $O(1)$ piece in the expansion of $\log \mathcal{Z} = \alpha 2 \pi  R - \gamma + o(1/R)$. We thus need to obtain the asymptotic behaviour of $\log \mathcal{Z}$ when $R \gg v \sim \sqrt{\rho_0}^{-1}$, namely when $q \rightarrow 1$. At first sight, this may not look obvious, because the sum diverges in that limit, however it is a very standard exercise in CFT. The trick is to use the modular properties of the characters, namely to express the character $\chi(q) = {\rm tr}_V \,  q^{L_0- \frac{c}{24}}$ as a function of $\tilde{q} = e^{- 2\pi \frac{2\pi R}{v}}$, and then take the limit $\tilde{q} \rightarrow 0$, which is well-defined. In general, the characters obey the following rule
\begin{equation}
	\chi_a(q) =  \sum_{b}  \mathcal{S}_{a}^{b}  ~\chi_{b}(\tilde{q})
\end{equation}
where $a$ and $b$ label the different irreducible modules over $\mathcal{A}$, and $\mathcal{S}_{a}^{b}$ is the so-called {\it modular $\mathcal{S}$-matrix} (note that this is not the same object as the shift operator $\mathcal{S}$ is section \ref{sec:screening_bc}). In the case when $V$ is the module of the identity, the leading contribution comes from the character of the identity module itself, noted $\chi_1$ (and the corresponding diagonal element of the modular $\mathcal{S}$-matrix noted $\mathcal{S}_{1}^1$)
\begin{eqnarray}
	\log \chi_1(q) & \underset{q \rightarrow 1}{\sim} & \log \left[ \sum_{a} \mathcal{S}_{1}^{a} ~ \chi_{a}(\tilde{q}) \right] \\
 & \sim & \log \left[ \mathcal{S}_1^1 ~\chi_{1} (\tilde {q}) \right] \\
 & \sim & -\frac{c}{24} \log \tilde{q} + \log \mathcal{S}_1^1  . 
\end{eqnarray}
Since $\log \mathcal{Z} = \log \chi_1(q) + \frac{c}{24}\log q$, we find in the end
\begin{equation}
	\log \mathcal{Z}\, =\, \alpha 2\pi R - \gamma + o(v/R)
\end{equation}
with $\gamma  = - \log \mathcal{S}_1^1$ as expected. The coefficient $\alpha$ is not universal, as it involves the inverse length $1/v$. The argument given here, based on modular manipulations, can be easily extended to the case when anyonic excitations are present in parts $A$ and $B$. More details can be found in \cite{KitaevPreskill, FFN_holo}.

We do not address the robustness of $\gamma$ here. Since it is a dimensioneless constant, it cannot be affected by irrelevant perturbations of the CFT. One can make some sanity checks in a few cases, for instance for a CFT with both a $U(1)$ sector and a statistics sector, which come with different velocities $v_{U(2)}$ and $v_{\psi}$. One finds that $\gamma$ is invariant indeed.

In conclusion, the topological term $\gamma$ must appear as the order $O(1)$ piece in {\it all the low-lying pseudo-energies in the ES}. Because of this, it also appears in the von Neumann entanglement entropy, but this is not specific to this entropy. It is a property of all the pseudo-energies independently, rather than a property specific to the von Neumann entropy only. In particular, one consequence of the presence of the term $\gamma$ in all the low-lying pseudo-energies is that all the Renyi entropies contain this term $\gamma$ as well. This fact was noted already in \cite{WenRenyi}.

\section{The exactly solvable case of $p_x \pm i p_y$ paired superfluids}
In this appendix we discuss the case of a $p_x - i p_y$ superfluid treated in BCS theory \cite{RG}, where the overlaps between the edge states and the real-space entanglement spectrum can be computed exactly. We reformulate some of the results of \cite{DR} in the language of the present paper, to illustrate the ideas developed in sections \ref{sec:blocks}, \ref{sec:screening} and \ref{sec:ES}.

For convenience, we consider the system on the cylinder $\mathcal{C}$. 
The BCS ground state of spinless particles is expressed in terms of the pairing function $g(w_1,w_2)$
\begin{eqnarray}
	\label{eq:BCSgs}
	&& \left|\left. \Psi \right>\right> \, = \, \frac{1}{\sqrt{Z}} \times \\
\nonumber	&& \exp \left( \frac{1}{2} \int_{\mathcal{C}} d^2 w_1 \int_{\mathcal{C}}d^2 w_2 ~ g(w_1,w_2)~ c^\dagger(w_1) c^\dagger(w_2) \right) \left|\left. 0 \right>\right>
\end{eqnarray}
where $w = x + i y$ is the complex coordinate on the cylinder ($y$ and $y+L$ are identified). For a weakly paired $p_x-i p_y$ superfluid, the pairing function expressed in momentum space behaves as
\begin{equation}
	g({\bf k}) \, \propto \, \frac{1}{k_x- i k_y}
\end{equation}
as $|{\bf k}| \rightarrow 0$ (see \cite{RG}). Coming back to real space, on the cylinder $\mathcal{C}$, the pairing function is
\begin{equation}
	g(w_1,w_2) \, = \, \frac{\mu/2\pi}{\frac{L}{2 \pi} \sinh \frac{2 \pi (w_1-w_2)}{L}}
\end{equation}
where $\mu$ is a complex parameter with the dimension of an inverse length, which is related to the density of particles. Without loss of generality, one can assume that $\mu$ is real and positive. For this particular choice of the pairing function, the BCS ground state (\ref{eq:BCSgs}) can be rewritten as
\begin{equation}
	\left| \left. \Psi \right>\right> \, = \, \frac{1}{\sqrt{Z}} \left< \exp \left( \sqrt{\frac{\mu}{2\pi}} \int_\mathcal{C} d^2w ~\psi(w)\otimes c^\dagger(w) \right) \right>_{\mathcal{C}} \left| \left. 0\right>\right>
\end{equation}
where $\psi(w)$ is a free Majorana field on the cylinder with propagator $\left< \psi(w_1) \psi(w_2) \right>_{\mathcal{C}} \, = \, 1/(\frac{L}{2 \pi} \sinh \frac{2 \pi (w_1-w_2)}{L})$. The chiral correlators $\left< \psi(w_1) \dots \psi(w_N) \right>_{\mathcal{C}}$ are evaluated using Wick's theorem. Introducing an anti-chiral copy $\overline{\psi}(\overline{w})$ of the Majorana fermion field, the normalization factor $Z$ can be expressed as a correlator in the non-chiral theory ${\rm Majorana} \otimes \overline{{\rm Majorana}}$
\begin{eqnarray}
\nonumber	Z & = &  \sum_{n \geq 0} \frac{\left(\frac{\mu}{2\pi}\right)^n}{(n!)^2} \int \prod_{i=1}^n d^2w_i d^2w'_i \\
\nonumber	&& \qquad  \left< \overline{\psi}(\overline{w_1})\dots \overline{\psi}(\overline{w_n}) \psi(w'_1) \dots \psi(w'_n) \right>   \\
\nonumber	&& \qquad \times \quad \left< \left<\left. c(z_n)\dots c(z_1) c^\dagger(z_1') \dots c^\dagger(z_n')  \right>\right>\right.	\\
 & = & \left< e^{\frac{i\mu}{2\pi} \int d^2 w ~ \overline{\psi(w)} \psi(w)} \right>_{\mathcal{C}} .
\end{eqnarray}
The factor $i$ in the last exponential appears because of the convention that the holomorphic/anti-holomorphic modes to anticommute $\{ \psi , \overline{\psi} \} =0$. We also use the fact that only the correlators with an {\it even} number of fields $\psi(z_i)$ are non-zero. The term in the last exponential should be viewed as a mass term in the Majorana fermion theory, with the standard quadratic euclidean action
\begin{equation}
	S \, = \,\frac{1}{2\pi} \int d^2 w \,\left[ \psi \partial_{\overline{w}} \psi + \overline \psi \partial_w \overline{\psi} + i\mu ~ \psi \overline{\psi} \right] \, .
\end{equation}

\subsubsection{Edge states}
We now focus on the $p_x+i p_y$ superfluid filling only the left half-cylinder $\mathcal{C}_l$ corresponding to $x<0$. A trial wavefunction for this system is
obtained by restricting the integration in the expression (\ref{eq:BCSgs}) to the domain $\mathcal{C}_l$,
\begin{equation}
	\label{eq:BCS_half}
	\left|\left. \Psi, \mathcal{C}_l \right>\right> \, =\, \frac{1}{\sqrt{Z_l}} \left< e^{\sqrt{\frac{\mu}{2\pi}} \int_{\mathcal{C}_l} d^2w ~ \psi(w) \otimes c^\dagger(w)} \right>_{\mathcal{C}} \left| \left. 0\right>\right> .
\end{equation}
Note that the chiral correlator is still evaluated on the full cylinder, but the particles fill the left half-cylinder only. The normalization factor then corresponds to the partition function of the Majorana fermion field on the cylinder, with a mass term which is switched on in the left side only
\begin{equation}  	
	S_l \, = \, \frac{1}{2\pi}\int_\mathcal{C} d^2 w \left[ \psi \partial_{\overline{w}} \psi + \overline \psi \partial_w \overline{\psi} \right] \, + \,  \frac{i\mu}{2\pi}  \int_{\mathcal{C}_l} d^2 w \, \psi \overline{\psi}  \; .
\end{equation}
From now on we make use of the translation invariance of the system in the $y$-direction, by using a $k_y$-momentum basis. On the cylinder the field $\psi(w)$ can be expanded in Fourier modes
\begin{equation}
	\psi(x,y) \, =\, \sqrt{\frac{2\pi}{L}} \sum_{k_y} e^{k_y (x+i y)} \psi_{-k_y}
\end{equation}
where the sum runs over all the admissible momenta, namely $k_y \in \frac{2 \pi}{L}(\mathbb{Z}+\frac{1}{2})$ for anti-periodic boundary conditions. The Fourier modes obey the canonical commutation relations $\{\psi_{k_y}, \psi_{{k'_y}}\} = \delta_{k_y+k_y',0}$. The ground state trial wavefunction (\ref{eq:BCS_half}) is then
\begin{eqnarray}
	&& \left| \left. \Psi, \mathcal{C}_l \right>\right> \, = \, \frac{1}{\sqrt{Z_l}} \times \\  \nonumber &&  \left< 0 \right| \mathcal{T}~ \exp \left(\displaystyle \sqrt{ \mu} \sum_{k_y} \int_{-\infty}^0 dx ~ e^{k_y x}~ \psi_{-k_y} \otimes c^\dagger_{k_y}(x)\right) \left|0\right> \,\left| \left. 0 \right>\right>
\end{eqnarray}
where the symbol $\mathcal{T}$ denotes time-ordering for the imaginary time $x$. The state $\left| 0\right>$ is the vacuum of the CFT, annihilated by all the mode $\psi_{k_y}$ with $k_y>0$ (note the difference with the physical fermionic vacuum $\left|\left. 0 \right>\right>$ annihilated by the $c(w)$'s). The time-ordering operator $\mathcal{T}$ implies that the Cooper pairs of particles with momenta $+k_{y}$ and $-k_y$ must appear in a specific order, as discussed in \cite{DR}.

As explained in section \ref{sec:edge}, one can write down trial states for the edge excitations by replacing the out vacuum $\left<0\right|$ by an excited state in the CFT, such as $\left<0 \right| \psi_{k_1} \psi_{k_2}$ for some $k_1,k_2>0$
\begin{eqnarray}
	&& \left| \left. \Psi_{\left<0\right| \psi_{k_1} \psi_{k_2}} \right>\right> \, = \, \frac{1}{\sqrt{Z_l}}  \times \\  \nonumber && \left< 0 \right| \psi_{k_1} \psi_{k_2}~ \mathcal{T}~ e^{  \sqrt{ \mu} \sum_{k_y} \int_{-\infty}^0 dx ~e^{k_y x}~ \psi_{-k_y} \otimes c^\dagger_{k_y}(x)} \left|0\right> \,\left| \left. 0 \right>\right>
\end{eqnarray}
The study of the overlaps between these states is equivalent to the one of correlators of of the fields $\psi$ and $\overline{\psi}$ in the right half-cylinder $\mathcal{C}_r$ ($x>0$), in the presence of the massive phase in the left half-cylinder $\mathcal{C}_l$ ($x<0$). In the limit $\mu \rightarrow \infty$, one is left with a conformal boundary condition at $x=0$ for the massless Majorana fields in $\mathcal{C}_r$
\begin{equation}
	\label{eq:bc_pip}
	\psi(0,y) \, =\, \overline{\psi}(0,y)
\end{equation}
which gives, in terms of the Fourier modes acting on the {\it boundary state} $\left| B \right>$
\begin{equation}
	\psi_{k_y} \left| B \right> \, =\, \overline{\psi}_{-k_y} \left| B\right>  \, .
\end{equation}
The boundary state $\left|B \right>$ can easily be written in term of the fermion modes
\begin{equation}
	\left|B \right> \, = \, e^{\sum_{k_y >0} \psi_{-k_y} \overline{\psi}_{-k_y}} \left| 0 \right> \overline{\left| 0\right>} .
\end{equation}
As explained in section \ref{sec:edge_conjecture}, this implies that the overlaps in the $\mu \rightarrow \infty$ limit are given by
\begin{eqnarray}
	\label{eq:edge_conjecture_pip}
 &&	\left< \left< \, \Psi_{\left< 0 \right| \psi_{k_1} \dots \psi_{k_p}} \,\left| \, \Psi_{\left<0\right|\psi_{k'_1}\dots \psi_{k'_p}} \,  \right>\right>\right. \\
\nonumber &&  = \, \left[ \overline{\left<0\right|\psi_{k_1}\dots \psi_{k_p}} ~\left<0\right|\psi_{k'_1}\dots \psi_{k'_p} \right] \left| B \right> \\
\nonumber &&= \, \left< 0 \right| \psi_{k'_1}\dots \psi_{k'_p} \psi_{k_p} \dots \psi_{k_1} \left| 0 \right> \, .
\end{eqnarray}
We have arrived at this result thanks to the fact that the massive phase of the theory ${\rm Majorana} \otimes \overline{{\rm Majorana}}$ in the domain $\mathcal{C}_l$ can be reformulated as the boundary condition (\ref{eq:bc_pip}). This is a particular case of the general argument of section \ref{sec:edge_conjecture}, which leads to the result (\ref{eq:edge_conjecture}).

Since the $p_x +i p_y$-paired superfluid wavefunction (\ref{eq:BCSgs}) is Gaussian, one can actually compute the overlaps (\ref{eq:edge_conjecture_pip}) exactly for any $\mu$, not only in the limit $\mu \rightarrow \infty$. This can be done as follows. For some given $k_y, k_y' >0$, we have
\begin{eqnarray}
	\label{eq:overlaps_pip_1}
\nonumber && \left<\left<  \Psi_{\left<0 \right| \psi_{k_y}}  \left| \Psi_{\left< 0 \right| \psi_{k_y'}}  \right>\right>\right. \\
\nonumber	&& =\, \left<\left< \Psi, \mathcal{C}_l \right.\right| \sqrt{\mu} \int_{-\infty}^0 dx~e^{k_y x}~c_{k_y}(x) \\
\nonumber	&& \qquad \times \; \sqrt{\mu} \int_{-\infty}^0 dx'~e^{k_y' x'}~c^\dagger_{k_y'}(x') \left|\left. \Psi,\mathcal{C}_l \right>\right>   \\ 
&& = \;  e^{-2 S_\mu(k_y)} ~\delta_{k_y,k_y'}
\end{eqnarray}
where
\begin{equation}
	S_\mu(k_y) \, = \, \log \left( \sqrt{1+\left(k_y/\mu\right)^2} + k_y/\mu \right)  .
\end{equation}
Here we have used a result of Ref. \cite{DR} (see formula (13) and (14) in that reference).
The case with more fermionic excitations follows simply from Wick's theorem: in general, the overlaps at finite $\mu$ are
\begin{eqnarray}
	\label{eq:overlaps_pip_2}
\nonumber &&	\left< \left< \, \Psi_{\left< 0 \right| \psi_{k_1} \dots \psi_{k_p}} \,\left| \, \Psi_{\left<0\right|\psi_{k'_1}\dots \psi_{k'_p}} \,  \right>\right>\right. \\ 
&& \qquad = \, \left< 0 \right| \psi_{k'_1}\dots \psi_{k'_p} e^{-S_b(\mu)} \psi_{k_p} \dots \psi_{k_1} \left| 0 \right> \; . \quad
\end{eqnarray}
where we have defined the boundary perturbation as
\begin{equation}
	\label{eq:Sb_pip}
	S_b(\mu) \, = \, \sum_{k_y>0} S_\mu(k_y) \psi_{k_y} \psi_{-k_y}
\end{equation}
Of course, in the limit $\mu \rightarrow \infty$, the boundary perturbation goes to zero, and we recover the formula (\ref{eq:overlaps_pip_1}).

\subsubsection{Real-space entanglement spectrum}
The real-space entanglement spectrum on the cylinder (where $A$ is the left half-cylinder $\mathcal{C}_l$ and $B$ is the right half-cylinder $\mathcal{C}_r$) can also be computed exactly, as pointed out in \cite{DR}. Because the left and right half-cylinders are symmetric here (such a symmetry was not assumed in Part \ref{sec:ES}), the boundary perturbations for part $A$ and part $B$ are both equal to $S_b(A)=S_b(B)=S_b(\mu)$ defined in (\ref{eq:Sb_pip}). Following the discussion in section \ref{sec:ES}, and in particular the formula (\ref{eq:eSBeSA}), we define
\begin{equation}
	e^{-\frac{S_{\rm ES}}{2}} \, = \, e^{-\frac{S_b(\mu)}{2}} e^{-\frac{S_b(\mu)}{2}} \, = \, e^{-S_b(\mu)}
\end{equation}
and the entanglement spectrum is directly given (up to an additive constant shift) by the spectrum of the operator $S_{\rm ES} = 2 S_b(\mu)$, which is a free fermion spectrum, generated by the set of single-particle pseudo-energy levels
$2 S_{\mu}(k_y)$ for the admissible $k_y>0$. The constant shift can be computed easily using this fact. For example, if we focus on the smallest pseudo-energy $\xi_0$, corresponding to the cut ground-states in $A$ and $B$ (that is, without any insertion of a $\psi_{k_y}$ mode), then we have the exact formula
\begin{equation}
	e^{-\xi_0/2} \, = \, \prod_{k_y>0} \frac{1}{1+e^{-2 S_\mu(k_y)}}
\end{equation}
where the product runs over all the admissible $k_y>0$. Then $\xi_0$ is given by the Euler-MacLaurin formula as $L \rightarrow \infty$
\begin{equation}
	\xi_0 \, = \, \alpha ~L - \gamma +\dots
\end{equation}
where the coefficient $\alpha$ is the integral $\int_0^\infty d k_y 2 \log (1+e^{-2 S_\mu(k_y)})$. 
Here the topological term is actually $\gamma=0$, in apparent contradiction with the claim made in section \ref{sec:ES} that this
term is the topological entanglement entropy. However, the topological entanglement entropy itself can be computed exactly by the same method, using a similar Euler-MacLaurin expansion, and the result would also be $0$, again in contradiction with the expected result $\gamma = \log 2$. This comes from the fact that we have not restricted the parity of the particle number in part $A$ (or equivalently $B$). The topological term can only appear in that case, and it does if we restrict the number of particles in $A$ to be, say, even (see also the discussion in \cite{DR}). Thus the result is indeed in agreement with the general claim in section \ref{sec:ES}. The set of pseudo-energies is then given by
\begin{equation}
	\xi_{k_1,k_2,\dots,k_p} \, = \, \xi_0 \,+\, \sum_{j=1}^p 2 S_\mu(k_j) + \dots 
\end{equation}
where $p$ is assumed to be even, and $\xi_0 \sim \alpha~L - \log 2 +\dots$. The same result holds if we restrict
the particle number in $A$ to be odd.

Finally, to conclude this appendix, we note that, by expanding the formula (\ref{eq:Sb_pip}) in powers of $k_y/\mu$, and going back to real space, the operator $S_{\rm ES}$ is indeed the integral along the cut between $A$ and $B$ of some local operators in the CFT. As expected from the discussion in section \ref{sec:ES}, the first term in $S_{\rm ES}$ is nothing but the stress-tensor itself
\begin{equation}
	S_{\rm ES} \, = \, \frac{2}{\mu} \int_0^L \frac{dy}{2\pi}~ T(0,y) \, + \, \dots
\end{equation}
and the length $1/\mu$, which is the only microscopic length scale in this model, appears as a coefficient. It is the extrapolation length discussed in the main text (up to some numerical factor).


\end{document}